\documentclass{aastex}          %% The manuscript based on AASTeX v5.x
\usepackage{spr-astr-addons}    %% mimicing ASTR journal style
\usepackage{url}
\usepackage{graphicx}	% Including figure files

\newcommand{\const}{\mathop{\text{const}}\nolimits}
\newcommand{\FAP}{{\text{FAP}}}
\newcommand{\sampleid}[1]{\texttt{#1}}

\begin{document}
\title{Fine-resolution analysis of exoplanetary distributions by wavelets: hints of an
overshooting iceline accumulation}

\shorttitle{Exoplanetary Distributions by Wavelets}
\shortauthors{R.V.~Baluev and V.Sh.~Shaidulin}

\author{Roman V. Baluev}%\altaffilmark{1,2}}
\affil{Central Astronomical Observatory at Pulkovo of Russian Academy of Sciences, Pulkovskoje
shosse 65/1, Saint Petersburg 196140, Russia}
\affil{Saint Petersburg State University, Faculty of Mathematics and Mechanics,
Universitetskij prospekt 28, Petrodvorets, Saint Petersburg 198504, Russia}
\email{r.baluev@spbu.ru} %% non-output
\and
\author{Vakhit Sh. Shaidulin}%\altaffilmark{2}}
\affil{Saint Petersburg State University, Faculty of Mathematics and Mechanics,
Universitetskij prospekt 28, Petrodvorets, Saint Petersburg 198504, Russia}

%\altaffiltext{1}{}
%\altaffiltext{2}{}
%\altaffiltext{3}{}

\begin{abstract}
We investigate 1D exoplanetary distributions using a novel analysis algorithm based on the
continuous wavelet transform. The analysis pipeline includes an estimation of the wavelet
transform of the probability density function (p.d.f.) without pre-binning, use of
optimized wavelets, a rigorous significance testing of the patterns revealed in the p.d.f.,
and an optimized minimum-noise reconstruction of the p.d.f. via matching pursuit
iterations.

In the distribution of orbital periods, $P$, our analysis revealed a narrow subfamily of
exoplanets within the broad family of ``warm jupiters'', or massive giants with $P\gtrsim
300$~d, which are often deemed to be related with the iceline accumulation in a
protoplanetary disk. We detected a p.d.f. pattern that represents an upturn followed by an
overshooting peak spanning $P\sim 300-600$~d, right beyond the ``period valley''. It is
separated from the other planets by p.d.f. concavities from both sides. It has at least
two-sigma significance.

In the distribution of planet radii, $R$, and using the California Kepler Survey sample
properly cleaned, we confirm the hints of a bimodality with two peaks about $R=1.3
R_\oplus$ and $R=2.4 R_\oplus$, and the ``evaporation valley'' between them. However, we
obtain just a modest significance for this pattern, two-sigma only at the best. Besides,
our follow-up application of the Hartigan~\&~Hartigan dip test for unimodality returns $3$
per cent false alarm probability (merely $2.2$-sigma significance), contrary to $0.14$ per
cent (or $3.2$-sigma), as claimed by Fulton~et~al.~(2017).
\end{abstract}

\keywords{methods: data analysis - methods: statistical - astronomical data bases: miscellaneous -
planetary systems - stars: statistics}

\section{Introduction}
A great diversity of extrasolar planets was discovered over the last two decades. After the
detection of the first ``hot jupiter'' orbiting 51 Pegasi \citep{MayorQueloz95}, the number
of known exoplanets grew with an increasing rate, and presently, according to \emph{The
Extrasolar Planets Encyclopaedia} \citep{Schneider11}, it exceeds the milestone of $3000$.

Such a number of objects represents a huge interest from the statistical point of view. The
exoplanetary statistics is capable to provide a lot of invaluable information, in
particular with a concern to planet formation and evolution theories \citep{CummingStat}.

There are several recognized statistical features in exoplanetary distributions that were
known long ago. For example, in the orbital periods distribution the following large-scale
patterns appear: (i) the subfamily of ``hot jupiters'' (HJ hereafter) at the short-period
edge of the distribution, (ii) the ``period valley'' (PV) that represents a wide depression
in the period range $P\sim 10-300$~d, (ii) the dominant maximum of ``warm jupiters'' (WJ)
with $P\gtrsim 300$~d. The latter one also incorporate more ``cold'' long-period objects
like Jupiter in the Solar System. These statistical patterns were more or less succesfully
reproduced in planet formation and migration models \citep{IdaLin04,IdaLin08}.

In particular, the WJ family is usually treated as an evidence in favour of the effect of
the iceline barrier that stimulates the planet formation near the frost edge, where the
water starts to condensate in ice in a protoplanetary disk. In the Solar protoplanetary
disk, the ice line was situated about $\sim 2.7$~AU from the Sun \citep{Hayashi81}, and
this would correspond to the Keplerian orbital period of $P\sim 1600$~d. This appears much
larger than the value of $P\sim 300$~d mentioned above, but giant planets, which are
predominant in that statistics, additionally undergo the type-II migration that brings
their final position inward relatively to the actual iceline \citep{IdaLin08,Schlaufman09}.

In multiple works many other apparent statistical features in the exoplanetary ensemble are
noticed sometimes. However, they often lack a good justification in terms of the
statistical significance. A particular goal of this paper is to advocate (and demonstrate)
that regardless of the large number of exoplanetary candidates already available, their
statistical analysis still need to be considerably more careful and conservative than
usually assumed in the literature.

Indeed, it is not easy to supply a formal statistical support (like e.g. a p-value) in
favour of e.g. a hypothesis that there is a submafily of objects with given
characteristics. The issue appears even before any analysis, since it is unclear what is a
``subfamily'' from the formal point of view. The analysis would likely require
sophisticated mathematics, something like cluster analysis with a rich statistical content.

Due to this difficulty, only the large-scale structures remain undoubtful in the
exoplanetary statistics, like those described above.

In this work, we set a goal to perform a \emph{fine-resolution} analysis of exoplanetary
distributions. That is, we seek to detect small-scale features in their statistics. For
this, we apply our newly-developed wavelet analysis algorithm \citep{Baluev18a}. It was
designed to detect patterns belonging to the following basic classes: quick density
gradients, convexities and concavities of the probability density function (p.d.f.). In a
less formal wording, the method would be helpful in revealing transitions between
statistical families, these families themselves, and separations between.

Basically, this paper also represents the first effort to perform a field-test of that
analysis algorithm. Our goal is to perform a systematic investigation of different
exoplanetary distributions and to decide in a uniform way, what p.d.f. details are
significant there and which are not.

The structure of the paper is as follows. In Sect.~\ref{sec_hack} we discuss some important
issues that decrease reliability of any statistical analysis, and how to treat them. In
Sect.~\ref{sec_alg} we provide a brief description and justification of the algorithm. The
Sect.~\ref{sec_testing} contains example ``teaching'' application of the method to
simulated datasets. In Sect.~\ref{sec_samples} we described how we constructed exoplanetary
samples to be analysed. Sect.~\ref{sec_res} presents results of the analysis for different
exoplanetary characteristics. In Sect.~\ref{sec_sig} we discuss some further pitfalls of
the correct significance determination, in view of the presented results.

\section{Dangers of data-dependent analysis, multiple testing, and hidden p-hacking}
\label{sec_hack}
Mathematical statistics provides us with great tools aimed to make scientific results more
trustable. This is achieved by filtering out insignificant conclusions likely inspired by
noise. However, in practice statistical methods may be used too formally or even wrongly,
violating their necessary prerequisite conditions of applicability. Such a violation is not
necessarily a result of a negligent research, because it is often remains implicitly
hidden.

In such a case, the results of the analysis would become statistically invalid, even though
still pretty convincing for a quick external view. After new data are acquired, previous
interpretations, even those that deemed significant, are often disavowed or transformed
drastically on the basis of new data. As a result, either the given statistical method, or
even the whole statistics itself, become undeservedly less trustable in view of other
researchers.

This phenomenon of the ``untrusted significance'' and its roots were detailedly reviewed by
\citet{GelmanLoken14}. They also give several practical examples (mainly from the medicine)
of how statistical tests can be misapplied. We advocate here that astronomy and
astrophysics are unfortunately not free from this negative phenomenon. This appears
especially important when performing analysis of various distributions, in particular the
distributions of exoplanets too.

According to \citet{GelmanLoken14} mistakes come from hidden degrees of freedom that are
processed either subconsciously or in some pre-analysis or pre-selection procedure. They
call it the ``data-dependent analysis''.

Rethinkering their argumentation and adapting it to our tasks here, it is likely that the
following would be a bad practice:
\begin{enumerate}
\item Use parametrized model in a test that was designed to handle either non-parametric
models or models with a smaller number of parameters. This includes cases when additional
parameters are estimated ``on the fly'' or ``subconsciously'', and left out of the formal
statistical test scheme.
\item Formulate comparison hypotheses or models \emph{post-hoc}, i.e. based on the empiric
data, and then process these same data with a test that remains ``unaware'' of such a
preanalysis (i.e., it was not designed to handle these specific conditions).
\item Perform a large number of similar tests, varying some parameters. In particular,
to vary them over a grid, since this is an implicit version of point (1).
\item Select a very specific test only to verify a similarly specific hypothesis emerged
after some pre-analysis or after just a visual inspection of the same data. Such a practice
might looks quite innocent at a glance, but it turns out to be a version of point (2).
\item Start the analysis without clear understanding of what question we ask to the data
and without predefining a clear set of possible hypotheses or models. It is admissible if
in the beginning of some new research we want to detect just ``something interesting'', but
then this ``something interesting'' must be more or less clearly formalized and the set of
possible outcomes should be clearly outlined (though it must remain wide). This need to be
done in advance of the analysis, or it would violate point (2).
\item Assume that a low false alarm probability guarantees the same low fraction of false
alarms in our output results (even if the test itself was applied correctly). The problem
is that usually only ``positive'' results (or ``alarms'') are published, and all other
data, that appeared not so attractive, remain hidden. E.g., if $N=100$ stars were
investigated, and $n=10$ of them were found to have a planet with a false alarm probability
of $\FAP=1$ per cent, this $1$ per cent refers to the \emph{entire} sample of $N$ objects.
In total we would expect about $N\,\FAP=1$ false positives, but this is already $10$ per
cent of the output list of $n$ candidates. This fraction is obviously larger than requested
originally ($\FAP=0.01$).
\end{enumerate}

We summarize our understanding of how to avoid unreliable statistical results via a
citation from \citep{GelmanLoken14}: ``For a p-value to be interpreted as evidence, it
requires a strong claim that the same analysis would have been performed had the data been
different.'' And what should be treated as a bad practice, is ``data-dependent analysis
choices that did not appear to be degrees of freedom because the researchers analyze only
one data set at a time.''

To finish this section, we note that although all the above discussion relied on such a
purely frequentist notion like p-value (or false alarm probability), Bayesian methods are
equally sensitive to these issues. The main difference between the Bayesian and frequentist
methods is in the metric that they utilize to order statistical decisions. But issues
listed above owe to a wrongly formulated statistical model that hiddenly involved more
degrees of freedom and more uncertainty than an analysis method was designed to handle.

\section{Description of the algorithm}
\label{sec_alg}
In \citep{Baluev18a}, a self-consistent pipeline of the statistical wavelet analysis is
described. In view of what is said in Sect.~\ref{sec_hack} concerning the danger of
data-dependent analysis and hidden degrees of freedom, we need to highlight some andvantages
of this pipeline.

Just like the most of other well-formulated statistical tools, our method is supposed to be
applied \emph{blindly}, i.e. without any preanalysis of the input data. But even if some
preanalysis was done, our algorithm scans all possible structures anyway, allowing for a
very wide set of their characteristics. In other words, its set of test hypotheses is
inherently wide, and relatively little space is left for hidden degrees of freedom. In some
sense, wavelets offer one possible formalization of a vague task ``to find something
interesting'' in the distribution, avoiding too narrow specification of this ``something
interesting''.

\begin{figure*}
\begin{tabular}{@{}l@{}l@{}}
\includegraphics[width=\linewidth]{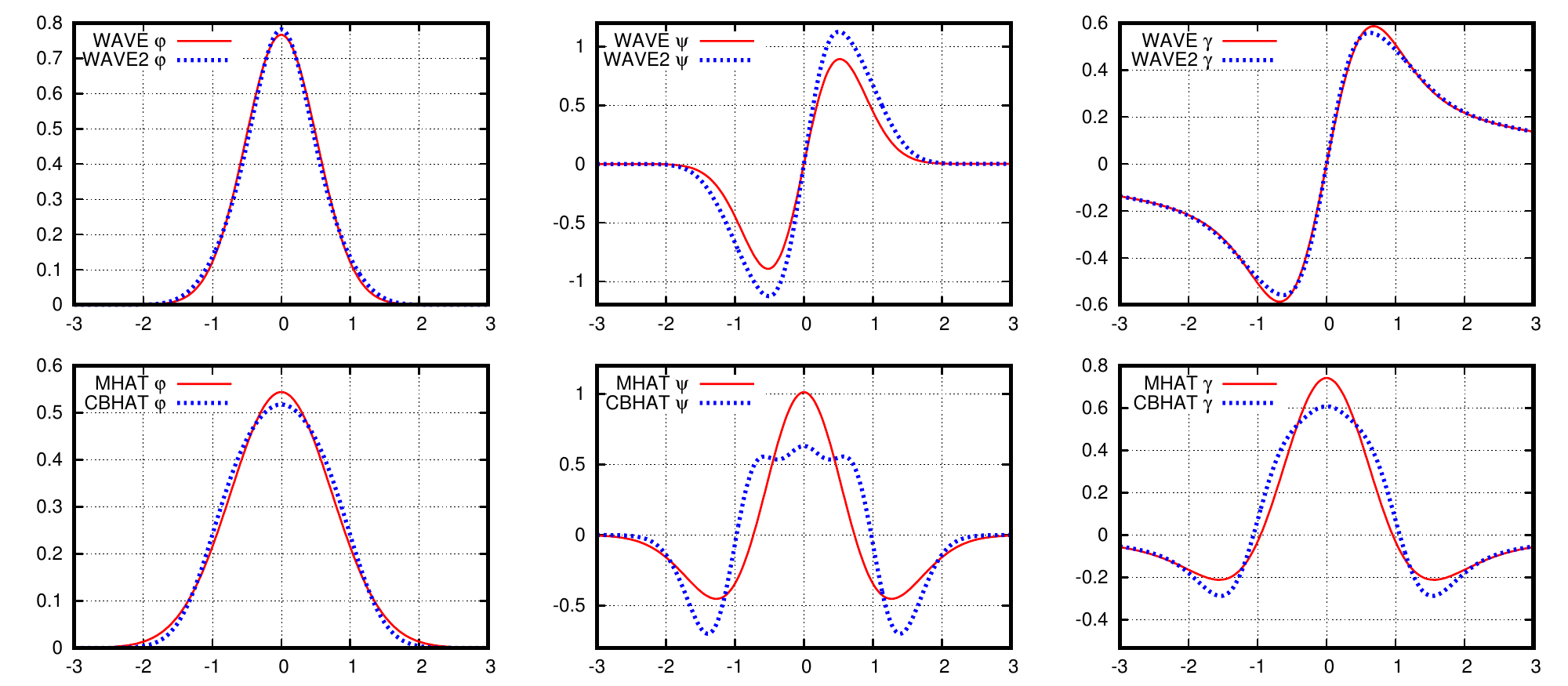}
\end{tabular}
\caption{Several analysing wavelets $\psi$, their generating functions $\varphi$, and the
associated optimal reconstruction kernels $\gamma$.}
\label{fig_wavs}
\end{figure*}

The algorithm is aimed to analyse only unidimensional random samples. The mathematical
details are given in \citep{Baluev18a}, and a brief step-by-step explanation of the method
follows below.
\begin{enumerate}
\item We use the wavelets called as CBHAT (``cowboy hat'') and WAVE2 that represent
optimized modifications of the classic MHAT (``Mexican hat'') and WAVE wavelets. Just like
WAVE and MHAT can be obtained by differentiating a Gaussian once or twice, the new optimal
wavelets $\psi$ can be produced as derivatives of bell-like generating functions $\varphi$:
\begin{equation}
\psi_{\rm WAVE2}(t) = \varphi_{\rm WAVE2}'(t), \quad \psi_{\rm CBHAT}(t) = \varphi_{\rm CBHAT}''(t).
\end{equation}
Such wavelet transforms would highlight zones where the first or second derivative of the
p.d.f. is large, revealing specific patterns like quick density gradients or p.d.f.
convexities and concavities. The latters can be associated with clumps and gaps in the
input sample.

\item The continuous wavelet transform (CWT) of the p.d.f. $f(x)$ is estimated from the
random sample $\left\{x_i \right\}_{i=1}^N$:
\begin{eqnarray}
& & Y(a,b) = \int\limits_{-\infty}^{+\infty} \psi\left(\frac{x-b}{a}\right) f(x) dx \nonumber\\
& & \text{estimated by} \quad \widetilde Y(a,b) = \frac{1}{N}\sum_{i=1}^N \psi\left(\frac{x_i-b}{a}\right).
\end{eqnarray}
Using similar formulae, the variance of $\widetilde Y$ is estimated too, in each point
$(a,b)$.

\item With $\widetilde Y$ and its variance estimate, and also based on a comparison
model $Y_0$ (possibly zero on start and updated iteratively), a normalized function
$z(a,b)$ is constructed. This $z$ basically represents the Student $t$-statistic that have
approximately constant noise properties over different $(a,b)$.

A particular value $z(a,b)$, computed at any fixed position $(a,b)$, is asymptotically
standard Gaussian, if $N\to\infty$. But this convergence is not uniform: for a particular
finite $N$, even a large one, the quantity $z(a,b)$ won't be Gaussian in some parts of the
$(a,b)$ plane. For example, points with too small scale $a$ and points with $b$ far outside
of the sample range must be removed from the analysis, as done by the next step.

\item Applying a criterion of Gaussianity on $z$, we determine the admissible work domain
$\mathcal D$ in the $(a,b)$-plane. The simplest version of such a criterion is to have many
enough sample objects $x_i$ that contribute in $\widetilde Y$ with a significant amount.
This requirement can be formalized as:
\begin{equation}
n(a,b) = \left. \sum_{i=1}^N \varphi\left(\frac{x_i-b}{a}\right) \right/ \int\limits_{-\infty}^{+\infty} \varphi(t) dt \geq n_\mathrm{min}.
\label{critn}
\end{equation}
The criterion guarantees that the distribution of $z(a,b)$ remains almost Gaussian and does
not degenerate inside $\mathcal D$. However, more subtle Gaussianity criteria should be
used in practice, and they are given in \citep{Baluev18a}.

Larger samples imply larger $N$, causing $\mathcal D$ to expand to smaller scales $a$,
while reducing $N$ causes $\mathcal D$ to shrink. Practical calculations revealed that for
$N\lesssim 100$ the domain $\mathcal D$ becomes (almost) empty, so our pipeline is useful
mainly for $N\gtrsim 100$.

\item Inside the work domain $\mathcal D$, we compute the maximum of $|z(a,b)|$. If this
maximum is large enough, than there are significant deviations between the model $Y_0$ and
the actual estimate $\widetilde Y$. The formal significance threshold can be estimated via
the false alarm probability (FAP) using an analytic formula, which is based on the
assumption that $z(a,b)$ is Gaussian:
\begin{equation}
\FAP(z) \simeq 2 W_{00} z e^{-z^2/2},
\label{fap}
\end{equation}
where the coefficient $W_{00}$ can be determined simultaneously with the computation of
$\widetilde Y$ and other quantities.

Setting some small threshold $\FAP_*$, e.g. $0.01$, and comparing the actual ``observed''
$\FAP(z_{\rm max})$ with it, we can suppress the level of false detections to the desired
low probability.

\item In practice, there are usually multiple places in the $(a,b)$ plane, where $|z(a,b)|$
rises above the significance threshold. Their location indicate the shift and scale
characteristics of the structures that we need to add to $Y_0$ to obtain a more matching
model. Peforming such a thresholiding, i.e. selecting $(a,b)$ points that satisfy
$\FAP(z(a,b))<\FAP_*$, we purge the noise from the $(a,b)$ plane, and leave only
significant structures.

\item After the noise thresholding, we can apply a generalized inversion formula
\begin{equation}
f(x) = \frac{1}{C_{\psi\gamma}} \int\limits_{-\infty}^{+\infty}\int\limits_{-\infty}^{+\infty} Y(a,b) \gamma\left(\frac{x-b}{a}\right) \frac{da db}{|a|^3}, \nonumber\\
\end{equation}
to the thresholded estimate $\widetilde Y(a,b)$. Here, the optimal kernel $\gamma$ depends
on the adopted wavelet $\psi$, and they both define the normalization constant
$C_{\psi\gamma}$. The inversion gives us an updated (or improved) comparison model $Y_0$
that includes all the significant structures detected at the current iteration.

\item The steps (3) and (5)-(7) are cycled over, until we remove all the significant
structures from the $(a,b)$ plane (i.e., until $|z(a,b)|$ is suppressed below the
significance threshold everywhere in $\mathcal D$). In the end we obtain a reconstructed
p.d.f. estimate $\tilde f(x)$ that contains only details that overpassed the desired level
of significance. In the process, we also obtain information about the FAPs for the
particular details detected.
\end{enumerate}

At a glance, the pipeline involves three big blocks algorithmically interleaved:
\begin{enumerate}
\item estimation of the wavelet transform and related stuff;
\item detecting significant structures;
\item iterative reconstruction of the p.d.f.
\end{enumerate}
We note that the last one, the reconstruction of the p.d.f., represents a version of the
matching pursuit method. From this point of view it becomes similar to the famous CLEAN
\citep{Roberts87} and CLEANest \citep{Foster95,Foster96a} algorithms of the spectral
deconvolution, which are also related to the matching pursuit \citep{Hara17}.

Probably the closest wavelet analysis method in other works was provided by
\citet{Skuljan99}. Let us list the main differences of our method from that one.
\begin{enumerate}
\item We avoid any binning of the data. The binning is an unnecessary and possibly
intervening procedure in this task. The sample is analysed ``as is''.
\item We consider more general formulations in many aspects: use of a general comparison
model, of two distincts wavelets, and paying attention to both positive and negative values
of the CWT. This enables us to detect a variety of distribution patterns. \citet{Skuljan99}
use zero comparison model, only the MHAT wavelet, and consider only positive deviations,
limiting themselves to only the clumps detection.
\item Our analysis relies on a normalized statistic that allows to equibalance the noise
over the shift-scale plane.
\item We corrected the significance testing approach to adequately treat the effect of the
``domain penalty''. We compute the false alarm probability ($\FAP$) based on the CWT
\emph{extreme values} within a work domain. \citet{Skuljan99} assumed wrongly \emph{a
single-value} distribution, triggering hidden multiple comparisons and p-hacking in
the final result. This effect is analogous to e.g. the well-known ``bandwidth penalty'' in
the periodogram analysis \citep{SchwCzerny98b,Baluev08a,Suveges14}, and it has a huge
effect on the significance. Without taking it into account, the significance of a detection
is dramatically overestimated, leading us to falsely confident conclusions.
\item The $\FAP$ is approximated in a completely analytic way, thus removing any need of
Monte Carlo simulations to estimate the significance.
\item Our method utilizes an objective criterion to determine the useful work domain in the
shift-scale plane.
\item We apply optimal wavelets that allow to improve the S/N ratio and to increase the
admissible work domain. In particular, the MHAT wavelet is basically useless in this task,
because of the large non-Gaussianity it generates \citep{Baluev18a}.
\item We built an optimized reconstruction kernel that allows to recover the p.d.f. from
a ``cleaned'' CWT in a minimum-noise way.
\item The main limitation of our algorithm is that it can process only 1D distributions. It
cannot analyse 2D distributions that were the main goal in \citep{Skuljan99}, or
distributions of higher dimension.
\end{enumerate}

\section{Demonstrative test applications}
\label{sec_testing}
\subsection{Pitfalls of the interpretation}
As outlined above, our wavelets allows to detect distribution patterns of the following two
types: (i) p.d.f. concavity/convexity, with CBHAT, and (ii) p.d.f. jump/fall, with WAVE2.
The first type is related to the second derivative $f''(x)$, meaning that this pattern
appears whenever $f(x)$ deviates from its local slope too much. The second type is related
to $f'(x)$, indicating the slope itself.

However, in practice it is usually impossible to imagine an isolated pattern of any of the
above type. This implies certain difficulties with the interpretation of the analysis, so
we need to consider some simulated teaching examples before we can proceed further.

In particular, it is necessary to acquire certain CWT ``reading skills'' before analysing
any real-world data. For example, if the p.d.f. contains a narrow peak layed over some
widespread background, it would imply the following patterns in the $(a,b)$-plane: (i)
concavity in the left wing~-- central convexity~-- concavity in the right wing (with
CBHAT), and (ii) a dipole-like uptrent--downtrend combination with WAVE2. A narrow
transitional edge separating two statistically different families would look like just a
single pattern with WAVE2 (say, an upturn), but with CBHAT it would imply a more
complicated dipole-like sequence: left-wing concavity~-- (the upturn itself unseen with
CBHAT)~-- right-wing convexity. Such illustrative examples are given in this section below.

The other difficulty comes from some deficiency of the p.d.f. reconstruction based on the
matching pursuit. In the most basic words, the matching pursuit is aimed to seek a
``sparse'' (that is, a constructively compact) approximation to some raw data. It offers an
easy and a quick solution, but not necessarily an accurate one. In particular, in the
time-series analysis by CLEAN/CLEANest method there are difficulties with treating the
aliases. A generally similar problem appears whenever wavelets mutually interfere in their
side lobes (although it did not become that much important here).

In practice, the problem appears when e.g. two structures approach each other closely. In
such a case, the reconstructed p.d.f. $f(x)$ may contain an artifact between them that does
not reveal itself in the wavelet transform (in $z(a,b)$). Such an effect likely have common
roots with the Gibbs phenomenon from the Fourier analysis.

Due to this effect, we currently consider the reconstructed $f(x)$ as just a useful hint
that helps to intuitively interpret the 2D wavelet map. But if it comes to the significance
estimate of any pattern, we must get back to $z(a,b)$.

\subsection{Histograms and their uncertainties}
The most basic and widely used tool of the distribution analysis is a histogram. We will
use histograms as a demonstrative reference to compare with our wavelet analysis method.
But first of all, we need to define some its formal characteristics and understand its
statistical uncertainties.

Mathematically, the histogram is a sequence of empiric frequencies $\nu_i=n_i/N$,
corresponding to $n_i$ objects falling inside an $i$th bin. An optimal bin size is given by
the \citet{FreedmanDiaconis81} rule:
\begin{equation}
h = 2 \frac{\text{IQR}}{N^{\frac{1}{3}}},
\label{hopt}
\end{equation}
where IQR is the interquantile range of the sample (counted from $0.25$ to $0.75$
quantile), and $N$ is sample size. However, in our cases this binning appeared
insufficiantly fine, and loosing too much small-scale details, so we adopted a more dense
grid. We actually set the total number of bins, $N_{\rm bin}$, rather than the bin width
$h$, and we set $N_{\rm bin}=20$ everywhere below. This usually resulted in a roughly twice
more dense binning than~(\ref{hopt}).

Since the distribution of each $n_i$ is binomial, the uncertainty of a given histogram box
can be represented by the standard deviation
\begin{equation}
\sigma_{\nu_i} = \sqrt{\frac{p_i(1-p_i)}{N}},
\label{hunc}
\end{equation}
where $p_i$ is the actual (true) probability corresponding to the bin,
$p_i=\int\limits_{{\rm bin}_i} f(x) dx$.

Since we need just a rough understanding of the histogram uncertaities, we will approximate
the distribution of $\nu_i$ by the normal one (assuming that $n_i$ is large) and substitute
the empiric $\nu_i$ in place of $p_i$ in~(\ref{hunc}).

The statistical deviation between some constructed model $\hat f(x)$ (hence model $\hat
p_i$) and the empiric histogram $\nu_i$ can be expressed in terms of the normal quantiles
$g_i=(\nu_i-\hat p_i)/\sigma_{\nu_i}$. Furthermore, considering a single histogram bin, the
normal quantile $g$ is tied to the false alarm probability, $\FAP$, by the equations
\begin{equation}
\FAP = 2 \left[1-\Phi(g) \right], \quad g = \Phi^{-1}\left(1-\frac{\FAP}{2}\right),
\label{gfap}
\end{equation}
where $\Phi(x)$ is the cumulative distribution function of a standard normal variate. Thus,
$g$ represents an absolute threshold such that a standard Gaussian random variable can
exceed $\pm g$ only with probability $\FAP$. The quantity $g$ is what is often understood
under a ``$g$-sigma'' significance level.

However, this $g$ is useful only if we considered just a single histogram bin, selected
prior to the analysis. It is illegal to rely on the usual $\sigma_{\nu}$ from~(\ref{hunc}),
if the entire histogram consisting of $N_{\rm bin}$ bins is considered, which is a typical
practical case. Statistically, different bins are almost independent from each other, and
the probability to make a mistake on multiple bins is therefore larger than on a single
bin. If we plainly set the $\FAP$ of each individual bin to the requred value, the
probability to make a false alarm in the entire histogram would then scale up as
\begin{equation}
\FAP' = 1 - (1-\FAP)^{N_{\rm bin}} \simeq N_{\rm bin} \FAP
\label{mt}
\end{equation}
This is the effect of multiple testing that leads to an implicit degradation of any
statistical accuracy. To compensate this effect, we should preventively set the $\FAP$ of
an individual bin to a reduced value $\FAP/N_{\rm bin}$. Then, taking into
account~(\ref{gfap}), the corrected normal quantile would be
\begin{equation}
g'(g) = \Phi^{-1}\left(1 - \frac{1-\Phi(g)}{N_{\rm bin}}\right)
\label{mtcorr}
\end{equation}
For example, for the one-sigma uncertainty, $g=1$ (hence $\FAP=31.7\%$), we obtain the
correction
\begin{equation}
g'(1) = \Phi^{-1}\left(1 - \frac{0.159}{N_{\rm bin}}\right) > 1
\end{equation}
Therefore, all the uncertainties~(\ref{hunc}) should be scaled up by multiplying them by
$g'$. For example, $N_{\rm bin}=20$ results in $g'(1)=2.4$. Only such pumped uncertainties
can be deemed as adequate $1$-sigma confidence intervals for the entire histogram.

To test the quality of the reconstructed p.d.f. models in an alternative manner, we also
apply the Pearson chi-square goodness-of-fit statistic. It is defined as
\begin{equation}
X^2 = N_{\rm bin} \sum_{i=1}^{N_{\rm bin}} \frac{(\nu_i-\hat p_i)^2}{\hat p_i}.
\end{equation}
Whenever our p.d.f. model is correct (all $\hat p_i = p_i$), the $X^2$ statistic
approximately follows the $\chi^2$ distribution with $N_{\rm bin}-1$ degrees of freedom. In
practice it is more convenient to approximate this $\chi^2$ distribution with the Gaussian
one. For example, the following quantity is close to being Gaussian:
\begin{equation}
\sqrt[3]{\frac{X^2}{N_{\rm bin}-1}} \sim \text{Normal}\left(\mu=1-\frac{2}{9N_{\rm bin}},\sigma^2=\frac{2}{9N_{\rm bin}}\right).
\end{equation}
Then, the standardized quantity
\begin{equation}
\tau = \left.\left[\sqrt[3]{\frac{X^2}{N_{\rm bin}-1}}-\left(1-\frac{2}{9N_{\rm bin}}\right)\right]\right/\sqrt{\frac{2}{9N_{\rm bin}}}
\label{pearson}
\end{equation}
can be deemed as a direct analogue of the normal quantile $g$ from~(\ref{gfap}), though
it corresponds to the entire histogram rather than to a single bin.

We use the Pearson test below mainly to demonstrate that our p.d.f. models agree with the
sample, even though sometimes it might seem that they are oversmoothed. Large positive
values of $\tau$ would indicate that our p.d.f. model (or $p_i$) disagrees with the sample.
But all results presented below, either demonstative or real-world ones, will have large
\emph{negative} $\tau$. This indicates an expected overfit, i.e. that we intentionally
seeked such p.d.f. model $\hat f(x)$ that approximates the sample well (even better than
the true $f(x)$ would do). But regardless of this aggregated overfit, our analysis remains
critical with respect to possibly noisy structural patterns, thanks to the strict
significance threshold~(\ref{fap}).

\subsection{Graphical representation of the results}
All figures that demonstarte results of our CWT analysis below, either of simulated or of
real-world data, follow the same layout scheme. They contain $4$ graphs for each
distribution analysed.

The \emph{top pair} in each quadruplet represents the 2D significance colormaps for
$z(a,b)$ for the CBHAT and WAVE2 wavelets. These colormaps are computed assuming the
comparison model $Y_0=0$, i.e. they correspond to the starting iteration of the matching
pursuit algorithm. Everything that appeared significant in these maps necessarily
contributes to the corresponding reconstructed p.d.f.

The significance function is named as ``EVD significance level'' (with EVD standing for the
Extreme Value Distribution). In absolute value, it equals to the Gaussian quantile $g$
from~(\ref{gfap}), where the $\FAP$ is calculated according to~(\ref{fap}). Formally the
function plotted in the 2D colormaps is $\pm g(\FAP(z(a,b)))$, where the sign is selected
accordingly to the sign of $z$.

This significance is considered only inside the normality domain $\mathcal D$, and
everything outside of $\mathcal D$ is hashed out. For the sake of an extra validation, we
additionally plot the simplified normality criterion~(\ref{critn}) as a dashed black curve.
We note that generally this curve traces the lower boundary of $\mathcal D$ rather well,
although important local deviations may appear.

In the \emph{bottom pair} of graphs of each quadruplet we show the corresponding histogram
for $N_{\rm bin}=20$, along with the p.d.f. models obtained for three levels of noise
tolerance, $g=1$ (green), $g=2$ (cyan), $g=3$ (yellow). The latter three curves differ only
in color, because even in a grayscale plot it remains obvious, which one is which (larger
smoothness means larger $g$).

It is necessary to draw attention to a couple additional details concerning the
interpretation of the plots.

Whenever some pattern is detected in the 2D significance map and barely achieves e.g. $g=2$
at its maximum, this pattern has to be \emph{purged completely} in the corresponding p.d.f.
plot (for $g=2$). This might seem a bit paradoxical, but a two-sigma detail should normally
disappear in a two-sigma p.d.f. model, though it should emerge, say, in the one-sigma curve
in such a case. In general, to appear in a $g$-sigma model, the pattern must rise
\emph{above} this $g$ in the 2D significance map. Moreover, even if a structure does rise
above the selected $g$-level, it still may appear practically invisible in the
reconstructed p.d.f. model just because it is small in absolute magnitude (regardless of
the high significance). Such a behaviour is normal and should not puzzle the reader.

\subsection{Detecting a spike}
We adopt the following double-Gaussian toy model:
\begin{equation}
f(x) = (1-p)\phi(x) + \frac{p}{\sigma} \phi\left(\frac{x-m}{\sigma}\right), \quad \phi(x)=\frac{e^{-\frac{x^2}{2}}}{\sqrt{2\pi}},
\label{dgauss}
\end{equation}
assuming that $\sigma$ is small.

We consider two cases: $p=0.1$ and $p=0.05$, with $m=0.5$ and $\sigma=0.05$ in the both.

\begin{figure*}
\begin{tabular}{@{}l@{}l@{}}
\includegraphics[width=0.5\linewidth]{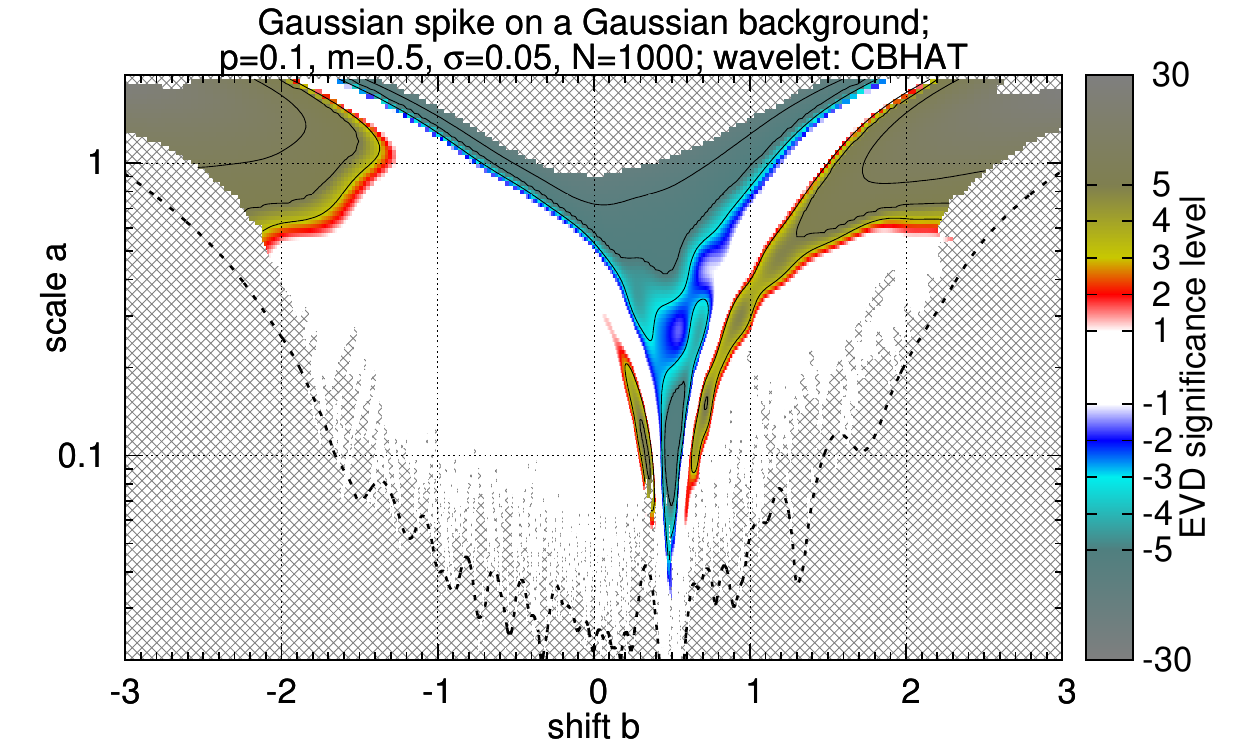} & \includegraphics[width=0.5\linewidth]{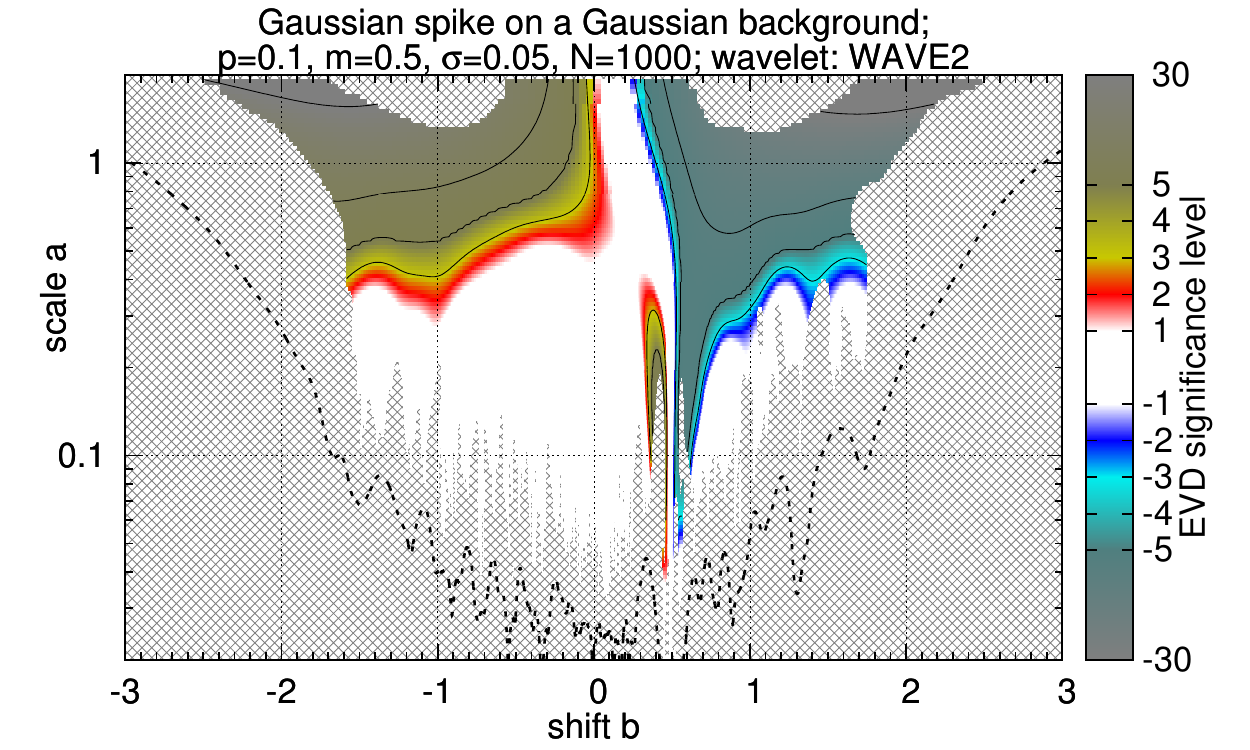}\\
\includegraphics[width=0.5\linewidth]{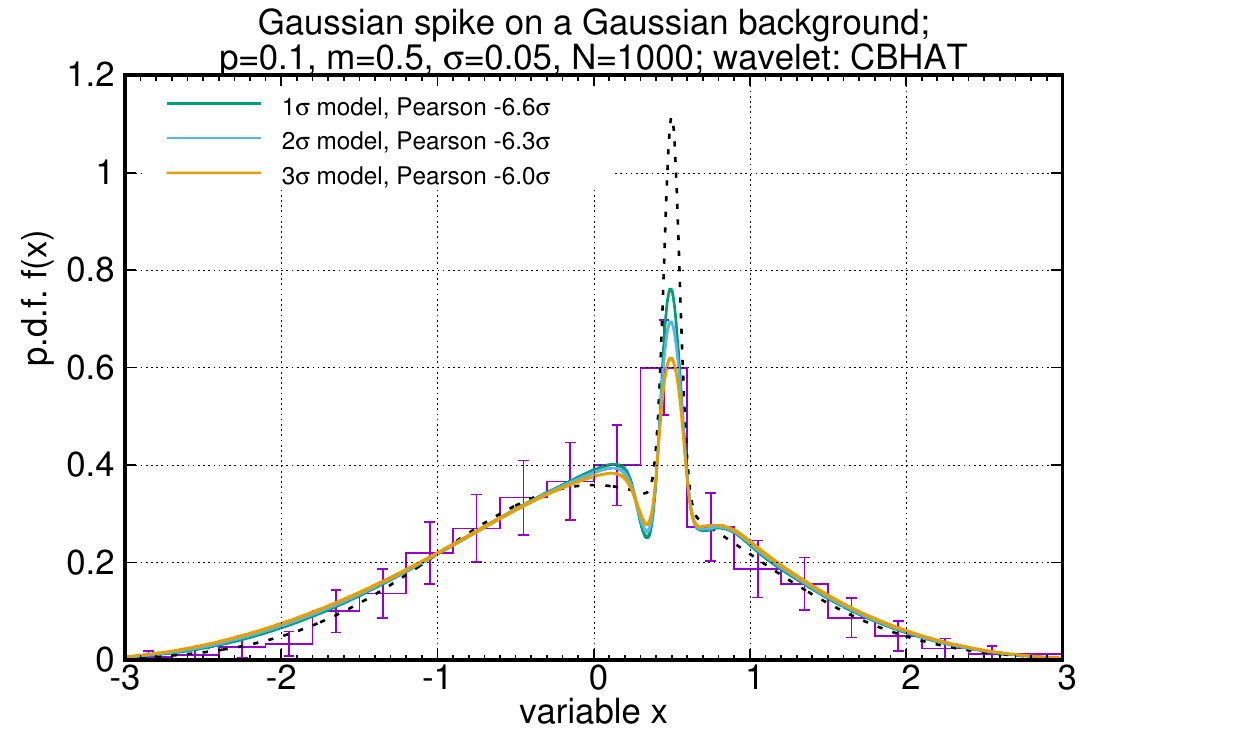} & \includegraphics[width=0.5\linewidth]{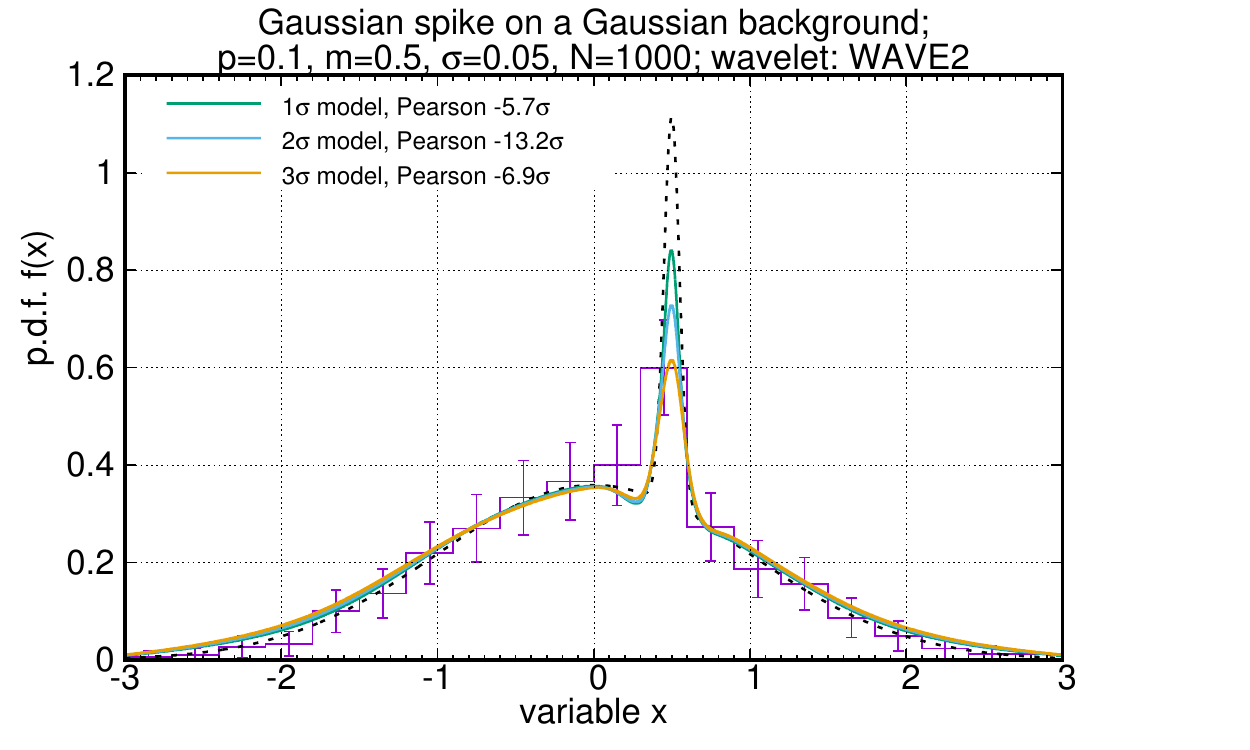}
\end{tabular}
\caption{Analysis of a simulated $N=1000$ sample: biased narrow Gaussian spike on a wide Gaussian
background~(\ref{dgauss}), with $p=10\%$, $\sigma=0.05$, $m=0.5$. Demonstrating
significance maps of the wavelet transforms (top) and reconstructed p.d.f. curves for three
levels of noise cleaning (bottom). The results obtained with the CBHAT and WAVE2 wavelets
are shown in the left and right panels, respectively. In the bottom panels, the true p.d.f.
is plotted by a dashed line, and the histogram of the sample is given, with the
uncertainties computed via~(\ref{hunc},\ref{mtcorr}). A ``standardized'' Pearson test
statistic $\tau$ from~(\ref{pearson}) for the synthesized p.d.f. models is also printed in
bottom panels. See text for further details and discussion.}
\label{fig_spike10}
\end{figure*}

The first case is shown in Fig.~\ref{fig_spike10}, and it demonstrates a robustly
detectable spike well above the significance thresholds. The 2D CBHAT map reveal a specific
small-scale pattern: a central p.d.f. convexity ($z<0$) accompanied by two smaller side
concavities ($z>0$). This is a typical pattern that indicates an isolated \emph{subfamily}:
the concavities can be treated as separations between this family and the background.

It is very important that without detecting both those side concavities we cannot say for
sure, whether the given structure is a statistically separated subfamily or it may
represent something different. Or, saying it in other words, the decisive component of a
subfamily is the p.d.f. concavities separating it from the rest of the sample, rather than
the central concentration itself.

\begin{figure*}
\begin{tabular}{@{}l@{}l@{}}
\includegraphics[width=0.5\linewidth]{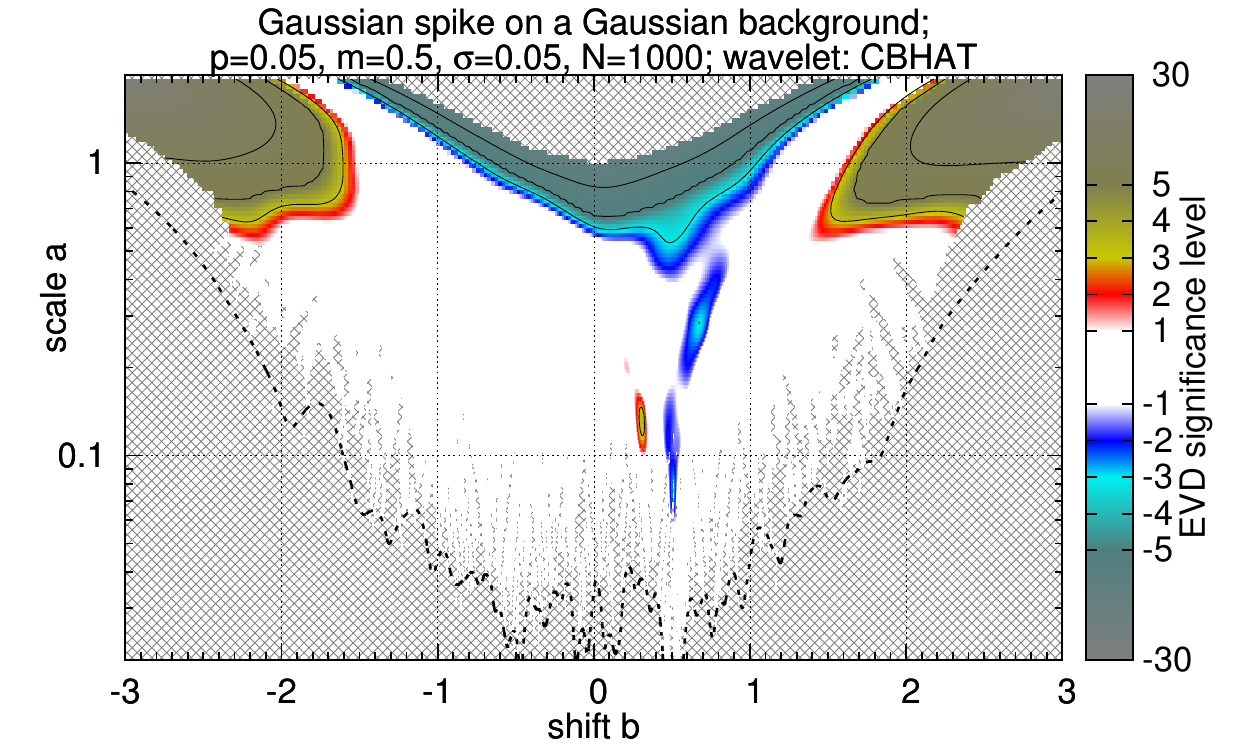} & \includegraphics[width=0.5\linewidth]{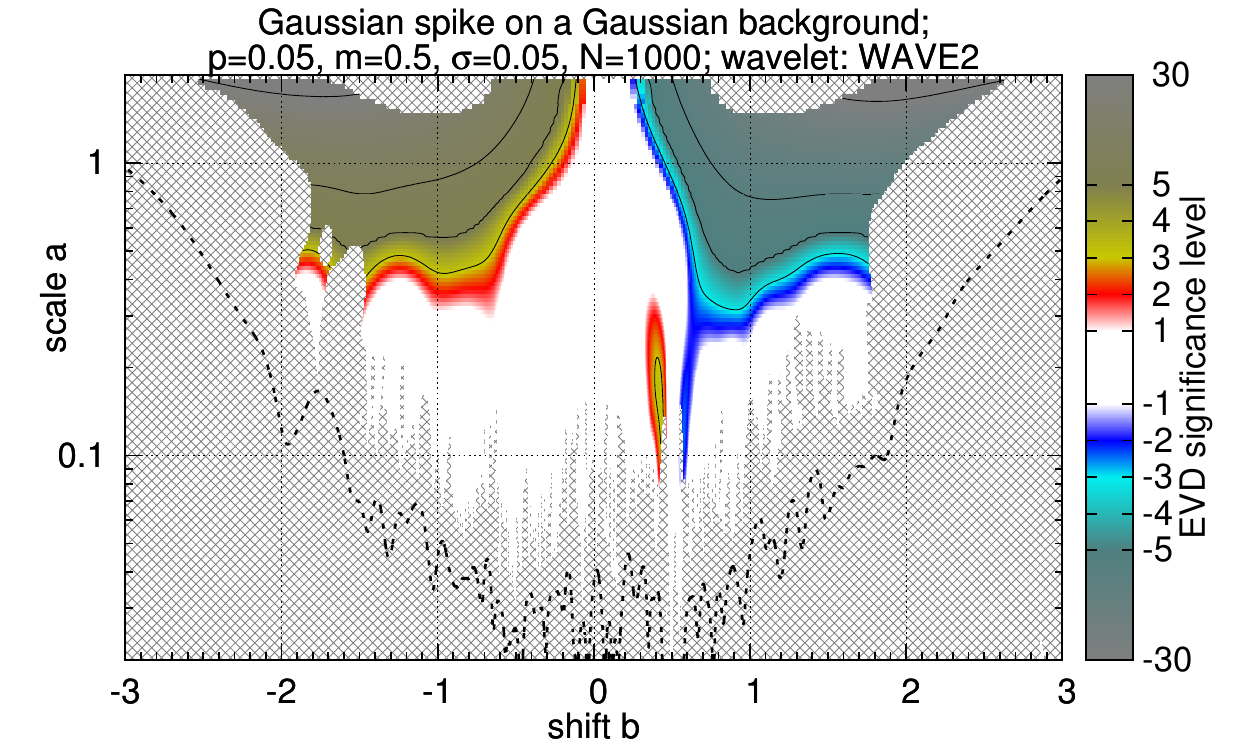}\\
\includegraphics[width=0.5\linewidth]{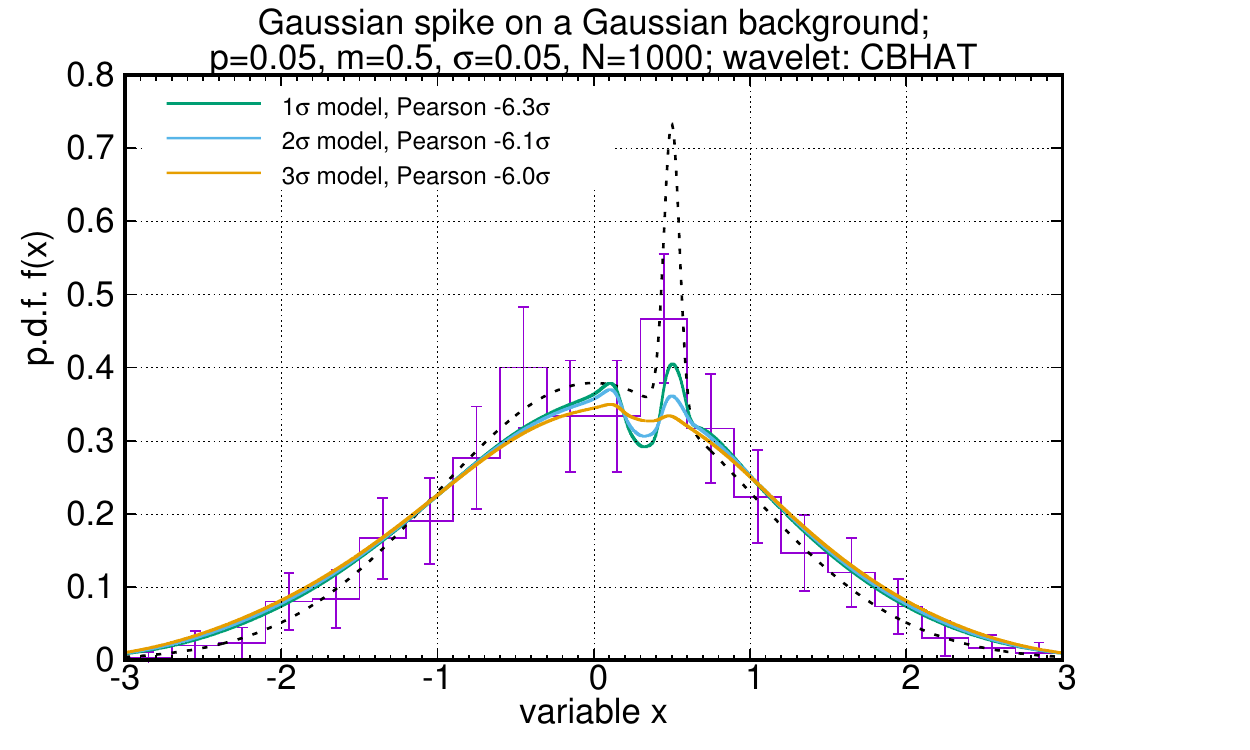} & \includegraphics[width=0.5\linewidth]{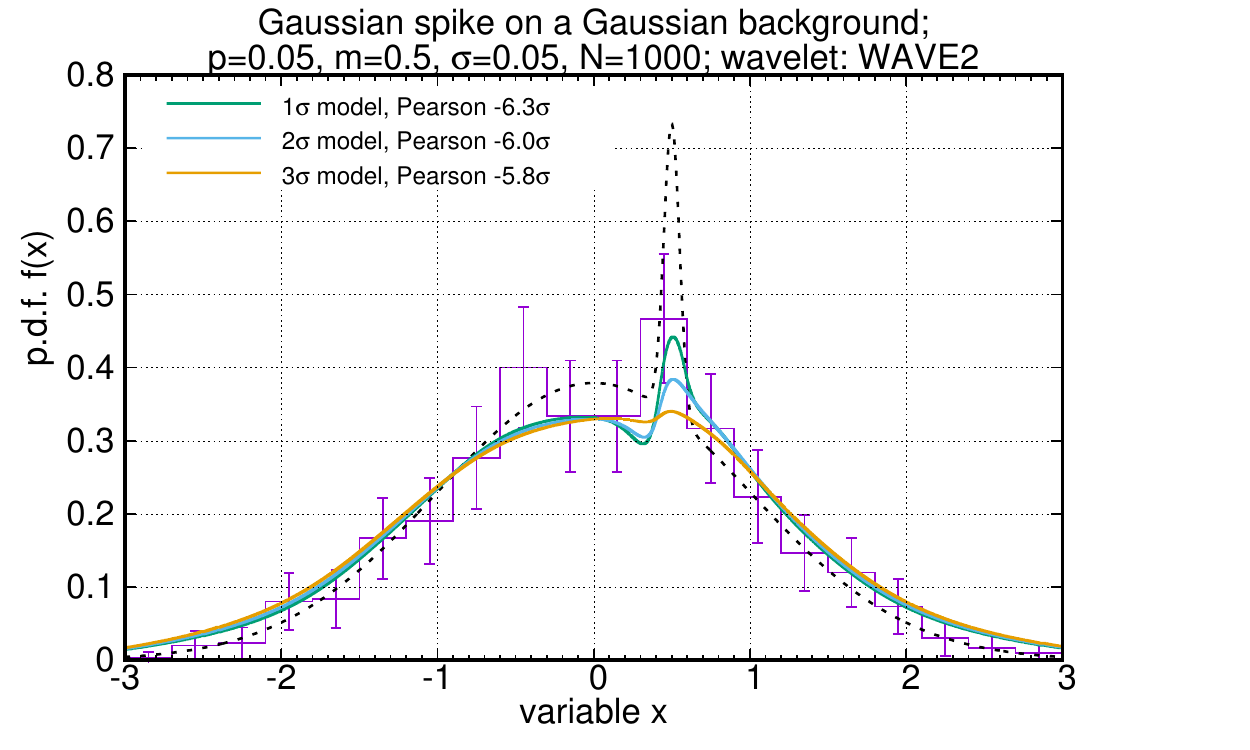}
\end{tabular}
\caption{Simulated example: biased narrow Gaussian spike on a wide Gaussian
background~(\ref{dgauss}), as in Fig.~\ref{fig_spike10} but with $p=5\%$, approximately
corresponding to the detection limit.}
\label{fig_spike05}
\end{figure*}

For example, in Fig.~\ref{fig_spike05} we present the analysis of a reduced $p=0.05$ spike
that corresponds to a detection threshold. In this case the pattern in the CBHAT
significance map is present only partially. We can see only the central convexity (rather
clobbered due to the noise), and the clear left concavity. But the right concavity appeared
insignificant because of the large noise. On the basis of only CBHAT analysis, we cannot
say definitely, whether our pattern indicates: (i) a spike (concavity~-- convexity~--
unseen concavity), (ii) a gap (unseen convexity~-- concavity~-- convexity), (iii) an abrupt
p.d.f. upturn (concavity~-- convexity), or even (iv) some more complicated wavy structure.
The reconstructed p.d.f. models look like an inconclusive mixture of these interpretations.

The WAVE2 analysis may appear useful in such ambiguous cases, because it may provide highly
complementary additional information. In the case of a spike substructire, the WAVE2
significance map reveals a diplole-like ``upturn--downturn'' structure. This enables an
alternative interpretation of the subfamily boundaries. However, in the low-SNR case of
Fig.~\ref{fig_spike05} it appeared that the right slope of the spike still has only a
moderate level of significance. Therefore, assuming a conservative detection threshold, our
interpretation should remain inconclusive in this case.

To end this subsection, we note that histograms seem to be largerly useless in our task.
Even in the high-SNR case (Fig.~\ref{fig_spike10}) their uncertainties appear too big for a
robust detection of the spike. Morever, when looking for patterns like ``spike'' or
``gap'', we must necessarily compare neighbouring histogram boxes. This becomes even more
unreliable and less obvious, because the comparison of two or multiple noisy values (values
with uncertainties) is a more intricate task than the comparison of a single noisy value
with a model one.

We may notice that mutual comparison of different histogram boxes is what is roughly
achieved with wavelets. According to Fig.~\ref{fig_wavs}, the CBHAT wavelet has such a
shape that it averages some portion in a p.d.f. $f(x)$ and then subtracts the average side
contributions. This is approximately the same as to subtract two side histogram bins from a
middle one, within a triplet, and perform such an operation for multiple bin triplets
(achieved by varying $b$) and for multiple histograms with different box width (achieved by
varying $a$). However, wavelets allow to perform such a ``brute force'' search in a
statistically safe way, taking into account the multiple testing penalty.

\subsection{Detecting an abrupt upturn}
Now we consider a piecewise-linear distributions $f(x)$, typically containing abrupt jumps:
\begin{equation}
f(x) = A x + \left\{\begin{array}{ll}
B, & x<0, \\
C, & x>0,
\end{array} \right.
\quad |x| \leq 2.
\label{pwl}
\end{equation}
The abrupt jumps occur at the boundary points $x=\pm 2$ and at $x=0$. The linear slope $A
x$ was added to~(\ref{pwl}) to make the model more tunable.

\begin{figure*}
\begin{tabular}{@{}l@{}l@{}}
\includegraphics[width=0.5\linewidth]{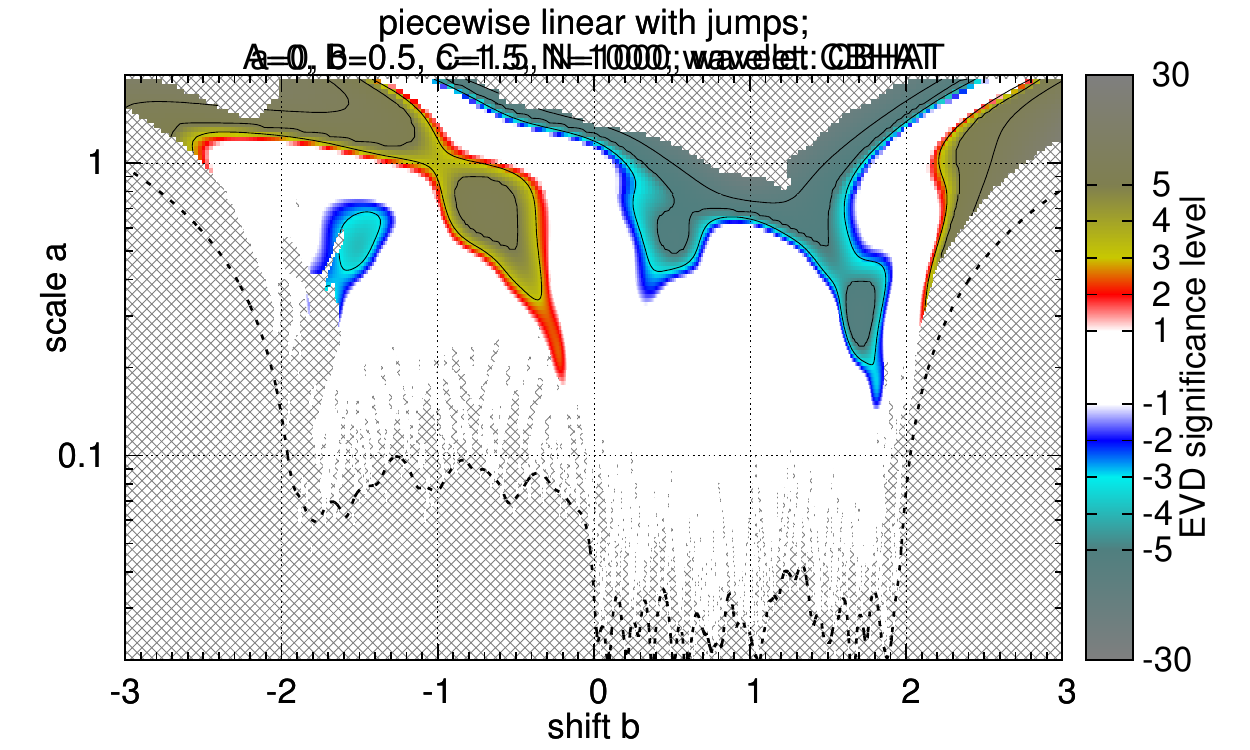} & \includegraphics[width=0.5\linewidth]{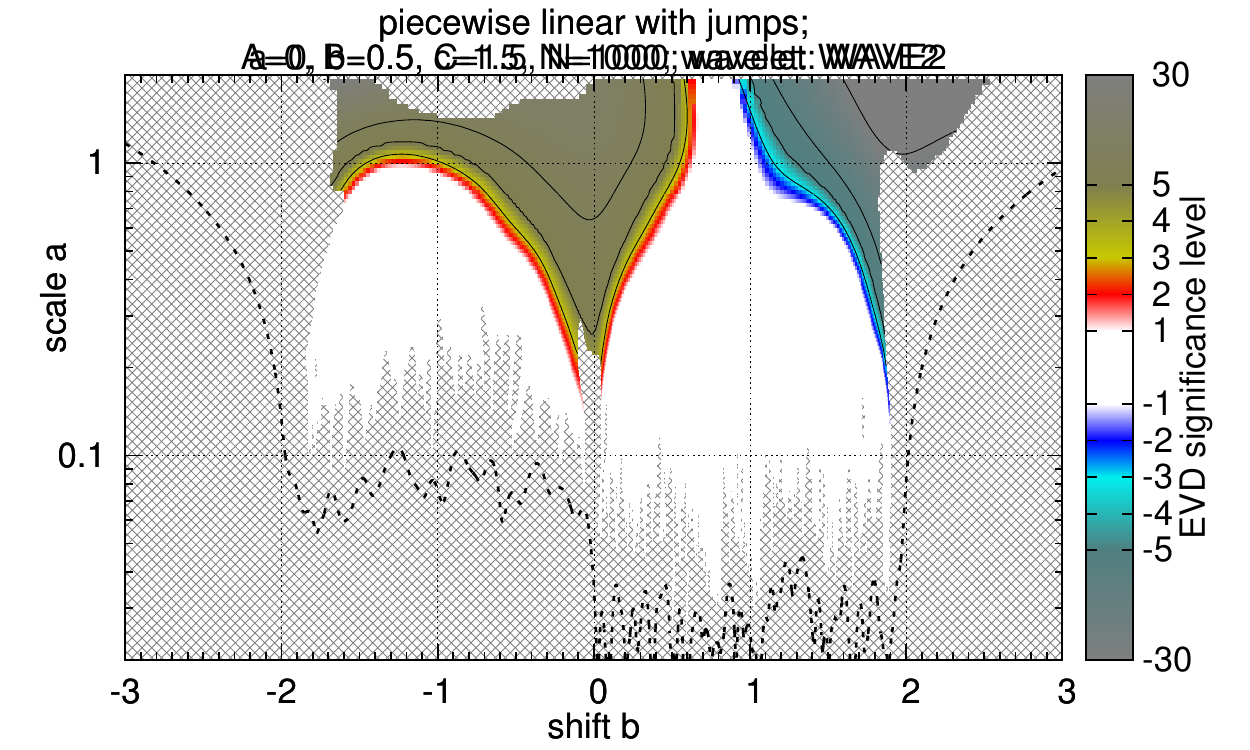}\\
\includegraphics[width=0.5\linewidth]{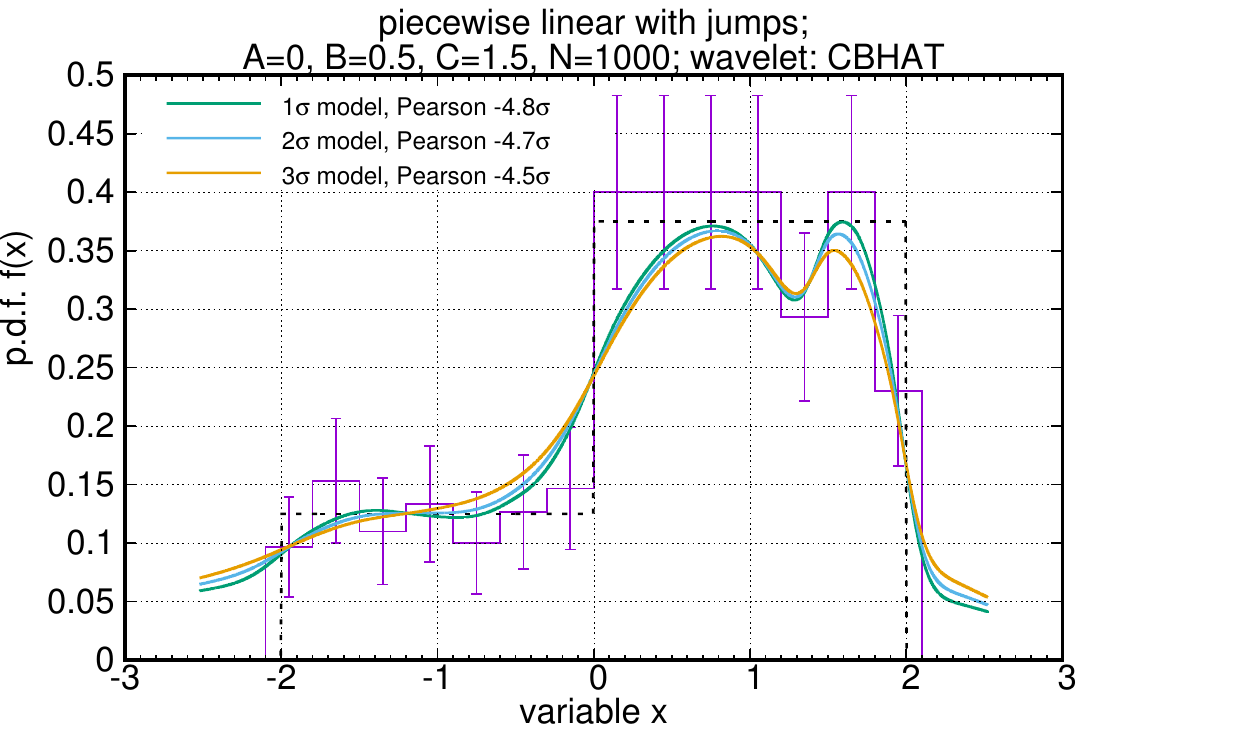} & \includegraphics[width=0.5\linewidth]{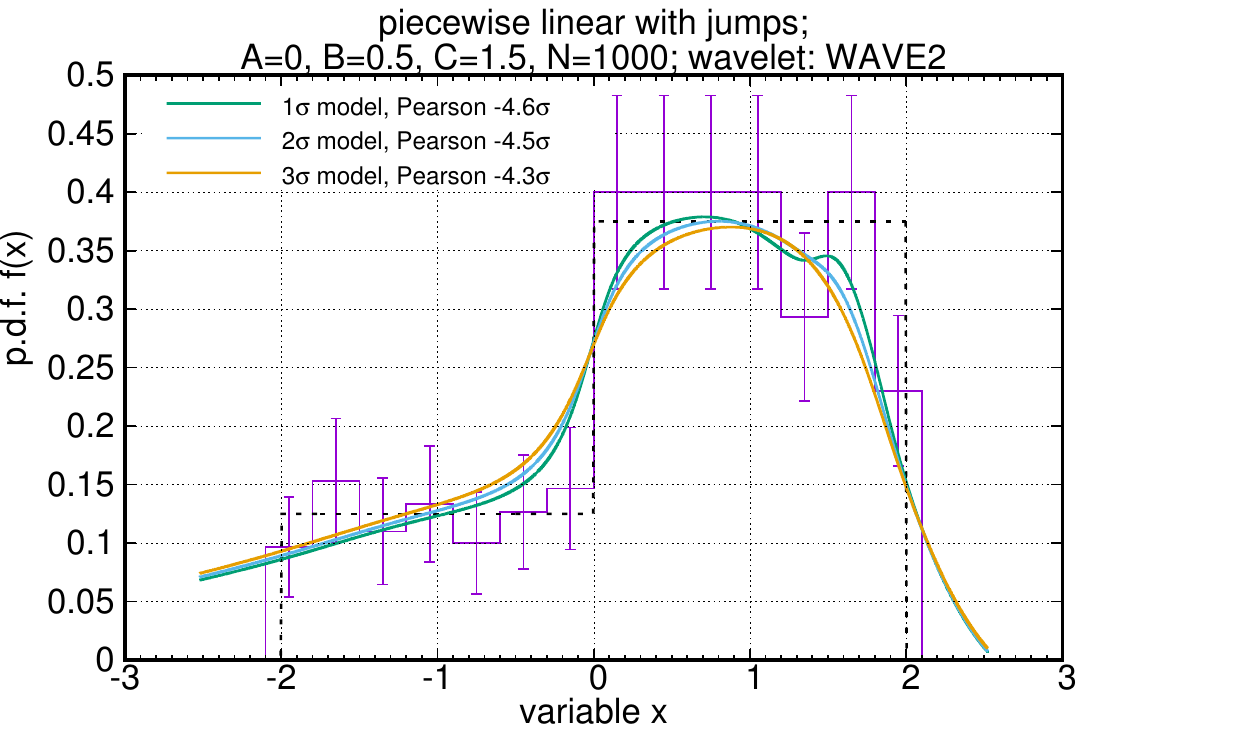}
\end{tabular}
\caption{Simulated example: piecewise linear distribution with jumps~(\ref{pwl}). Layout
similar to Fig.~\ref{fig_spike10}. Notice a reconstruction artifact, appearing as a local
drop in the p.d.f. at $x=1.3$, which is completely absent in the CWT significance map.}
\label{fig_pwl1}
\end{figure*}

\begin{figure*}
\begin{tabular}{@{}l@{}l@{}}
\includegraphics[width=0.5\linewidth]{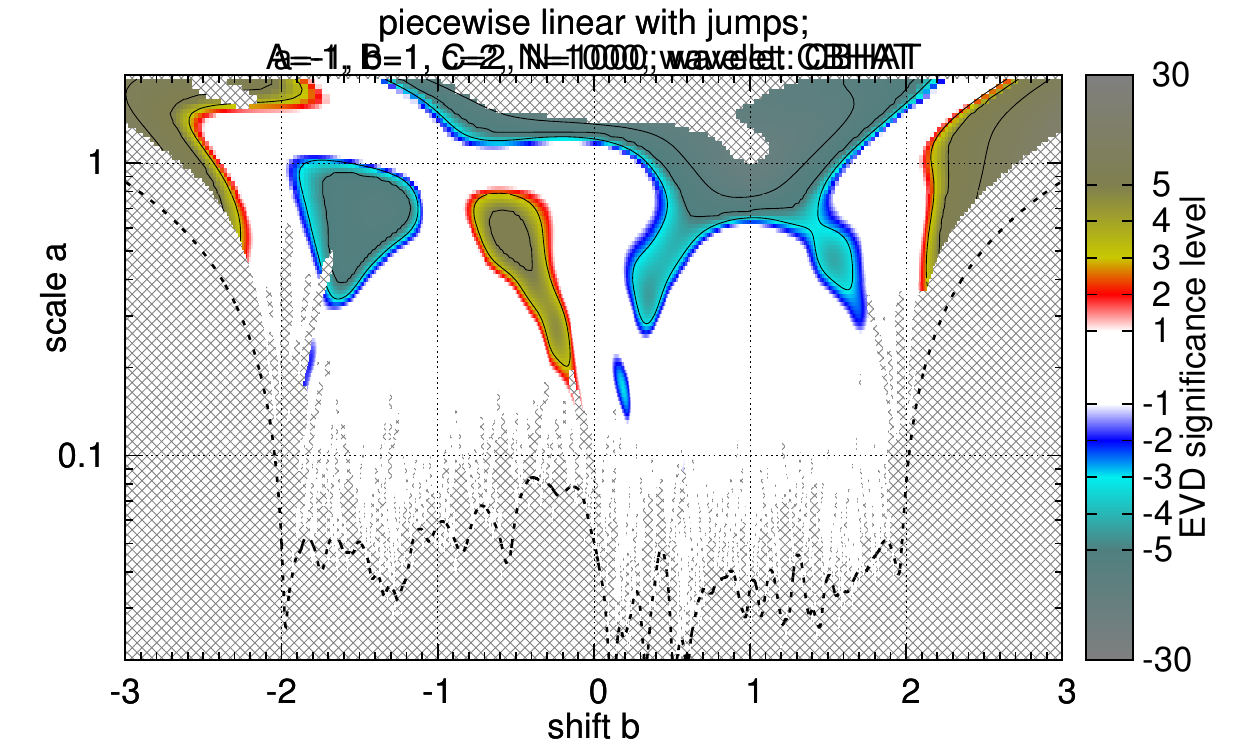} & \includegraphics[width=0.5\linewidth]{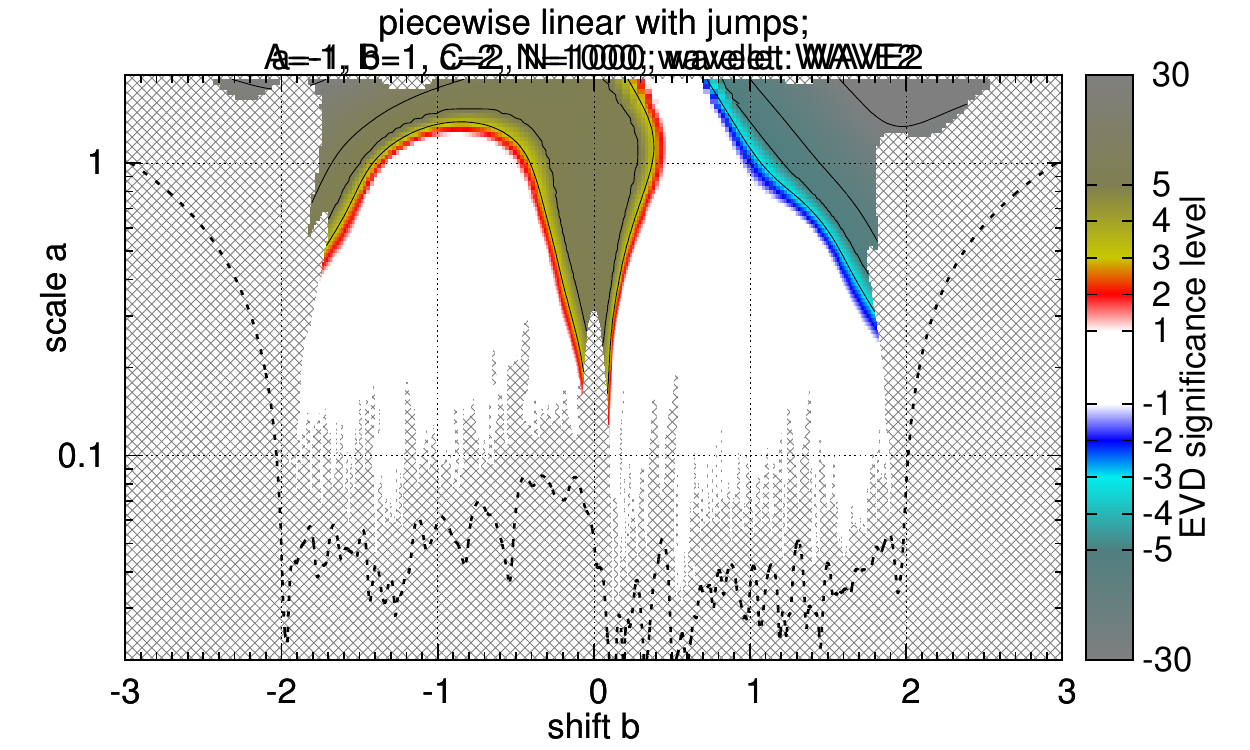}\\
\includegraphics[width=0.5\linewidth]{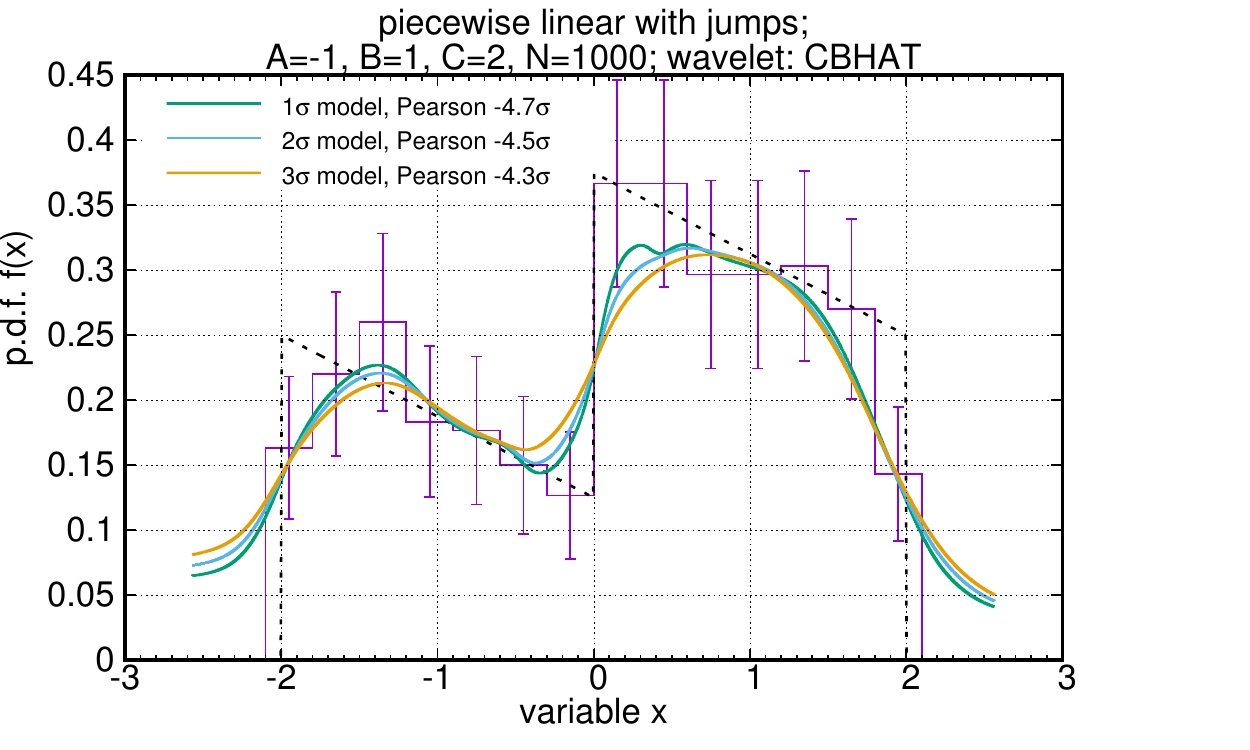} & \includegraphics[width=0.5\linewidth]{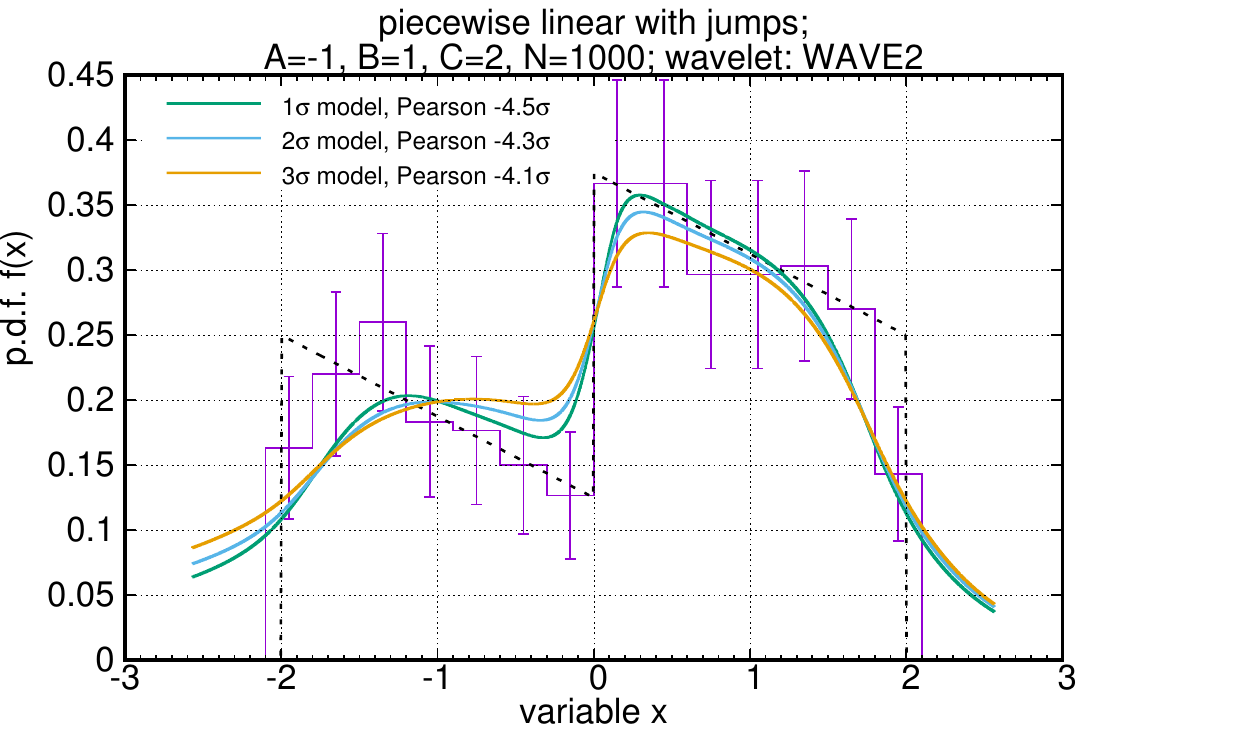}
\end{tabular}
\caption{Simulated example: piecewise linear distribution with jumps~(\ref{pwl}). Layout
similar to Figs.~\ref{fig_spike10}. Notice a Gibbs-like wavy artifact in the CBHAT p.d.f.
reconstruction (left-bottom panel). These waves (in particular the concave minimum at
$x=0.3$) are not present in the corresponding CWT significance map.}
\label{fig_pwl2}
\end{figure*}

\begin{figure*}
\begin{tabular}{@{}l@{}l@{}}
\includegraphics[width=0.5\linewidth]{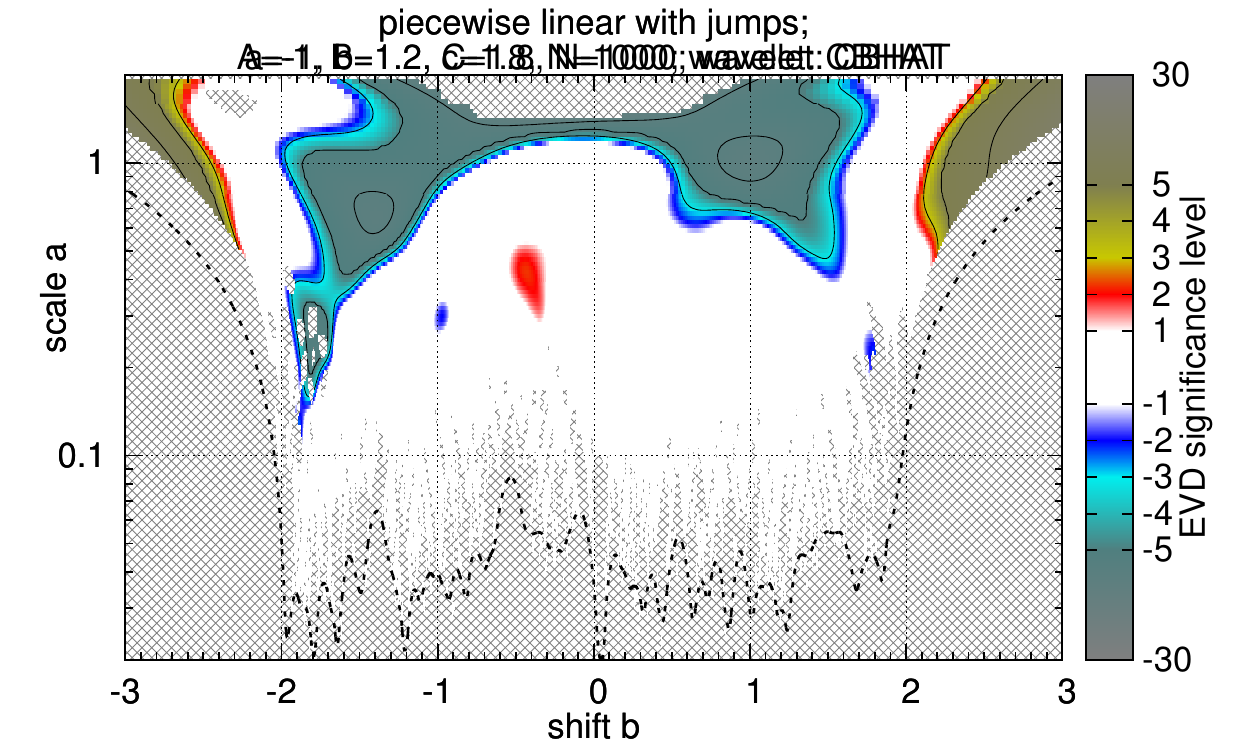} & \includegraphics[width=0.5\linewidth]{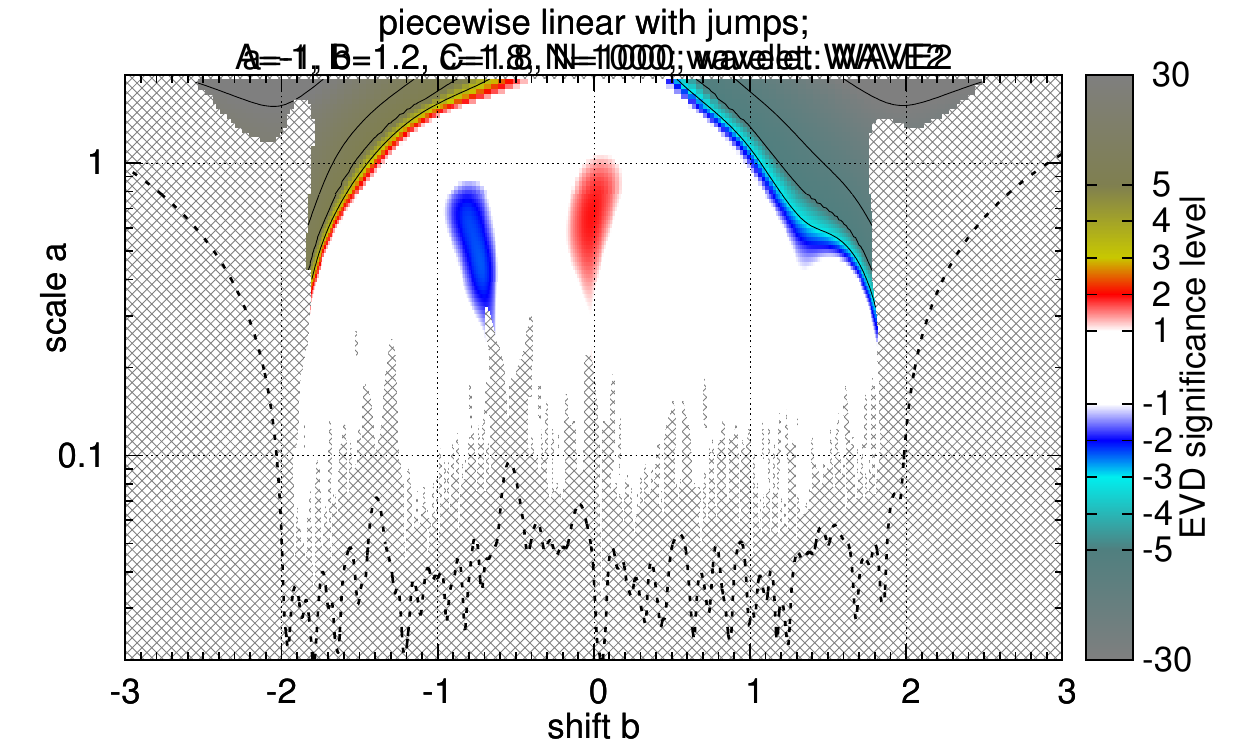}\\
\includegraphics[width=0.5\linewidth]{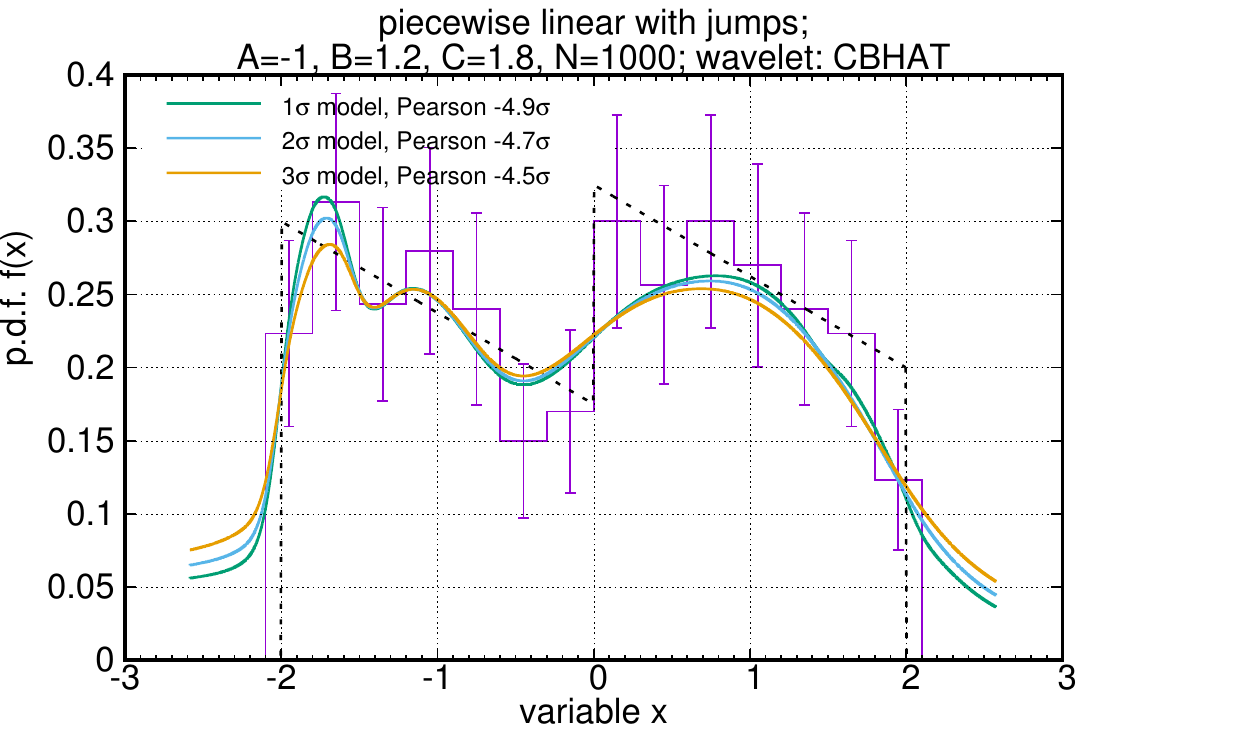} & \includegraphics[width=0.5\linewidth]{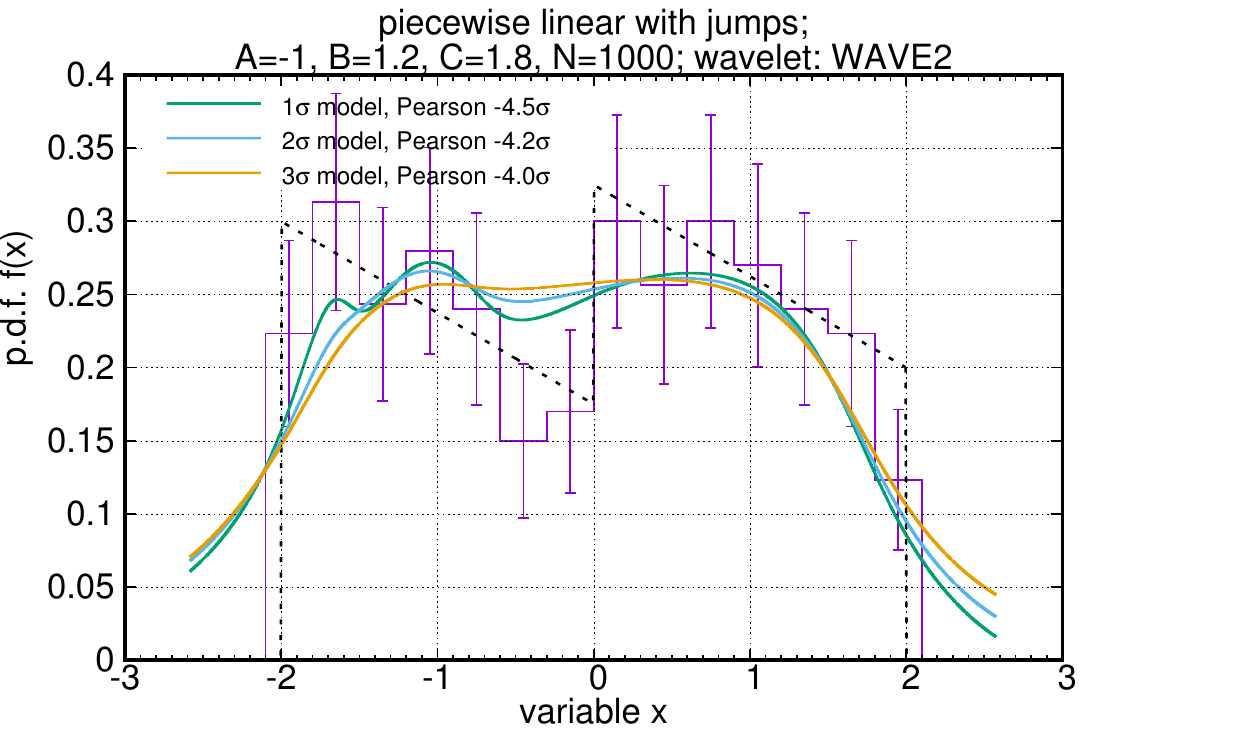}
\end{tabular}
\caption{Simulated example: piecewise linear distribution with jumps~(\ref{pwl}). Layout
similar to Figs.~\ref{fig_spike10}. Notice a Gibbs-like wavy artifacts in the p.d.f. plots,
not confirmed by the CWT significance maps.}
\label{fig_pwl3}
\end{figure*}

With the WAVE2 wavelet, an abrupt upturn may appear just like a single pattern (spot) in
the small-scale range of the $(a,b)$ plane. This behaviour is demonstrated in the top-right
panel of Fig.~\ref{fig_pwl3}. However, the small-scale spots may also merge with
large-scale structures (if they have the same sign), so in practice it frequently looks as
if the large-scale domain smoothly descended to smaller scales. This alternative behaviour
is revealed in Figs.~\ref{fig_pwl1} and~\ref{fig_pwl2}.

But we demonstrated above that upturns or downturns may also appear as parts of a gap or
spike, rather than like a standalone structure. To separate such cases from each other, the
CBHAT wavelet may be useful here. A ``pure'' upturn should ideally contain the following
sequence: a starting concavity (seen with CBHAT), the upturn itself (seen with WAVE2), the
final convexity (seen with CBHAT). This implies a dipole-like pattern in the CBHAT
significance map, such that the WAVE2 structure is projected in its middle. Any more
complicated configuration indicate hints of a more intricate structure than just a single
upturn.

This is demonstrated in Fig.~\ref{fig_pwl1} and~\ref{fig_pwl2}: the WAVE2 maps contain
highly-significant structures near $b=0$, and they are accompanied by a CBHAT structures at
$b=\pm 0.5$. Similar configurations are likely present at $b=\pm 2$, albeit they are less
clear due to regions of non-Gaussianity.

These test examples also reveal that our p.d.f. reconstruction algorithm is affected by a
Gibbs-like effect: abrupt jumps of high magnitude induce wavy artifacts. The matching
pursuit algorithm assumes that the most ``simple'' p.d.f. model is simultaneously the most
``economic'' one in terms of the number of non-zero wavelet coefficients. But this most
``economic'' p.d.f. model is not always the least varying one. In some cases it might be
necessary to include some statistically insignificant points from the $(a,b)$ plane, in
order to obtain a more smooth p.d.f. reconstruction.

Summarizing this, we caution the reader against blind interpretation of the p.d.f. models
presented below. They are offered as possibly helpful, but not decisive, interpretation of
the wavelet significance maps. The p.d.f. artifacts do not affect the significance maps
shown here. So, whenever it comes to the significance of a particular structure in $f(x)$,
it is necessary to get back to the relevant 2D map that represents the main source of the
significance information.

\section{Constructing exoplanetary samples to be analysed}
\label{sec_samples}
We utilize mainly the public database of the Extrasolar Planets Catalog \citep{Schneider11}
available at \url{www.exoplanet.eu}. This database currently includes information and data
about more than $3000$ exoplanets. However, this big ensemble appears too heterogeneous and
needs to be split into several subsamples before performing any processing.

Since our analysis does not currently consider any formal treatment of the detection bias,
we organize exoplanetary samples on the basis of their detection method. We construct two
big subsamples: (i) planets detected by radial velocities, and (ii) planets detected by
transits. These subsamples are not completely homogeneous yet, because they mix exoplanets
discovered either in distinct Doppler or distinct transit surveys that differ in their
accuracy and time base. This implies some differences in the corresponding ``detection
windows'' that overlap with each other in a complicated way. We should bear this in mind,
although it is unlikely that such a mixing effect would distort \emph{small-scale}
distribution details.

Our method can currently handle only unidimensional distributions, and this implies further
complications. There are multiple parameters associated to an exoplanet, and different
parameters reveal correlations forming complicatedly shaped structures in the
multidimensional space. Considering just single-dimensional distributions would mean to
look at 1D projections of these structures. This infers information losses, in particular a
decrease in sharpness. To suppress this effect to a certain degree at least, we can deal
with various slices of the parametric space. We construct them by cutting the full
exoplanetary sample with respect to some crucial parameters. These `cut parameters' are
different from the `target parameter' that defines what distribution is investigated.

Simultaneously, we should take care about the sizes of the resulting samples when
manipulating with them. In practice it appeared that samples smaller than $100$ objects are
useless in the analysis, because entire or almost entire shift-scale plane becomes
non-Gaussian. Moreover, at least $300$ objects are needed to obtain some informative
non-trivial results.

Our main samples that we used in the analysis, are listed in Table~\ref{tab_samples}. Note
that some values were missed for some planets in the database, so when considering the
distribution of particular parameters, the actual sample size can appear smaller than shown
in the table.

\begin{table}
\centering
\caption{Basic exoplanetary samples used in the work.}
\label{tab_samples}
\begin{tabular}{llp{40mm}}
\hline
 sample id & $N$ & description \\
\hline
 \sampleid{rv} & $708$ & Planets detected by radial velocity \\
 \sampleid{rv.FGK} & $468$ & As above, but FGK stars only ($M_\star \in [0.6,1.4] M_\odot$) \\
 \sampleid{rv.FGK.hmass} & $373$ & As above, but only giant planets ($m \sin i \geq 0.3 M_\mathrm{Jup}$) \\
\hline
 \sampleid{pt} & $2730$ & Planets detected by primary transit \\
 \sampleid{pt.FGK} & $2193$ & As above, but FGK stars only ($M_\star \in [0.6,1.4] M_\odot$) \\
\hline
 \sampleid{pt.CKS} & $737$ & California Kepler Survey sample cleaned similarly to \citep{Fulton17},
 see Sect.~\ref{sec_rad}. \\
\hline
\end{tabular}
\end{table}

We did not limit ourself to just drawing data from \emph{The Extrasolar Planets
Encyclopaedia}. Additionally we tried to reproduce the California Kepler Survey (CKS)
sample that was used by \citet{Fulton17} in their analysis of exoplanetary radii
distribution. Using the data from \citep{Petigura17,Johnson17}, we followed the same steps:
removed known false positives, removed stars with $K_p>14.2$, put the constraints on the
orbital period $P<100$~d and on the impact parameter $b<0.7$, removed giant stars using the
same criterion as in \citep{Fulton17}, and limited the temperature range by $4700-6500$~K.

Unfortunately, we failed to reproduce the final sample precisely. The most problematic
stage was filtering $b<0.7$. The cited works lack the values for $b$ and do not mention
were they taken from, so we decided to extract $b$ from the NASA Exoplanet
Archive\footnote{\url{http://exoplanetarchive.ipac.caltech.edu/}}. It appeared, however,
that the final sample shrinked to $N\sim 700$, which is smaller than $N\sim 900$ in
\citet{Fulton17}. We therefore decided that these values of $b$ from the NASA archive might
be less trustable, and then softened the threshold a bit, $b<0.8$. But to compensate for an
increased fraction of possibly unreliable radii estimates, we added an extra constraint
that the relative uncertainty in the planet radius must be below $0.1$. Thus we had $N=737$
planet candidates left in the sample.

\section{Results of the analysis}
\label{sec_res}
\subsection{Orbital periods and semimajor axes: fine-structured iceline accumulation?}
\label{sec_psma}
\begin{figure*}
\begin{tabular}{@{}l@{}l@{}}
\includegraphics[width=0.5\linewidth]{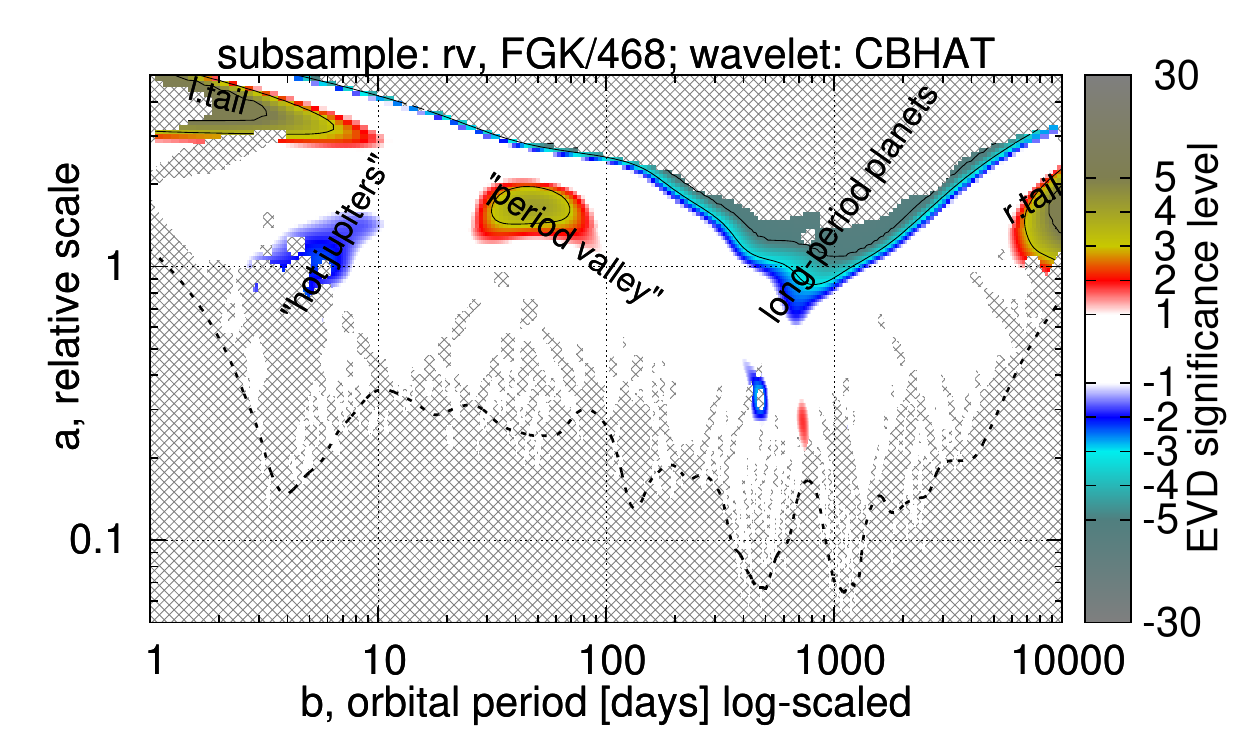} & \includegraphics[width=0.5\linewidth]{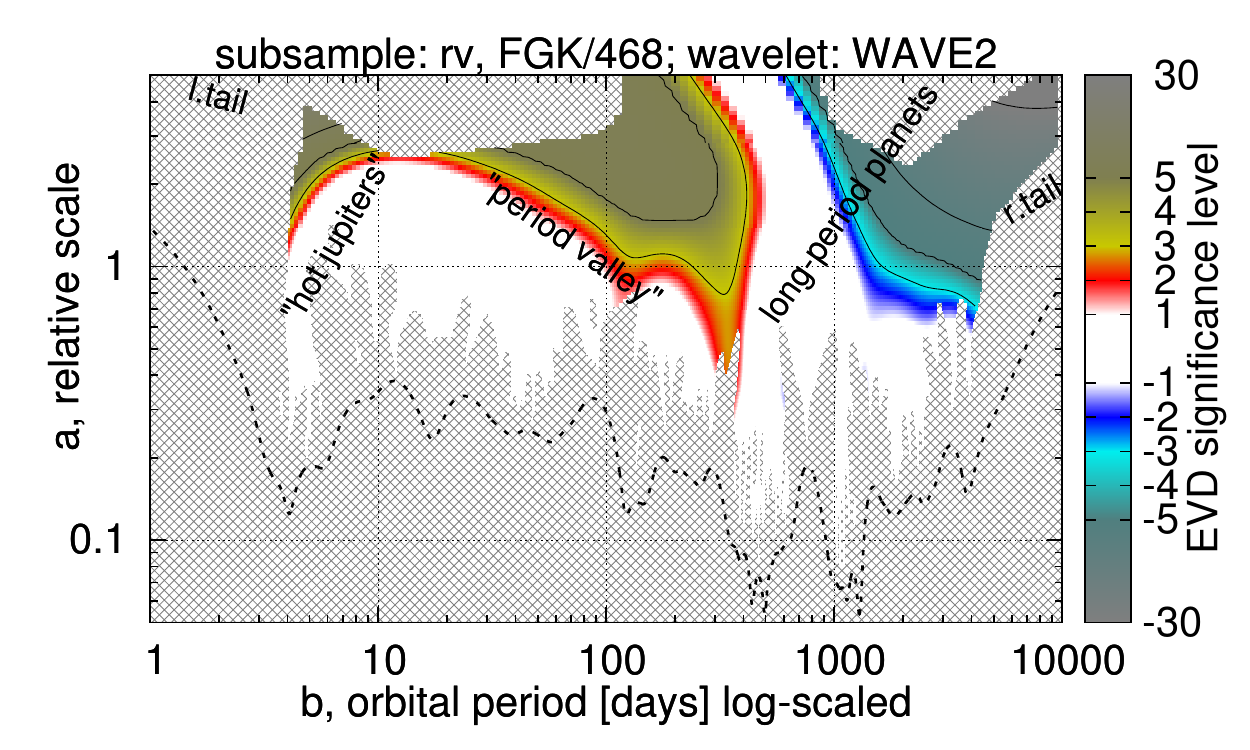}\\
\includegraphics[width=0.5\linewidth]{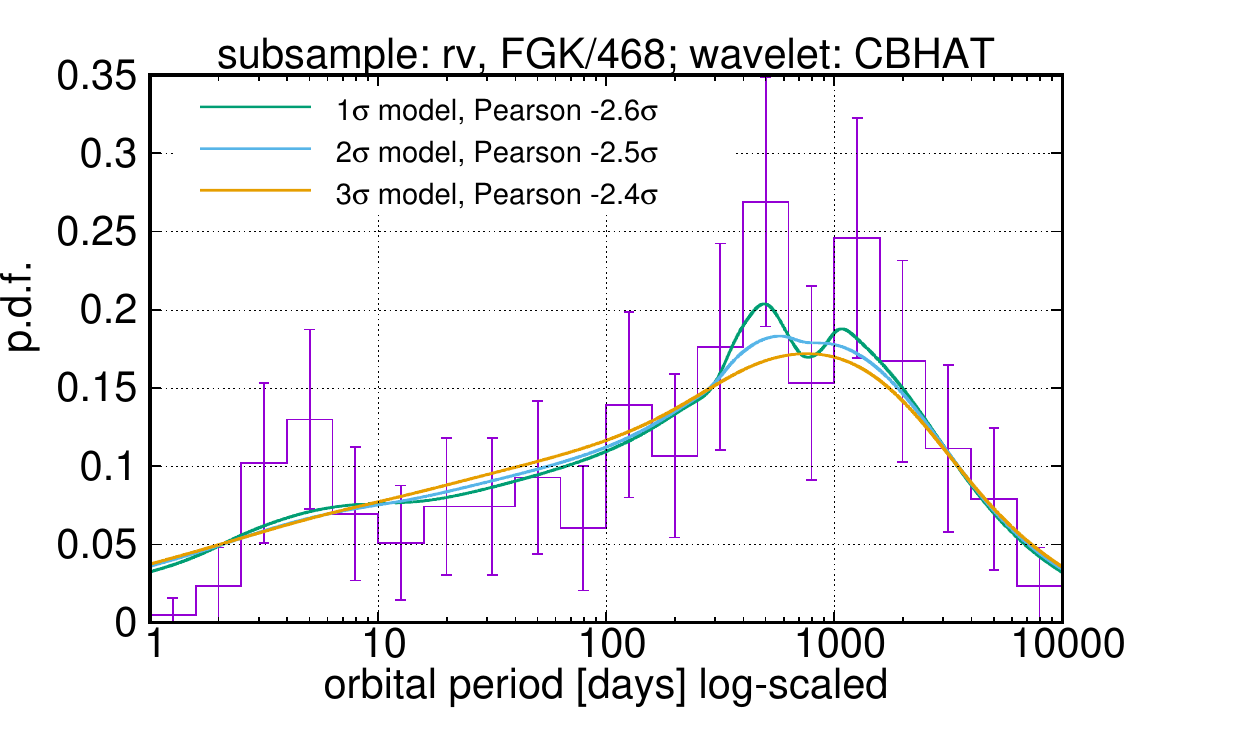} & \includegraphics[width=0.5\linewidth]{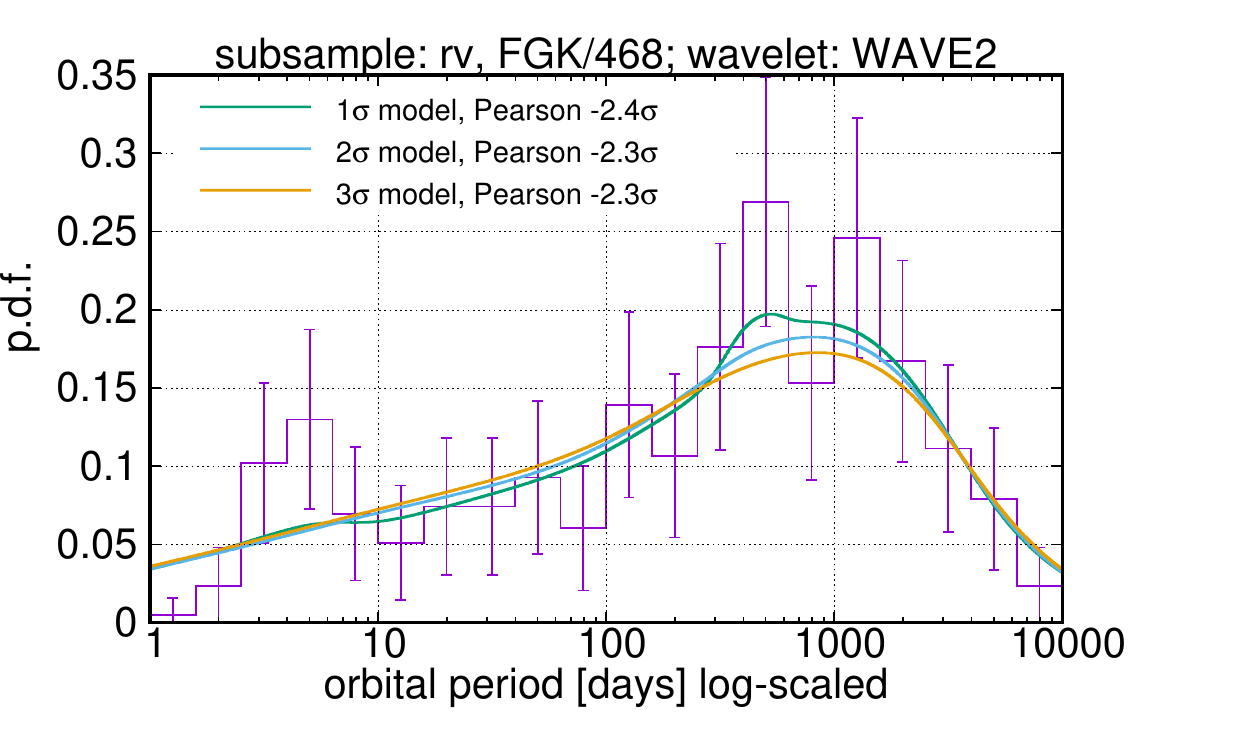}
\end{tabular}
\caption{Wavelet analyis of known exoplanetary candidates. Subsample: RV detection method,
FGK-type host stars, $N=468$. Variable: orbital period. The figure layout is the same as in
Fig.~\ref{fig_spike10}.}
\label{fig_rv_prd}
\end{figure*}

\begin{figure*}
\begin{tabular}{@{}l@{}l@{}}
\includegraphics[width=0.5\linewidth]{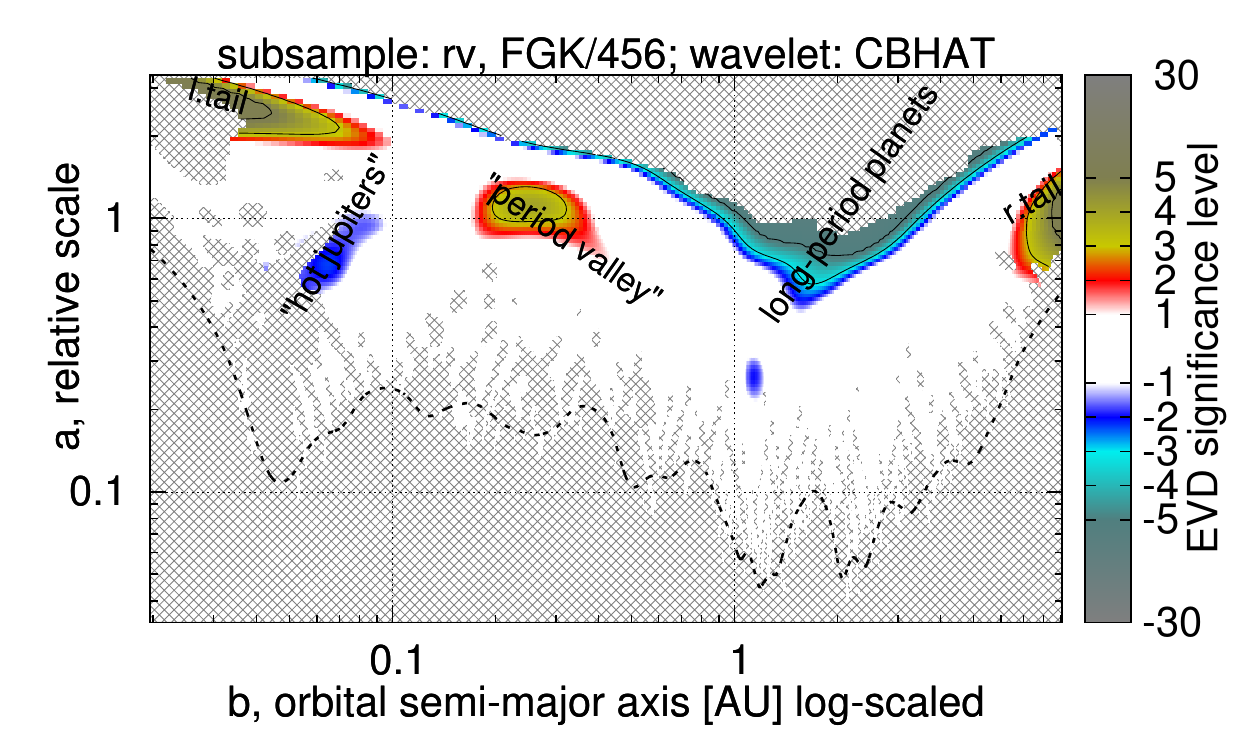} & \includegraphics[width=0.5\linewidth]{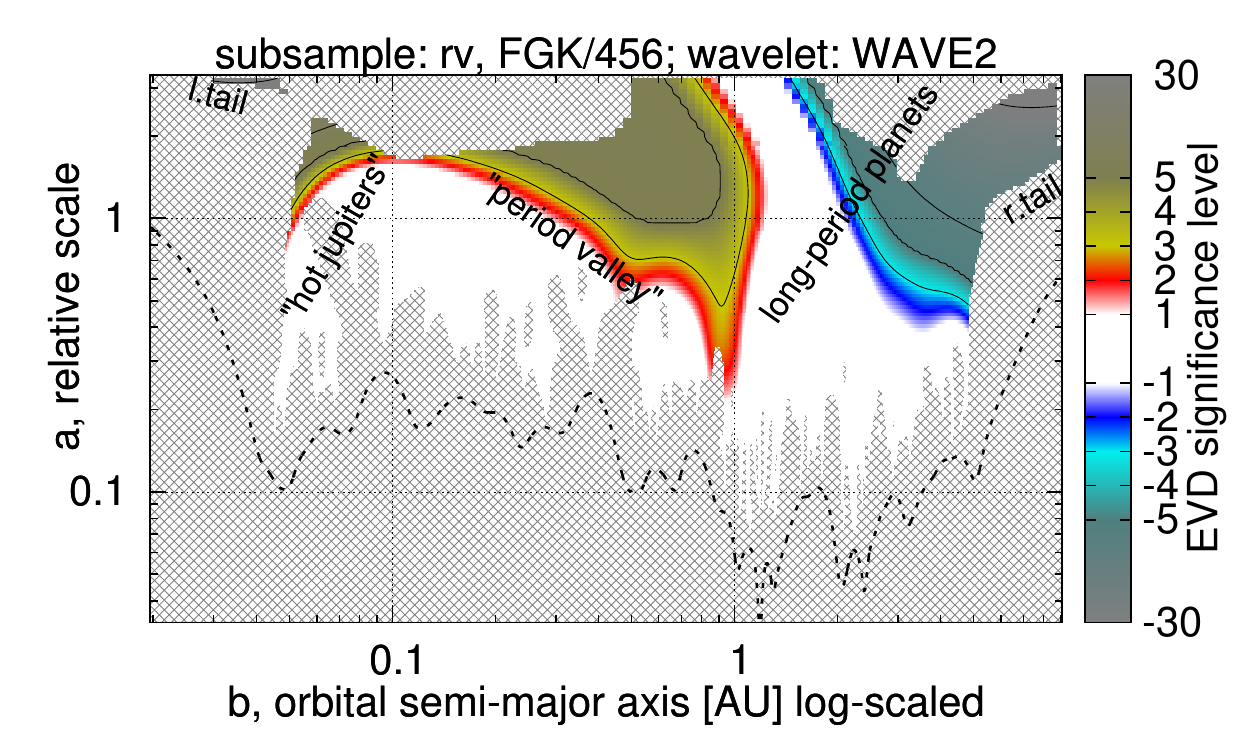}\\
\includegraphics[width=0.5\linewidth]{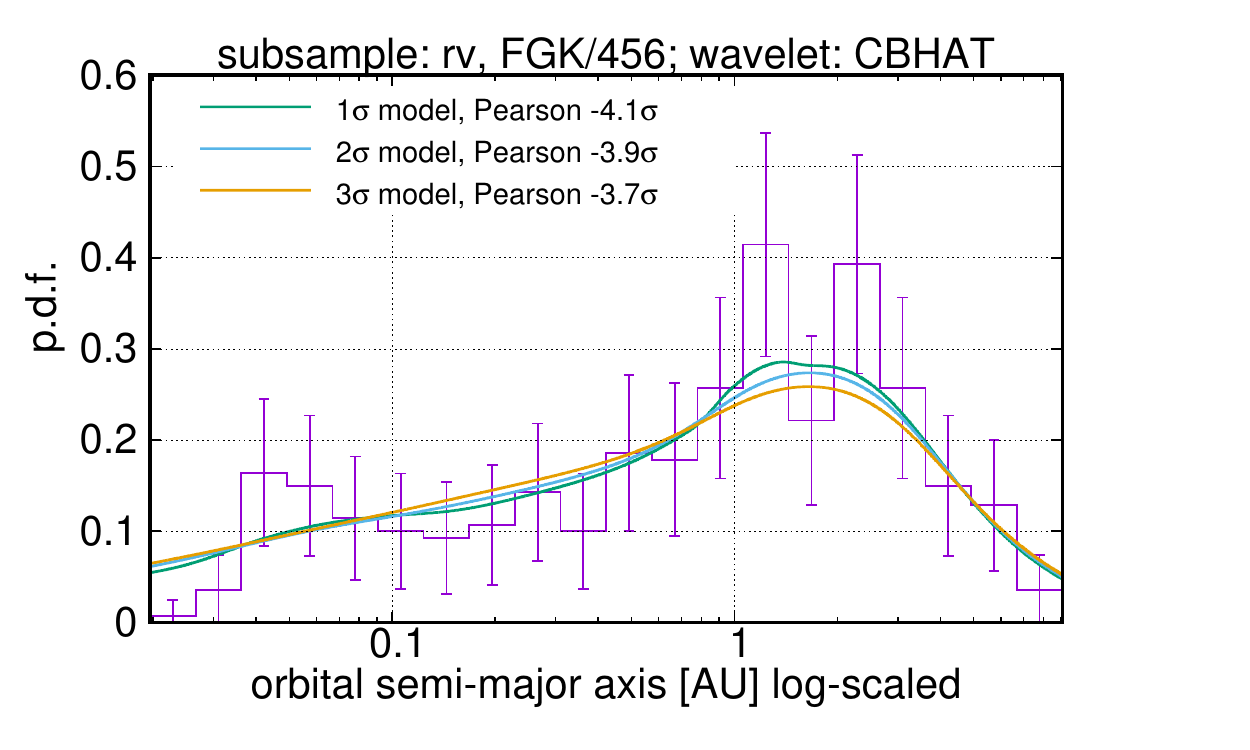} & \includegraphics[width=0.5\linewidth]{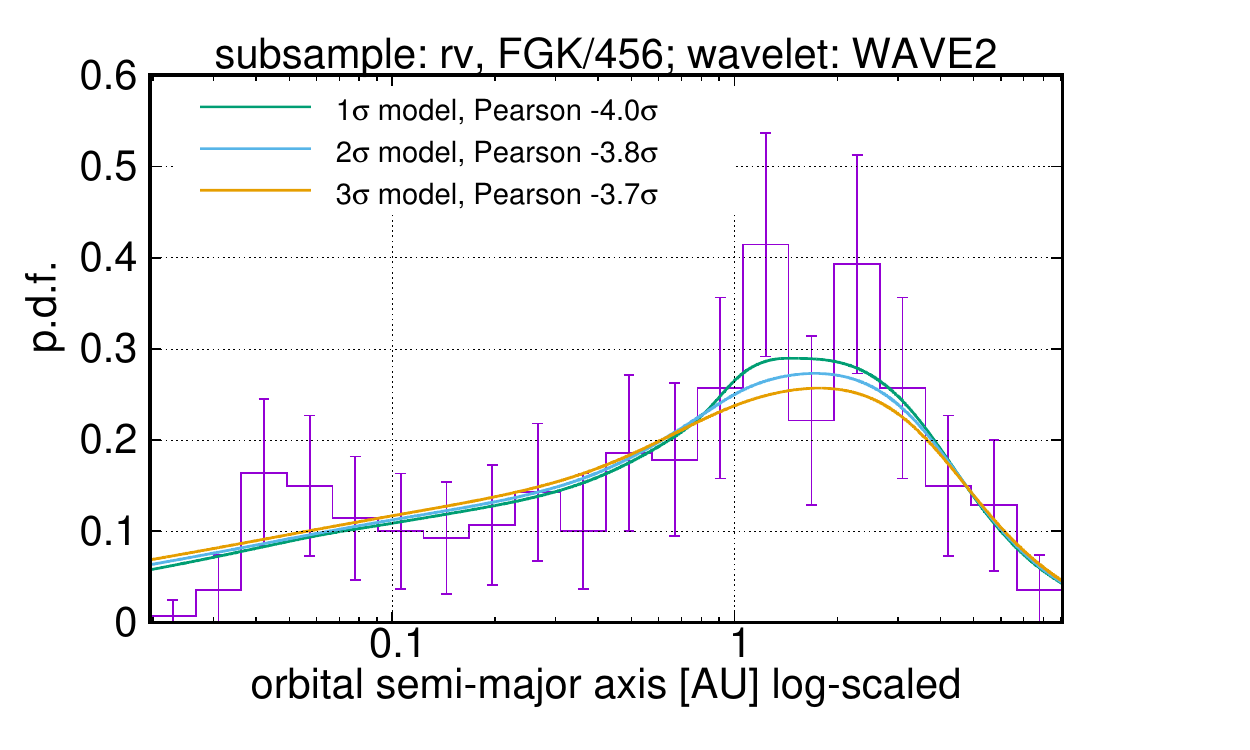}
\end{tabular}
\caption{Wavelet analyis of known exoplanetary candidates. Subsample: RV detection method,
FGK-type host stars, $N=456$. Variable: orbital semimajor axis. The figure layout is the
same as in Fig.~\ref{fig_spike10}.}
\label{fig_rv_sma}
\end{figure*}

Results for the period and semimajor axis distributions (both in logarithmic scale) are
shown in Fig.~\ref{fig_rv_prd} and~\ref{fig_rv_sma}. They refer to the \sampleid{rv.FGK}
subsample. It is well-known that these distributions should behave similarly, thanks to the
third Kepler law, so we consider them here together.

First of all, we can see the large-scale structure: a highly-populated wide peak of
long-period ``warm jupiters'', a shallow concentration of ``hot jupiters'' at short
periods, and a wide ``period valley'' between them. These are well-known features, more or
less succesfully explained by contemporary simulation works on planet formation and
migration, see e.g. \citet{IdaLin04,IdaLin08} and many more.

In particular, the predominant WJ maximum is attributed to the effect of the so-called
iceline accumulation. It was explained in \citep{IdaLin08} and extensively studied in many
works after \citep[e.g.][]{Schlaufman09,Hasegawa13}. Near the snow line physical processes
of ice particles migration, evaporation, outward diffusion and recondensation interact so
that the density of the material increases. This favours to planet formation just beyond
the ice line. Moreover, if a protoplanet grows to form a giant planet (above Saturn or
Jupiter mass), it then migrates modestly, so that the corresponding concentration of giant
planets gets shifted inward.

The inner boundary of the WJ family corresponds to the semimajor axis $A\sim 1$~AU or
orbital period $P\sim 300$~d.

In Fig.~\ref{fig_rv_prd}-\ref{fig_rv_sma} we can also see a fine-structured pattern, in the
period distribution especially. It appears as a smaller-scale local maximum layed over the
large-scale one, close to the upper boundary of the PV depression. In the reconstructed
p.d.f. curves this substructure looks like an overshoot peak, maybe even causing a
bimodality. However, we must be careful with these curves due to possible artifacts of the
matching pursuit. To gain more accurate understanding of what actually happens there, we
must refer to the corresponding wavelet transforms.

Let us focus on the period distribution at first, Fig.~\ref{fig_rv_prd}.

In its CBHAT wavelet transform, we can see a remarkable negative (cyan) spot at $P\sim
450$~d, with a relatively small scale $a\sim 0.3$.\footnote{Since we analyse the
distribution of $\log P$ actually, $a$ has the meaning of a relative scale. That is,
$a=0.3$ infers about $\pm 15$ per cent range from a given $P$.} It reaches $3$-sigma level
that confirms its high significance. Such a spot indicates that the p.d.f. attains a high
curvature at this location (with upward convexity). This does not necessarily indicate a
local maximum, but at least a zone of relatively abrupt transition between the PV and WJ
domains.

Such an abrupt transition is confirmed by the WAVE2 transform, where we can see a
high-significe domain near $P\sim 350$~d that descends to the same small scales.

All this suggests the following interpretation concerning the $\log P$ distribution of the
\sampleid{rv.FGK} sample:
\begin{enumerate}
\item there is a quick density gradient near $P\sim 350$~d, corresponding to the PV/WJ
boundary;
\item this gradient ends with a highly-curved p.d.f. convexity near $P\sim 450$~d that
further turns into the wide maximum of long-period planets.
\end{enumerate}
At this stage we still cannot say, if there is indeed an additional local maximum at $P\sim
450$~d or any separated subfamily of exoplanets.

To ensure that we talk about a subfamily, we would need to detect some side
\emph{separations} from the remaining sample of planets. That is, the small-scale spot at
$P\sim 450$~d in the CBHAT wavelet map should be accompanied by a couple of counter-signed
spots on both sides. They would indicate concave zones in the p.d.f., meaning that the
central convexity represents a standalone formation.

In fact, we do note some secondary spot in Fig.~\ref{fig_rv_prd}, but unfortunately its
significance is too small, below $g=2$.

Now we can use theoretic results to decide what sample to analyse further. Simulations of
the iceline barrier effect from e.g. \citep{IdaLin08} predict that in the range of
interest, $P\sim 300-1000$~d, less massive planets would only represent an interfering
background that blurs the giant planets subsample. The positions for the ``Warm Jupiters''
and ``Warm Neptunes'' maxima should be displaced considerably, because their migration
histories are different. Therefore, it might be reasonable to analyse the subsample of
giant planets, \sampleid{rv.FGK.hmass}. This subsample should reveal the iceline accumulation
effect in a more clean manner.

\begin{figure*}
\begin{tabular}{@{}l@{}l@{}}
\includegraphics[width=0.5\linewidth]{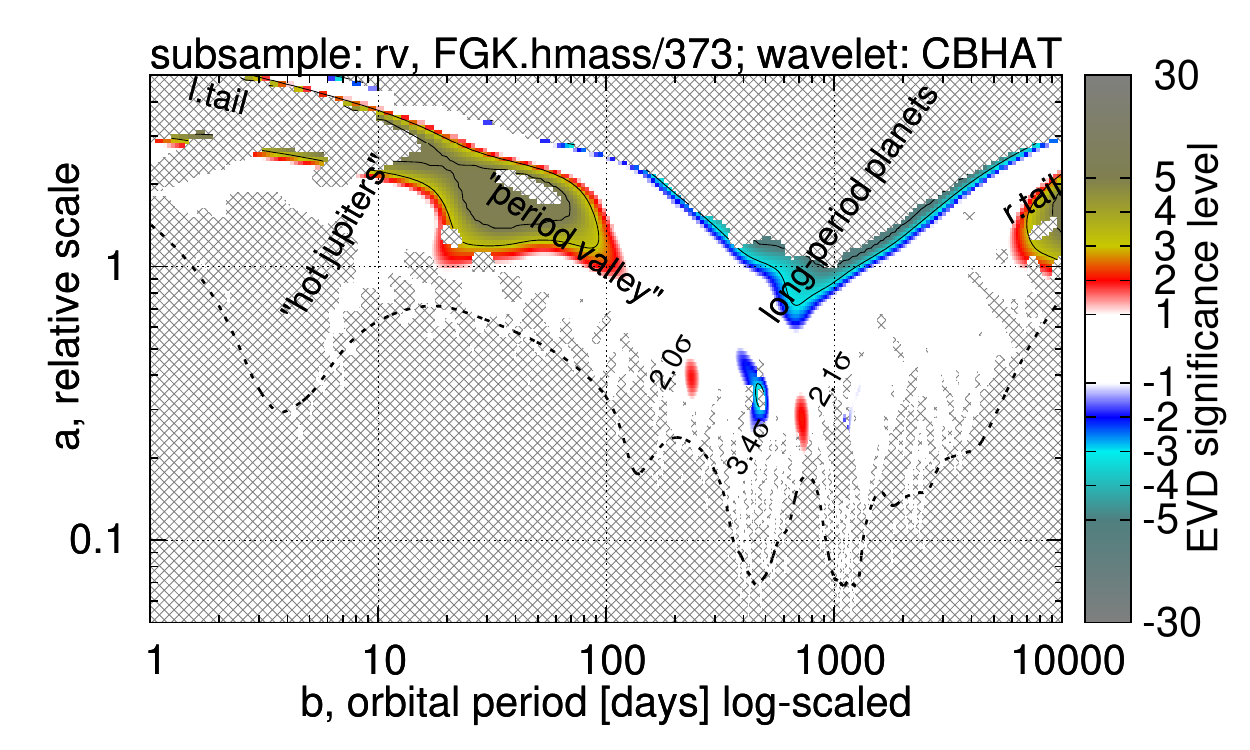} & \includegraphics[width=0.5\linewidth]{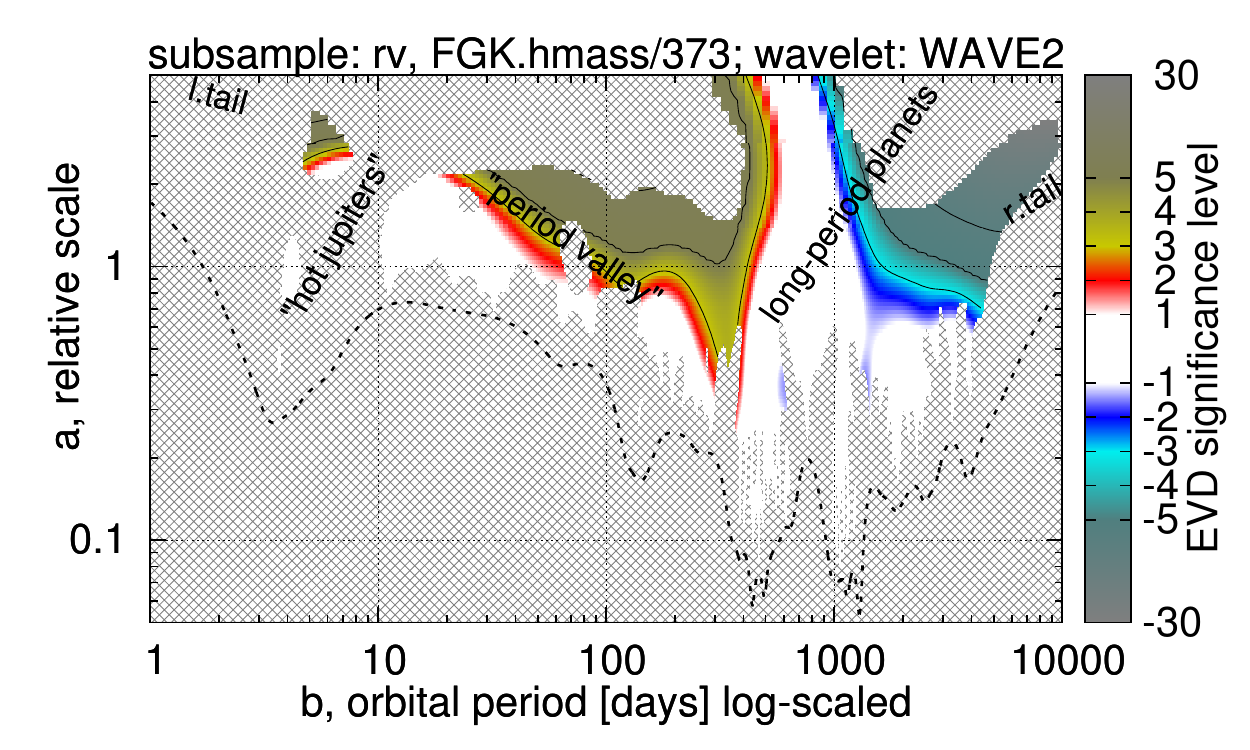}\\
\includegraphics[width=0.5\linewidth]{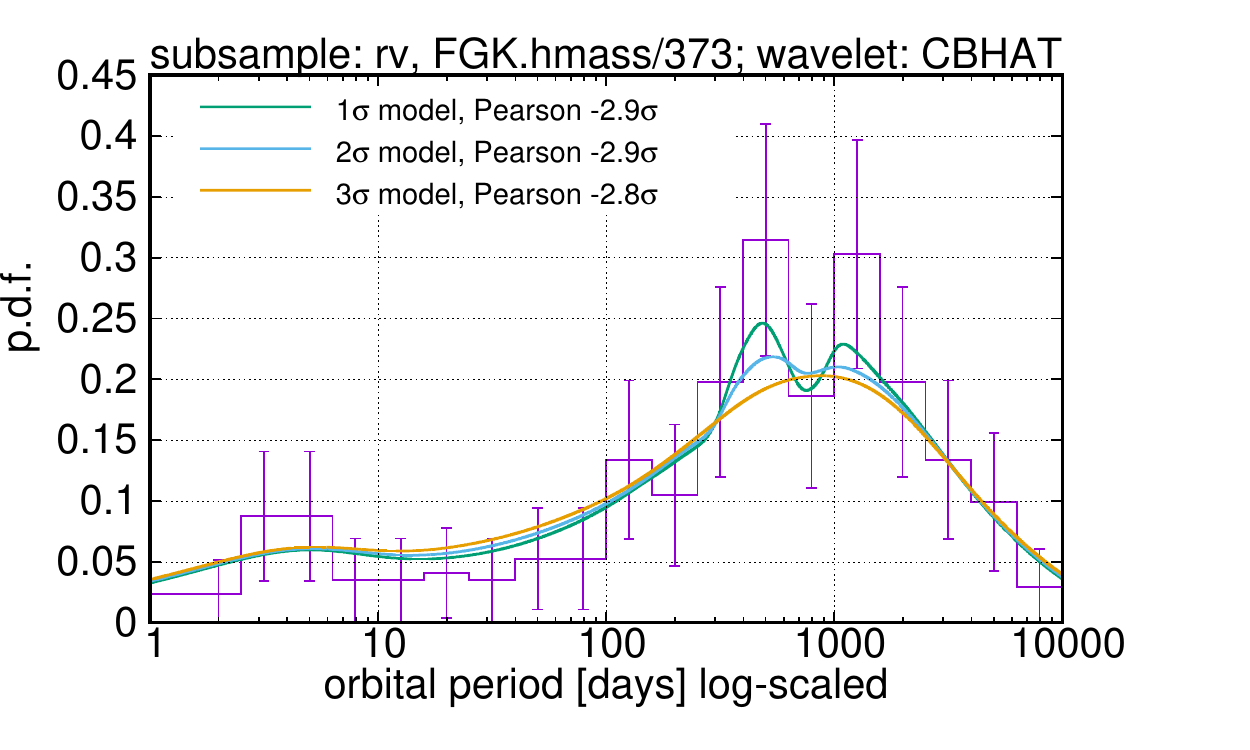} & \includegraphics[width=0.5\linewidth]{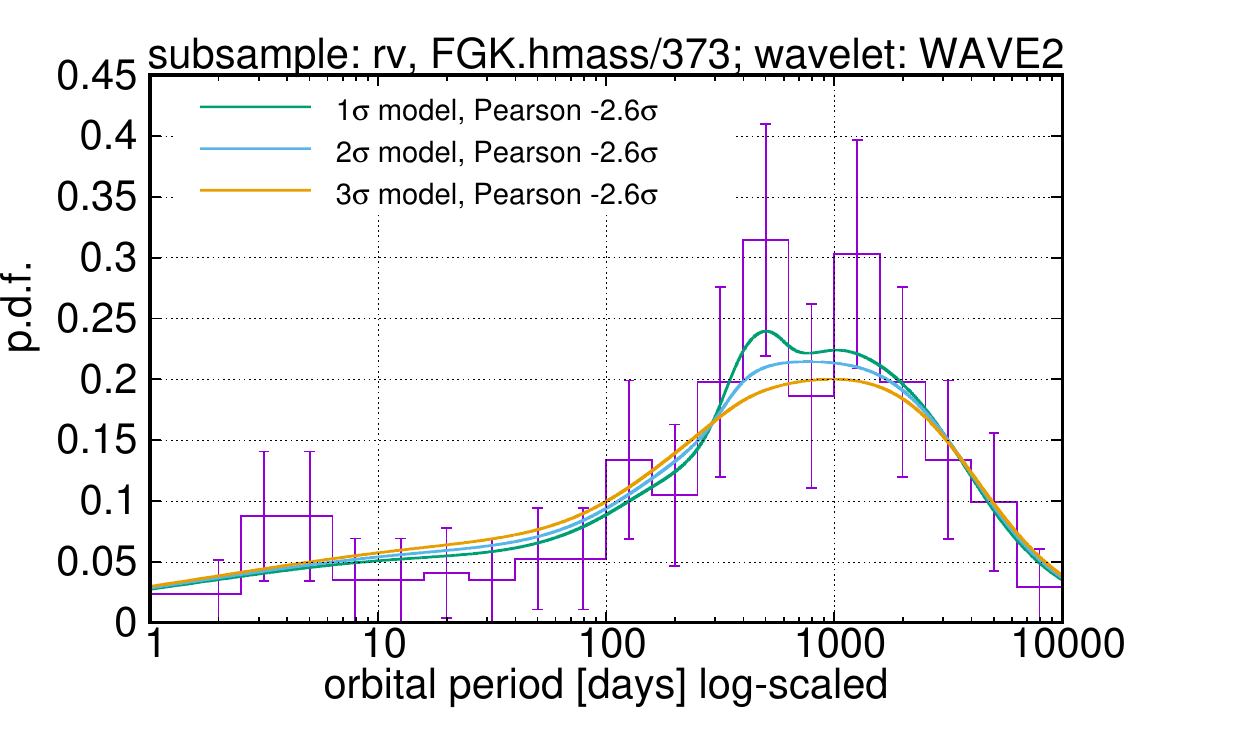}
\end{tabular}
\caption{Wavelet analyis of known exoplanetary candidates. Subsample: RV detection method,
FGK-type host stars, higher-mass planets ($m \sin i>0.3 M_{\rm Jup}$), $N=373$. Variable:
orbital period. Fine-scale structures are labelled with the significance quantile $g$. The
figure layout is the same as in Fig.~\ref{fig_spike10}.}
\label{fig_rv_prd_hm}
\end{figure*}

\begin{figure*}
\begin{tabular}{@{}l@{}l@{}}
\includegraphics[width=0.5\linewidth]{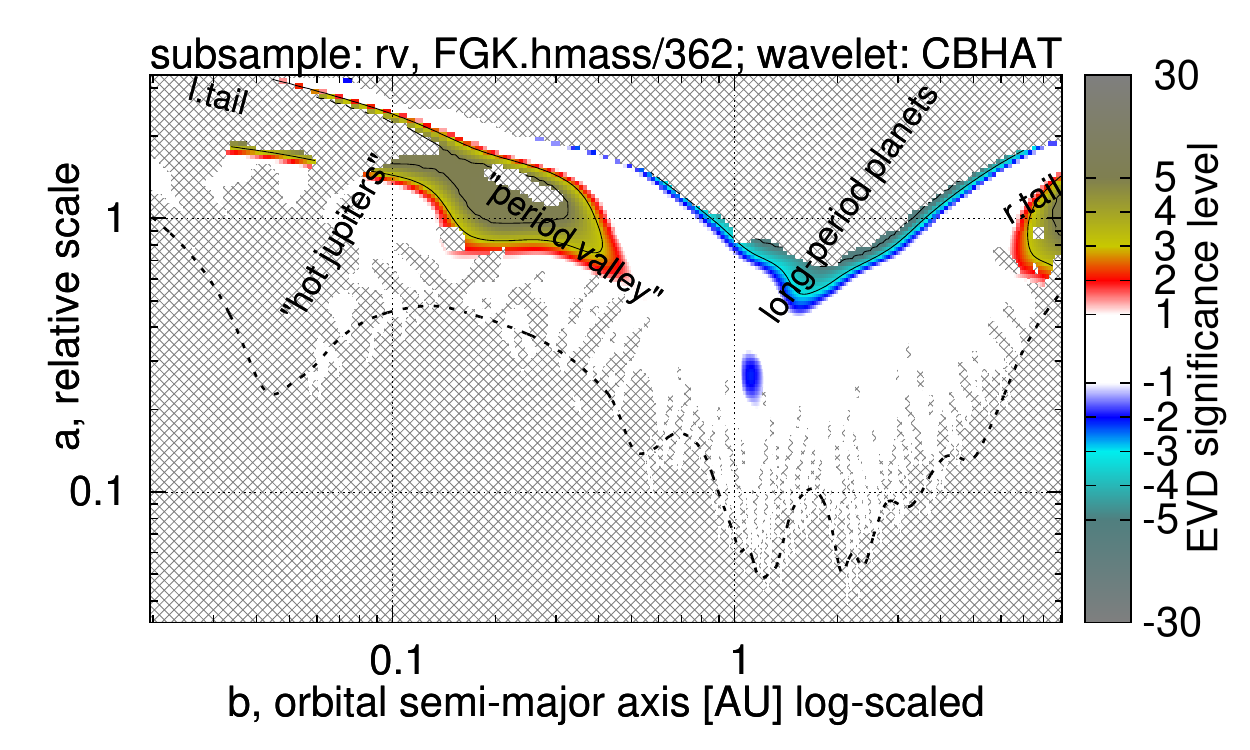} & \includegraphics[width=0.5\linewidth]{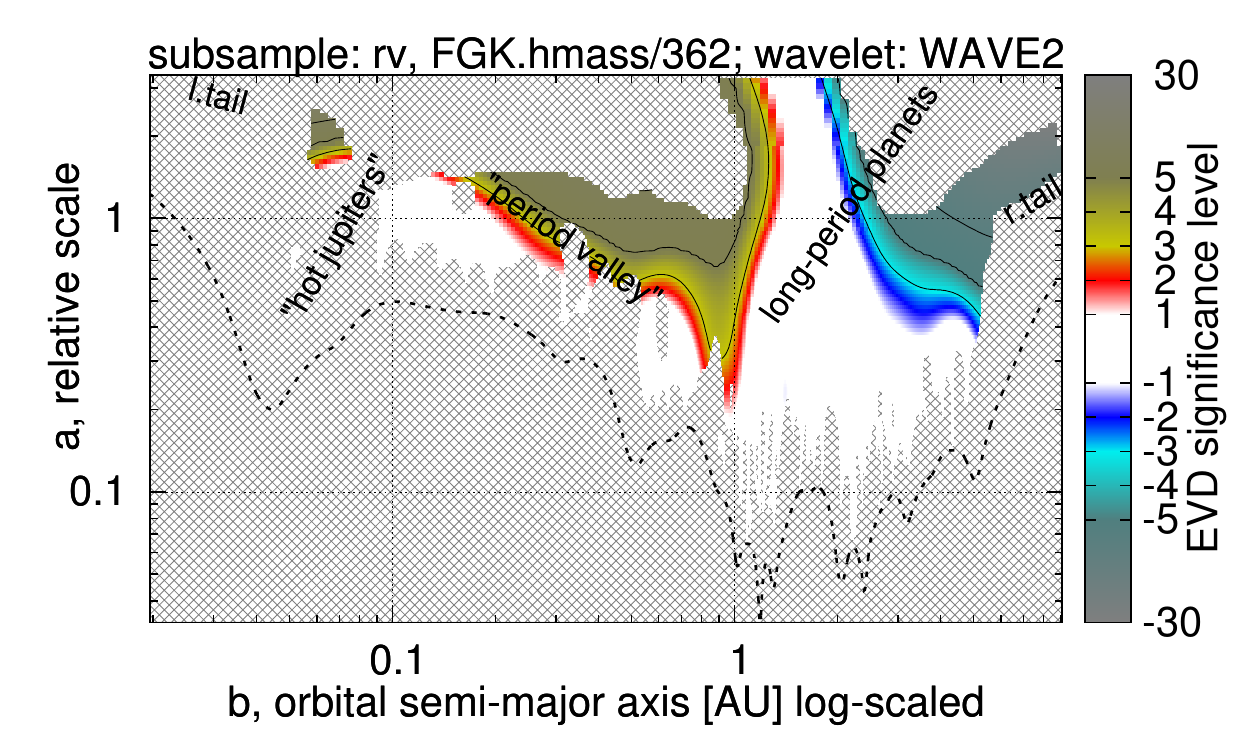}\\
\includegraphics[width=0.5\linewidth]{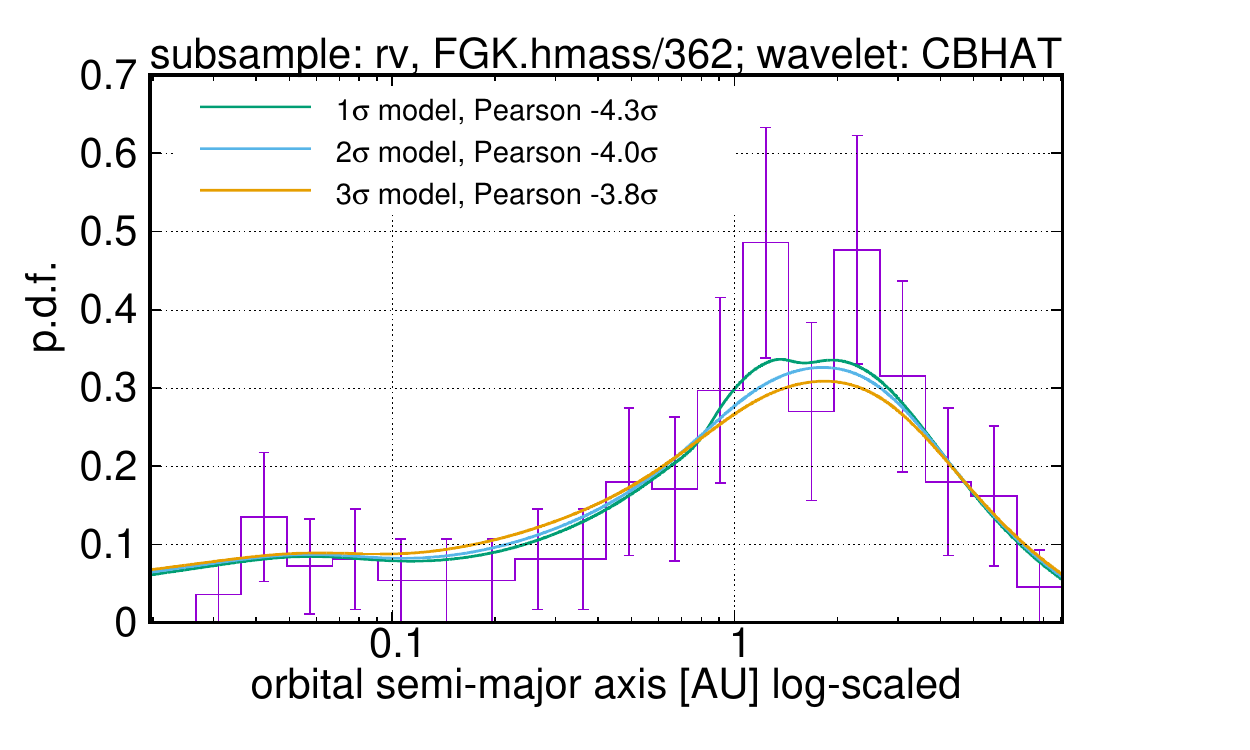} & \includegraphics[width=0.5\linewidth]{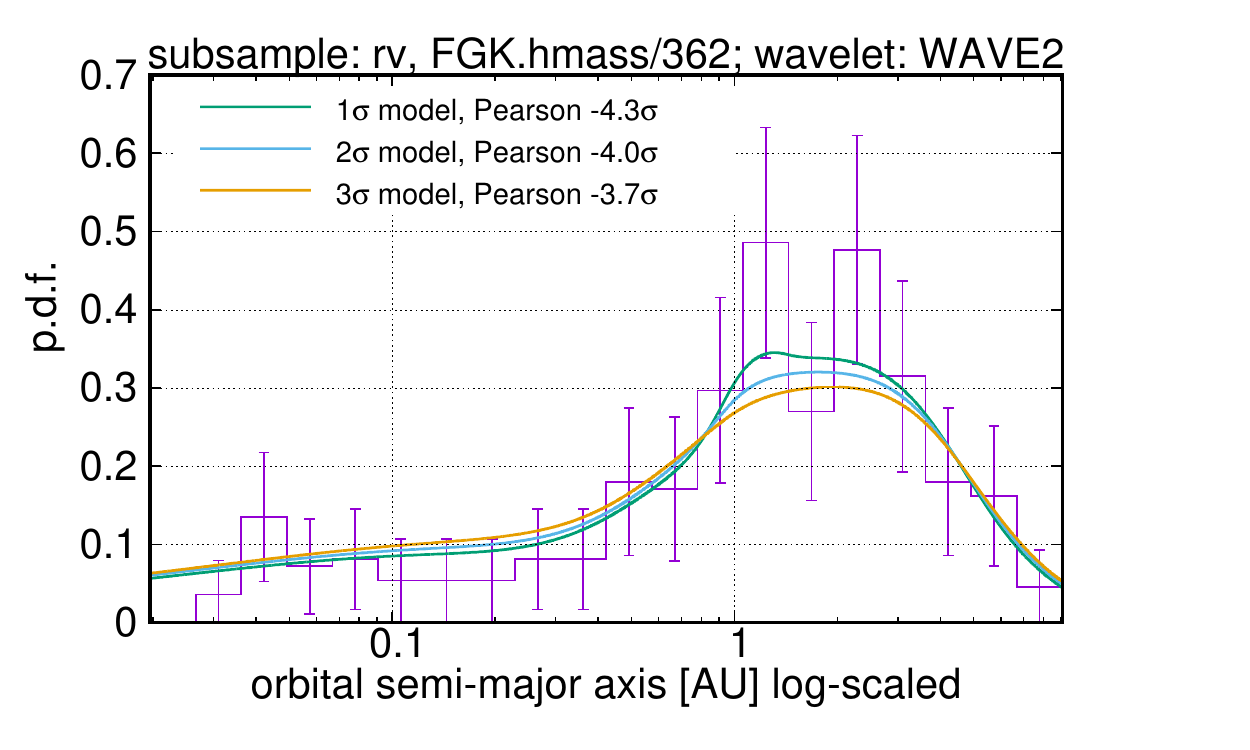}
\end{tabular}
\caption{Wavelet analyis of known exoplanetary candidates. Subsample: RV detection method,
FGK-type host stars, higher-mass planets ($m \sin i>0.3 M_{\rm Jup}$), $N=362$. Variable:
semimajor axis. The figure layout is the same as in Fig.~\ref{fig_spike10}.}
\label{fig_rv_sma_hm}
\end{figure*}

These results are presented in Fig.~\ref{fig_rv_prd_hm}-\ref{fig_rv_sma_hm}. We can see
that all the fine-scale structures within the range $P\sim 300-1000$~d indeed gained more
magnitude. In the CBHAT wavelet transform, we can see the primary (negative) spot at $P\sim
450$~d that got the significance $g=3.4$, accompanied by a pair of the counter-signed
secondary spots at $P\sim 700$ with $g=2.1$, and at $P\sim 250$~d with $g=2.0$. Such a
triplet pattern highlights a statistically separated narrow subfamily of giant planets in
the period range $\sim 300-600$~d. In the p.d.f. graph it looks like an overshooting peak
right after an upturn.

This significance increase was achieved by restricting the exoplanetary sample from {\sc
rv.FGK} to \sampleid{rv.FGK.hmass}, but this is not a p-hacking or data-dependent analysis. It
would be those, if we just searched many subsamples in the hope to finally obtain a desired
significant result. But our decision of how to restrict the sample relied on the previous
theoretic knowledge, regardless of any statistical analysis already done. The second sample
could equally decrease the significance of the result rather than increase it. In such a
case, this subsequent analysis attempt does provide additional evidence.

In either case, there is no significant hints of bimodality within the WJ maximum. Although
some of the p.d.f. plots, e.g. in Fig.~\ref{fig_rv_prd_hm}, suggest bimodality, it looks
like a reconstruction artifact appearing due to inherent weaknesses of the matching pursuit
algorithm.

So far we considered only the period distribution, but orbital semimajor axes are
apparently more appropriate concerning the iceline barrier effect. Disappointedly, this
distribution does not reveal so much subtle details
(Fig.~\ref{fig_rv_sma},\ref{fig_rv_sma_hm}). With the CBHAT wavelet, we can see just a
single small-scale spot at $A\sim 1.1$~AU indicating a local convexity. In the subsample of
giant exoplanets, its significance barely reaches the two-sigma level, and it becomes even
worse in the larger \sampleid{rv.FGK} sample.

The semimajor axis $A$ is merely a derived parameter, while the primary ``observed''
quantity is the orbital period $P$. Knowing $P$, we can recover $A$ using the third Kepler
law:
\begin{equation}
P \propto M_\star^{-\frac{1}{2}} A^{\frac{3}{2}}, \quad A \propto M_\star^{\frac{1}{3}} P^{\frac{2}{3}},
\label{KeplerIII}
\end{equation}
where $M_\star$ is the star mass. However, the value of $M_\star$ is not the same for all
exoplanets. In this case~(\ref{KeplerIII}) defines a statistical correlation between $P$
and $A$ rather than a strict binding. That is, the $P$- and $A$-distributions should look
generally similar, but small-scale details would be different due to the scatter in
$M_\star$. If a detail appeared sharp in the $P$-distribution, it gets blurred in the
$A$-distribution, and vice versa. It cannot remain sharp in the both.\footnote{Note that
details in the $A$-distribution should be always blurred additionaly, because $M_\star$
often involves remarkable uncertainties. But this effect is smaller than the scatter of
$M_\star$ in the sample.}

Our analysis revealed that small-scale subfamily discussed above appears sharply in $P$.
But then it should necessarily fade out in $A$. Since this is a necessary and predictable
behavior, it does not make the detection less reliable from the statistical point of view.

But then an interesting question emerges: should the iceline barrier effect generate more
sharp details in $a$- or $P$-distribution? Unfortunately, simple considerations did not
provide us with a definite answer. By combining~(\ref{KeplerIII}) with the inverse-squares
law, it is easy to assess the distance of the ice line, $a_\mathrm{ice}$, and the
corresponding Keplerian period, $P_\mathrm{ice}$, as a function of the star luminosity:
\begin{equation}
L_\star A_\mathrm{ice}^{-2} = \const \implies \quad A_\mathrm{ice} \propto L_\star^{\frac{1}{2}}, \quad
P_\mathrm{ice} \propto M_\star^{-\frac{1}{2}} L_\star^{\frac{3}{4}}.
\end{equation}
And by applying the mass-luminosity relation, $L_\star\sim M_\star^4$, we obtain
\begin{equation}
A_\mathrm{ice} \sim M_\star^2, \qquad P_\mathrm{ice} \sim M_\star^{-\frac{5}{2}}.
\label{iceline}
\end{equation}
But then it follows that \emph{both} the $A$- and $P$-distributions should be blurred
approximately equally, because $A_\mathrm{ice}$ and $P_\mathrm{ice}$ depend on $M_\star$
via almost the same power degree ($2$ or $-2.5$, the sign does not matter).

However, the formulae~(\ref{iceline}) do not take into account the migration of planets.
The migration rate and its distance should correlate with $M_\star$ somehow, but this
correlation depends on multiple factors. It is difficult to predict without performing
simulations like those available in \citep{IdaLin08}. Depending on the sign of the
correlation, it may sharpen either of the two distributions. In principle, this can be used
to verify the iceline nature of the exoplanetary subfamily for $P\sim 300-600$~d.

At last, we recall that the database of \emph{The Extrasolar Planets Encyclopaedia} is
heterogeneous, mixing planet candidates discovered by multiple independent Doppler
programmes. Such mixed samples might demonstrate spurous effects due to an overlap of
different detection thresholds. However, most radial-velocity surveys are very aged
already, and they are pretty complete in the period range of $300-600$~d (for giant planets
at least). Moreover, the orbital period $P$ is one of the best-determined exoplanetary
characteristics. It is therefore unlikely that observational selection, sample
heterogenity, or possibly inaccurate data could cause any unusual statistics in this period
range.

\subsection{Transiting planets radii: confirming evaporation valley}
\label{sec_rad}
The radii distribution among Kepler exoplanetary candidates demonstrates signs of a
bimodality \citep{Fulton17}. There are two local maxima of the density at $R=1.3R_\oplus$
and $R=2.4R_\oplus$, while planets with $R$ in the intermediate range $1.5-2.0 R_\oplus$
are comparatively rare. This effect was named the ``evaporation valley'', and it has got a
detailed theoretic treatment already \citep{OwenYanqin17}. In a few words, higher-mass
planets are capable to preserve their dense atmospheres even if they are exposed to intense
irradiation from the star, but low-mass planets loose most of their atmospheres eventually,
reducing their sizes to even smaller values than they initially were.

\begin{figure*}
\begin{tabular}{@{}l@{}l@{}}
\includegraphics[width=0.5\linewidth]{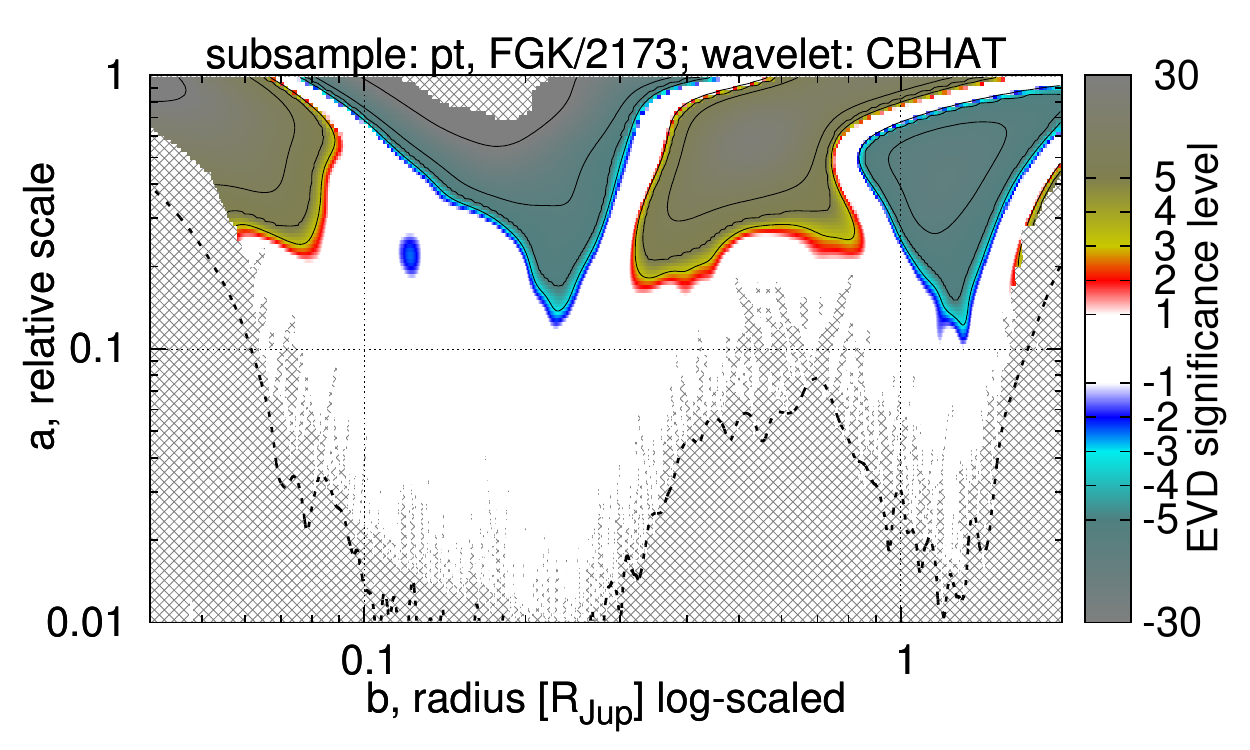} & \includegraphics[width=0.5\linewidth]{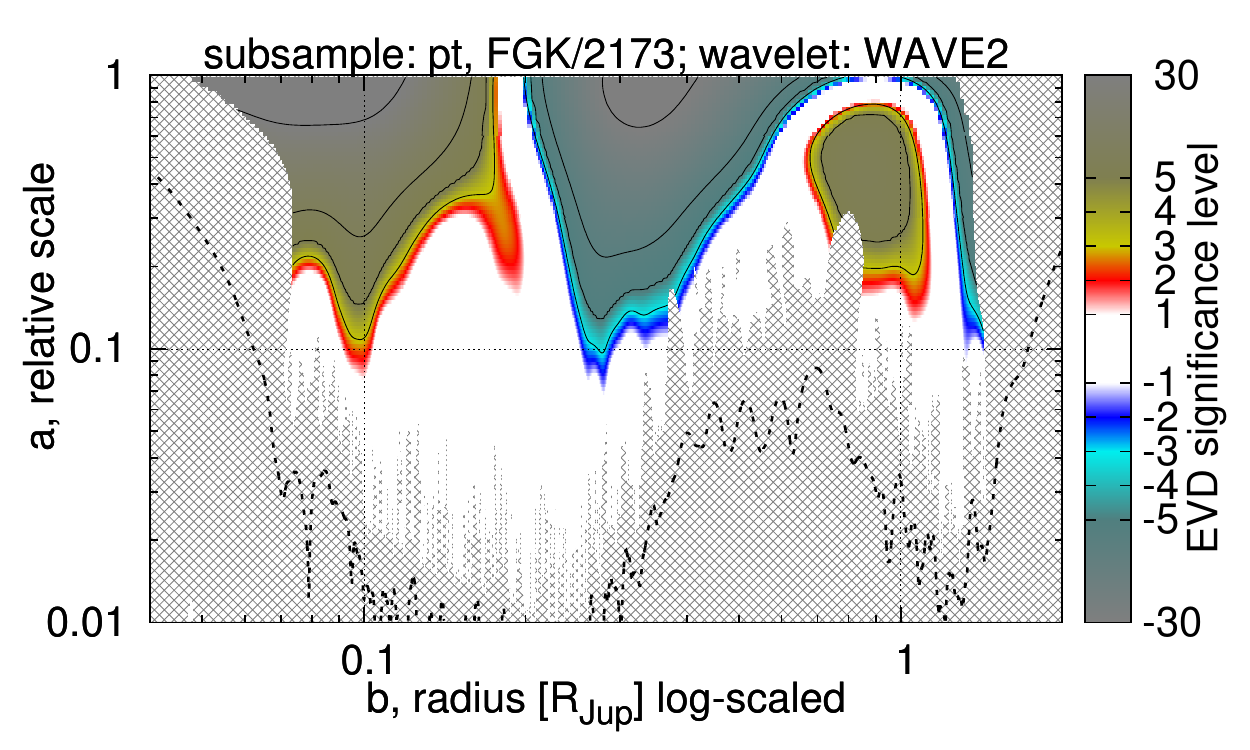}\\
\includegraphics[width=0.5\linewidth]{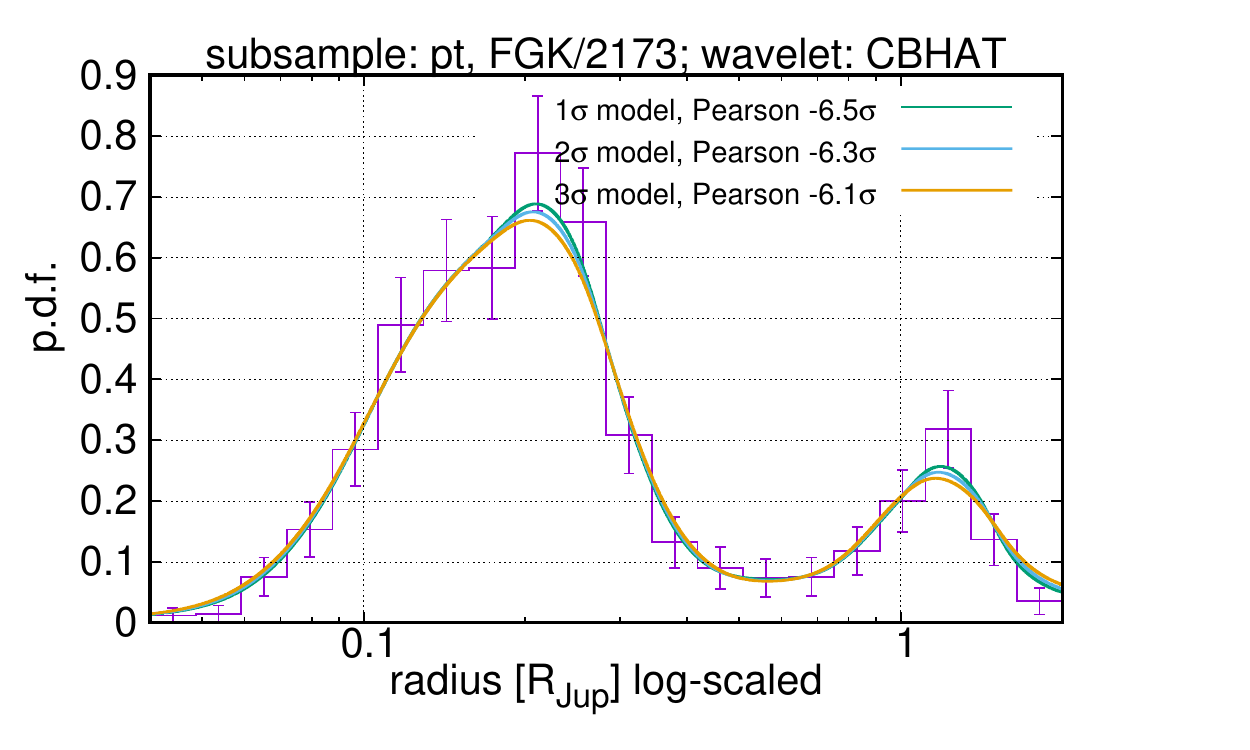} & \includegraphics[width=0.5\linewidth]{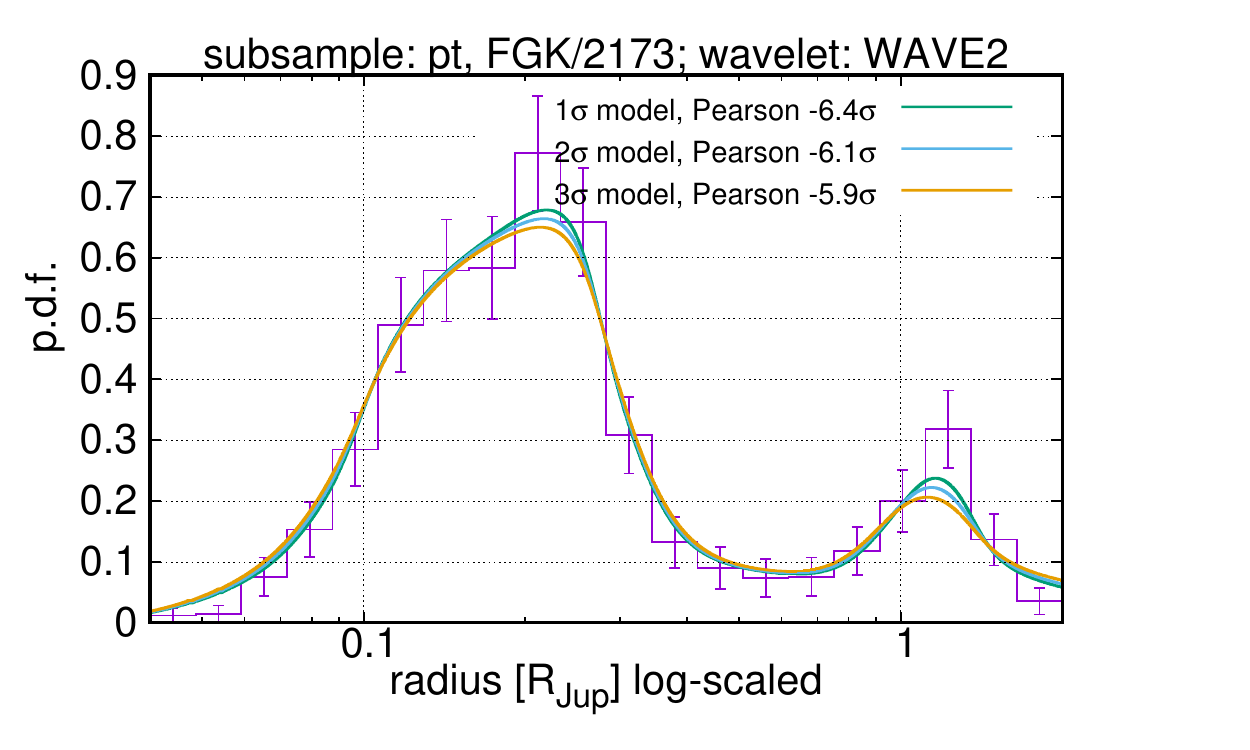}
\end{tabular}
\caption{Wavelet analyis of known exoplanetary candidates. Subsample: Primary transit
detection method, FGK-type host stars, $N=2173$. Variable: planet radius. The figure layout
is the same as in Fig.~\ref{fig_spike10}.}
\label{fig_pt_rad}
\end{figure*}

We were interested to apply our algorithm to this case. First of all, we followed our
common protocol set in advance, i.e. we analysed the transit sample \sampleid{pt.FGK}, drawn
from the database of \emph{The Extrasolar Planets Encyclopaedia}. This sample includes
Kepler candidates, as well as those detected by more common but less sensitive ground-based
surveys. The results for the planet radii distribution are shown in Fig.~\ref{fig_pt_rad}.

A bimodality is obvious in the reconstructed p.d.f., though this is unrelated to the
``evaporation valley''. The peaks are for the families of giant Jupiter-sized planets and
of the Earth-/superearth-sized ones. They are separated by a wide minimum in the range
$0.4-0.8 R_\mathrm{Jup}$. These major families with an intermediate desert are undoubtful,
and were predicted quite early, at least by \citet{IdaLin04}.

Apparently nothing related to the ``evaporation valley'' can be spotted in this p.d.f., but
we can see in the CBHAT wavelet transform a small negative (blue) spot located at $R\sim
0.12 R_\mathrm{Jup}$ or $1.3 R_\oplus$. It represents a hint of a smaller-scale p.d.f.
convexity. Its significance is above two-sigma, but the absolute magnitude appears so small
that practically nothing can be seen in the reconstructed p.d.f., or even in the histogram,
at this location.

In the WAVE2 wavelet transform there is a fancy anomaly at $R=0.2 R_\mathrm{Jup}$ and
$a=0.2-0.3$, indicating a relatively more quick p.d.f. upturn at this location, though it
is not obvious in the reconstructed model.

Likely, these subtle hints is all what is left from the ``evaporation valley'' in this
sample. The database in \citep{Schneider} is extremely heterogeneous for transiting
planets. And contrary to the previous subsection, we are now interested in the range where
many surveys appear incomplete. Transit programmes have different detection limits, and
their selection functions get overlayed, imposing spurous distortions to the statistics.
Besides, it seems that this catalog is quite ``dirty'' with respect to false positives and
unreliable values. It is not very much surprising that the ``evaporation valley'' could not
be reliably reproduced with these data, although our method was able to detect some
remaining hints.

\begin{figure*}
\begin{tabular}{@{}l@{}l@{}}
\includegraphics[width=0.5\linewidth]{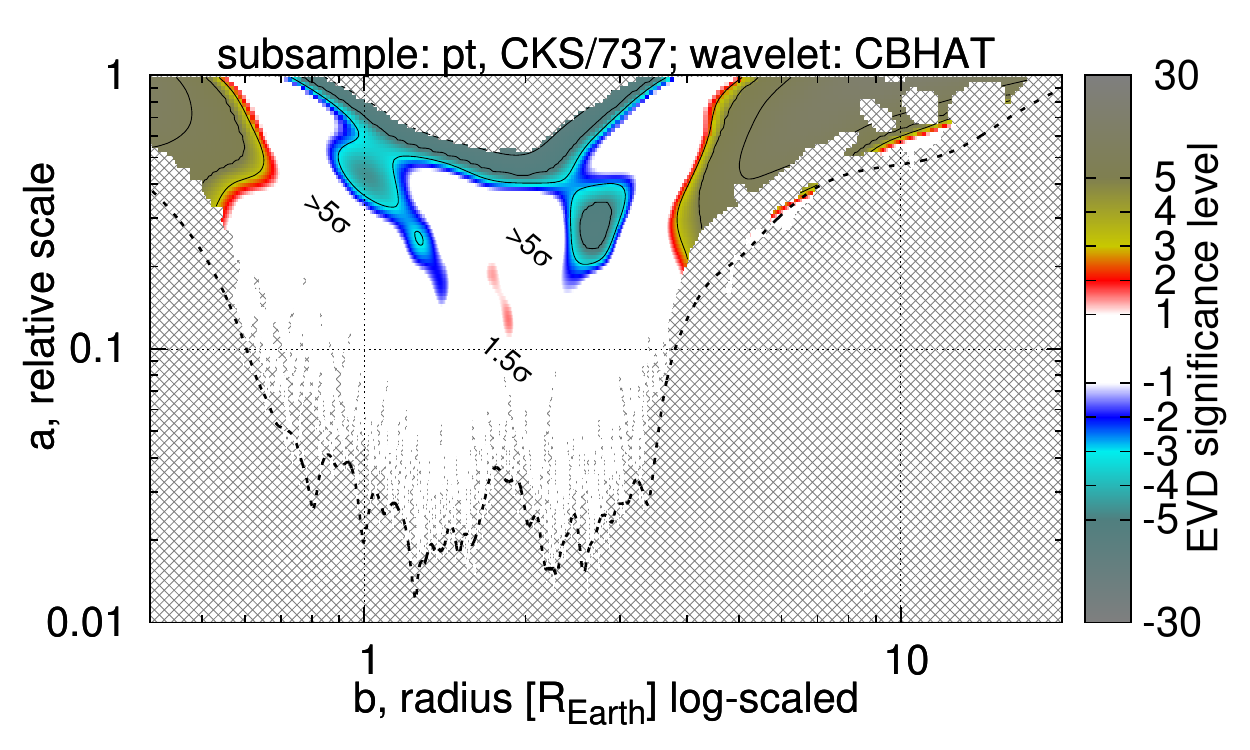} & \includegraphics[width=0.5\linewidth]{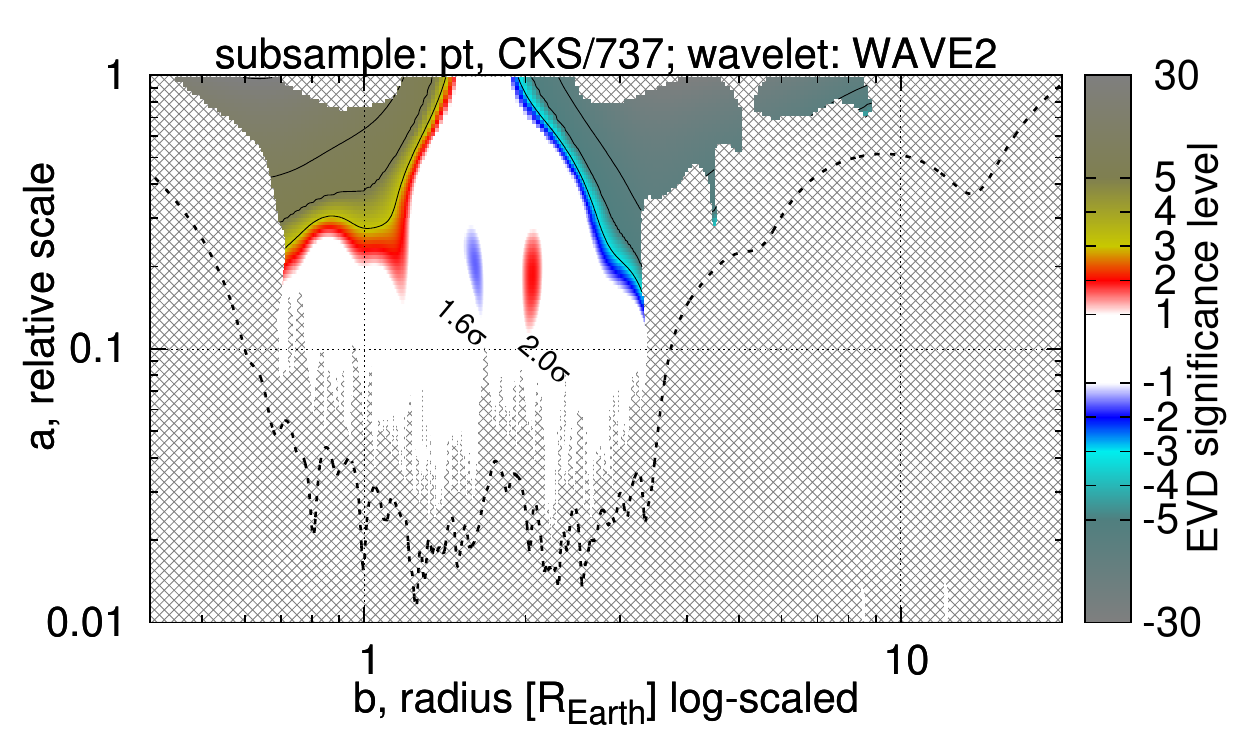}\\
\includegraphics[width=0.5\linewidth]{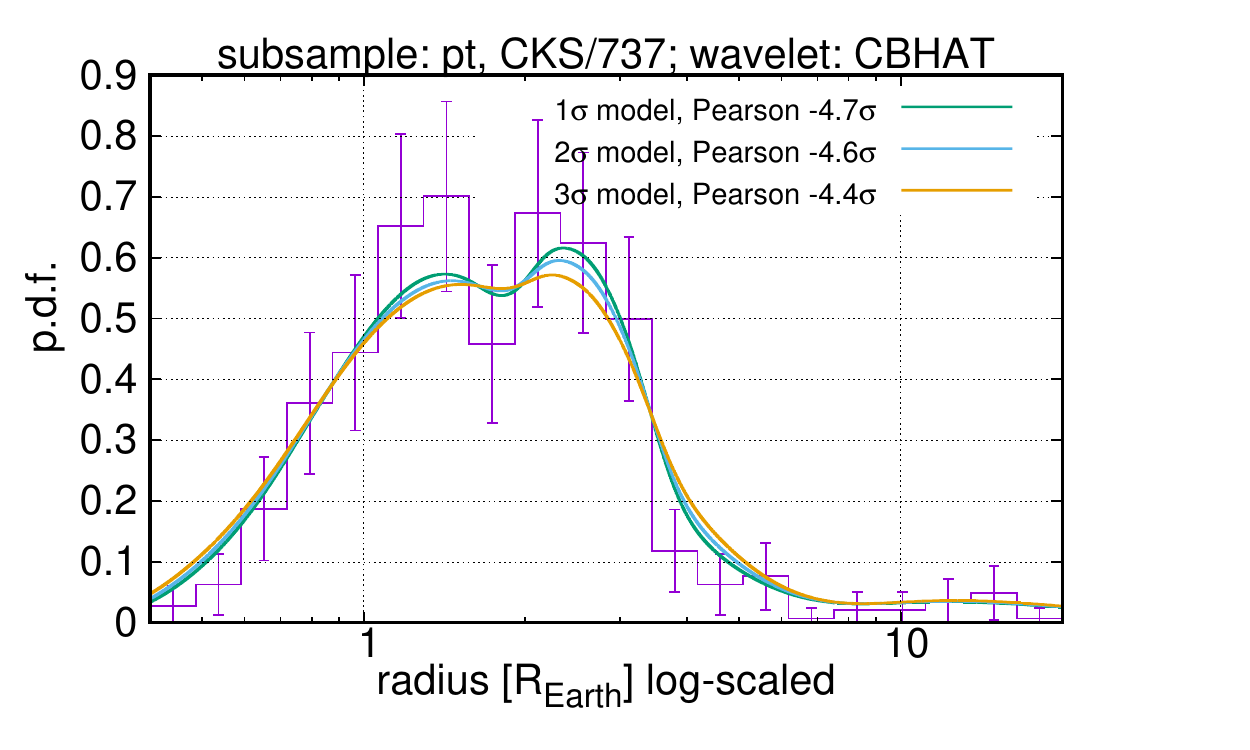} & \includegraphics[width=0.5\linewidth]{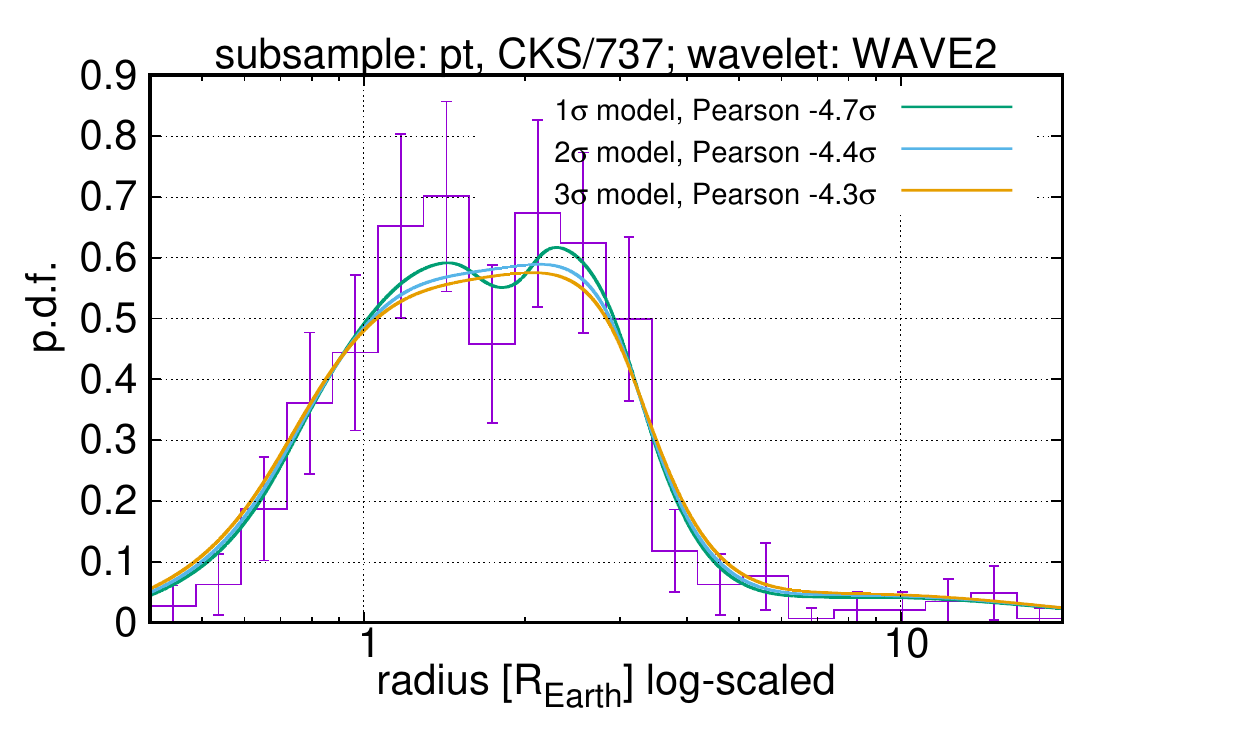}
\end{tabular}
\caption{Wavelet analyis of known exoplanetary candidates. Subsample: California Kepler
Sample (CKS) with filters (see text), $N=737$. Variable: planet radius. Fine-scale
structures are labelled with the significance quantile $g$. The figure layout is the same
as in Fig.~\ref{fig_spike10}.}
\label{fig_CKS_rad}
\end{figure*}

After that, we applied our analysis to the more homogeneous CKS data cleaned as described
in Sect.~\ref{sec_samples}. The results are shown in Fig.~\ref{fig_CKS_rad}. Now we can
clearly see a bimodal radii distribution with maxima at $1.3 R_\oplus$ and $2.6 R_\oplus$,
and a gap between. Though our sample is not strictly the same as in \citep{Fulton17}, the
histogram does not reveal any obvious visual difference.

But contrary to \citet{Fulton17}, our analysis does not provide so strong statistical
support to the bimodality. Although the CBHAT wavelet clearly reveals a pair of convexity
zones at $1.3$ and $2.4 R_\oplus$, the concavity between them does not reach enough
significance. The convexities are undoubtful, but on themselves they do not yet tell
anything in support of the bimodality and do not confirm standalone exoplanetary families.
Without a clear intermediate gap, the density function may have just a plateau-like shape
with more abrupt sides. That would mean just a single density maximum without any
submafilies.

Ensuring the bimodality necessarily requires to detect a gap, or at least an intermediate
concavity. However, this feature does not appear statistically significant with the CBHAT
wavelet (only $g=1.5$). A concavity and even a local minimum is obvious in the
reconstructed p.d.f. even for $g=3$, but this looks like an artifact of the matching
pursuit algorithm, since such a significance level is not supported by the wavelet
transform. So, the CBHAT wavelet fails to detect the ``evaporation valley''.

The WAVE2 transform in Fig.~\ref{fig_CKS_rad} appears a bit more promising: it resolves a
dipole-like standalone pair of downturn/upturn patterns, and the upturn part at
$R=2R_\oplus$ appears moderately significant, $g=2$.

Being puzzled by such an inconsistency with \citet{Fulton17}, we investigated the
argumentation from that work.
\begin{enumerate}
\item \emph{Non-parametric Kolmogorov--Smirnov and Anderson--Darling tests to verify the
agreement with a loguniform distribution.} Even if the $R$-distribution does deviate from
the loguniform one significantly, this conclusion tells us nothing in favour of possible
bimodality. There are many unimodal distributions that could be different from the
loguniform one. The K--S and A--D tests are incapable to provide any information concerning
multimodality. They are not decisive in this task.

Besides, the application of those tests seems illegal, because the loguniform model was not
completely specified a priori. The left and right cutoff positions were unknown to the
authors \emph{a priori} and were guessed from the same data, which is a condition for
data-dependent analysis.

\item \emph{The dip test of unimodality by \citet{Hartigan85}.} The test is well-known and
useful in itself, but its application nonetheless seems invalid due to a hidden
pre-analysis. It is likely that the authors decided to apply this particular very specific
test only because the bimodality was initially suspected by visual inspection of the
histogram for the same Kepler data. If the data were different then such a suspection would
not appear, and the question of bimodality would not arose at all. In such a case, some
other pecularity could be spotted, triggering the use of another specific test designed to
detect patterns of that type. Again, we can clearly see a condition for a data-dependent
analysis: the hypothesis was formulated from the same data that were then tested as if this
hypothesis was given in advance.

We nonetheless applied the \citet{Hartigan85} test to our version of the CKS sample to
compare the results. We failed to reproduce the same high significance level as
\citet{Fulton17}. Instead of $\FAP=0.14$ per cent (or $g=3.2$), for the $\log R$ data we
obtained merely $\FAP=3$ per cent or $2.2$-sigma significance. If it is not a software bug
in either work\footnote{We used a public {\sc python} implementation of the test, available
at \url{https://github.com/alimuldal/diptest}.} then the test itself might be not very
robust. We considered other CKS subsamples, filtering the data at different levels of
tolerance. The best significance from the dip test was $\FAP=0.6$ per cent, if we preserved
the radii with uncertainties greater than $10$ per cent. However, even this significance is
much worse than \citet{Fulton17} reported.

\item \emph{Fitting a spline model and spline deconvolution.} The main conclusion was that
because of a $\sim 12$ per cent uncertainty in the planet radii, the ``observed''
$R$-distribution is merely a smoothed version of the original, ``underlying'', distribution
of the precise radii. Hence, the true distribution must have a deeper bimodality gap than
it seems from the observation.

There is no doubts on this conclusion, but it does not offer us any additional evidence in
favour of the bimodality, because it has nothing to do with the statistical significance.
The true gap should be deeper only if it actually exists, i.e. if the observed one is not a
noise artifact. Otherwise, the gap just does not exist in the both distributions.
\end{enumerate}

The conclusion is that the hints of the gap in the radii distribution are not ignorable,
but its significance remains rather marginal. Its reliability is much worse than
\citet{Fulton17} claimed. We definitely cannot give it any ``strong support''. An
optimistic significance estimate is about two-sigma level.

\subsection{Orbital eccentricities}
\label{sec_ecc}
We treated the orbital eccentricity $e$ as a symmetric parameter, extending its
distribution to the negative range, $e \Leftrightarrow -e$. This was necessary to avoid an
artificial distribution cutoff at $e=0$, in order to process the low-eccentricity domain
more adequately. To reach this goal, we duplicated every $e$ value in the sample with $-e$.
Thus we artificially increased the sample size to $2N$, but when computing the significance
and related stuff, we substituted the correct size of just $N$, so that such a manipulation
should not lead to an overestimated significance. Also, we consider our work domain
$\mathcal D$ only for $e\geq 0$, so the coefficients in the $\FAP$ estimate~(\ref{fap}) do
not get doubled (this would lead to an underestimated significance otherwise).

We recognize that even with such corrections the analysis might be not fully rigorous,
because instead of the pure CBHAT and WAVE2 we basically use wavelets of a different shape,
symmetrized about zero. The actual remaining effect is not obvious, but we did not try to
treat it better, because we did not obtain any remarkable results with this distribution.

\begin{figure*}
\begin{tabular}{@{}l@{}l@{}}
\includegraphics[width=0.5\linewidth]{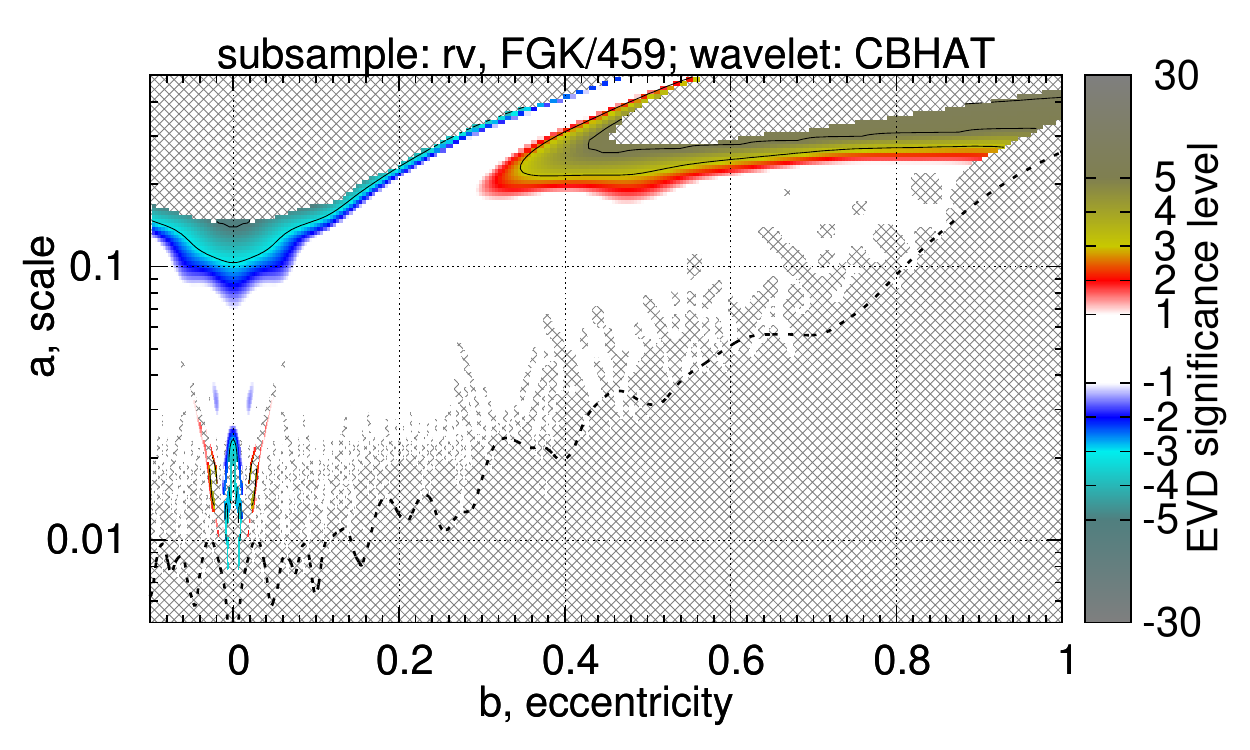} & \includegraphics[width=0.5\linewidth]{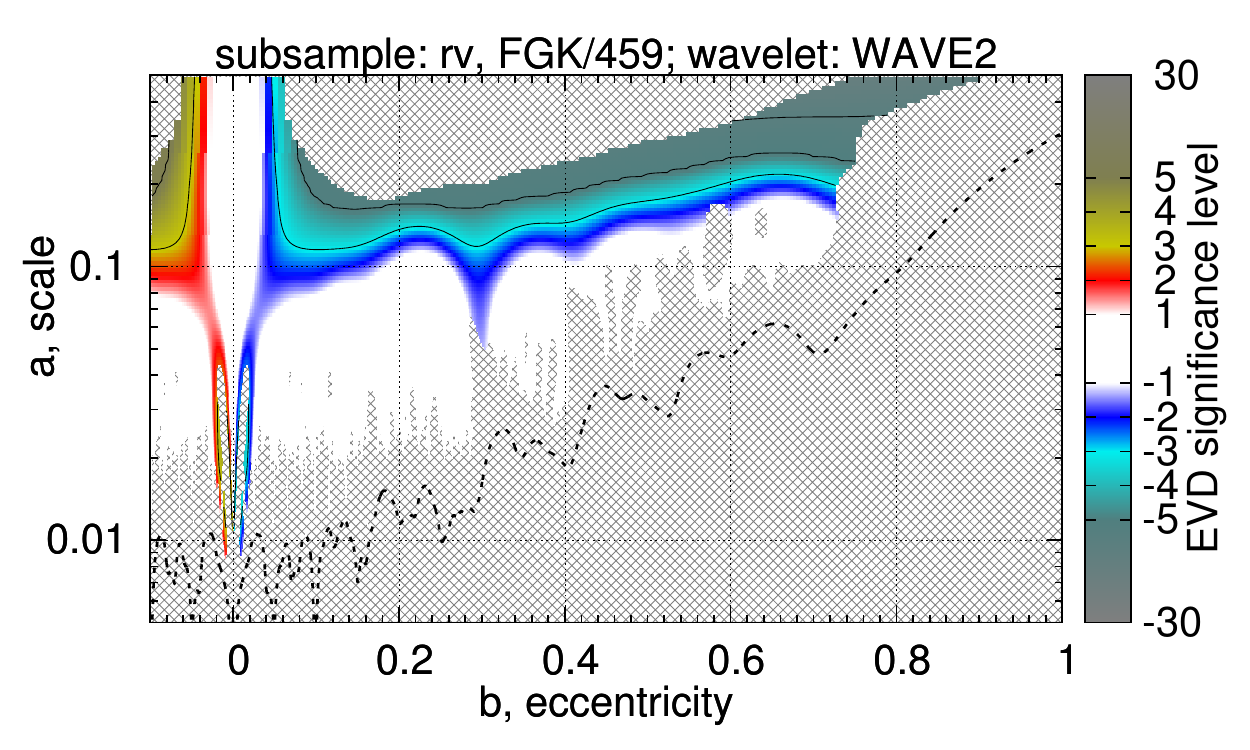}\\
\includegraphics[width=0.5\linewidth]{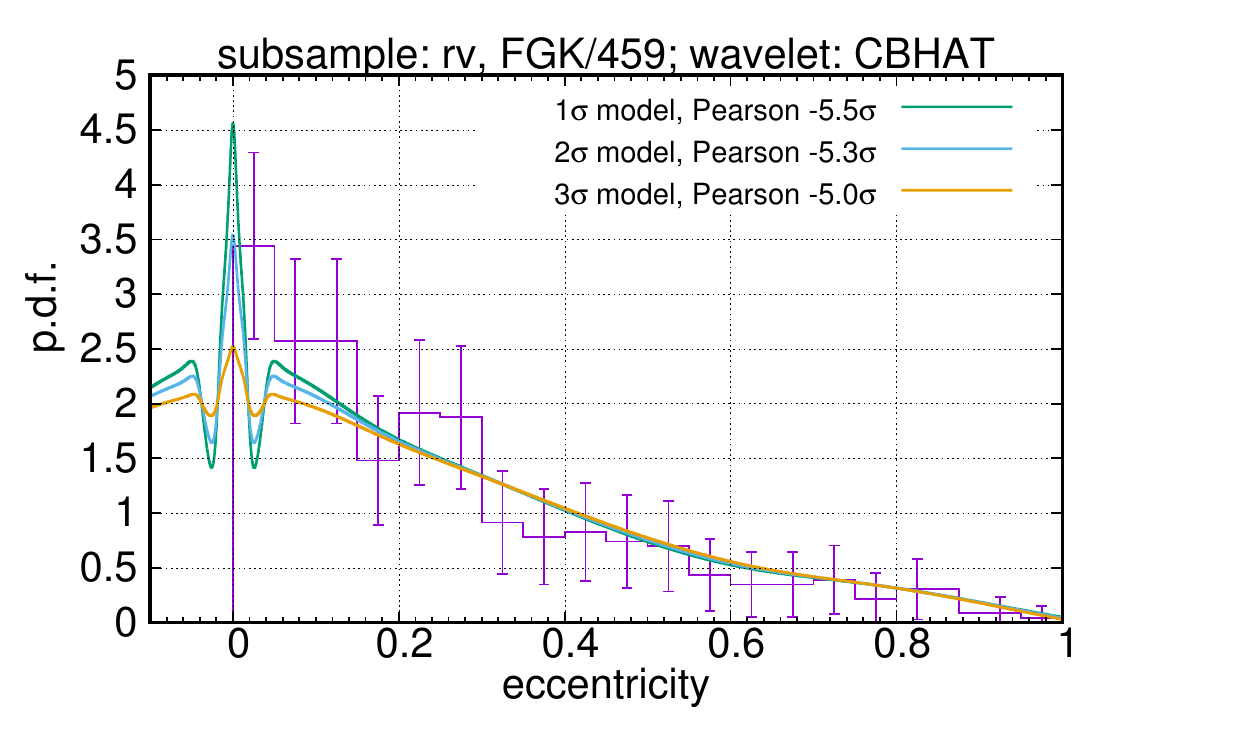} & \includegraphics[width=0.5\linewidth]{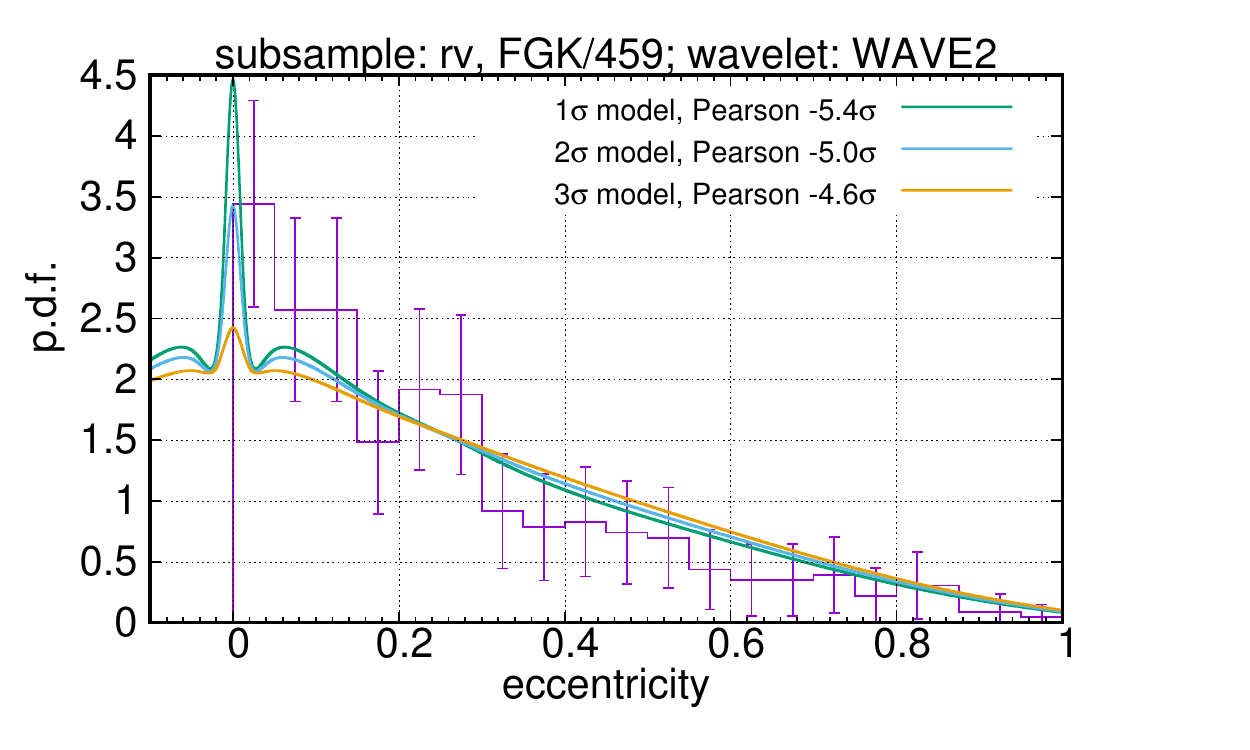}
\end{tabular}
\caption{Wavelet analyis of known exoplanetary candidates. Subsample: RV detection method,
FGK-type host stars, $N=459$. Variable: orbital eccentricity. The figure layout is the same
as in Fig.~\ref{fig_spike10}.}
\label{fig_rv_ecc}
\end{figure*}

The results for the \sampleid{rv.FGK} sample are presented in Fig.~\ref{fig_rv_ecc}. We can see
an intricate small-scale structure near $e=0$ that consists of a very narrow peak
accompanied by narrow side gaps. Such a structure near $e=0$ is not surprising and it can
be explained by the following obvious sources:
\begin{enumerate}
\item Deficient RV data are usually fitted with the eccentricity fixed at zero. This leads
to an increased height of the $e=0$ peak, because some exoplanets with non-zero $e$ are
reset to zero.
\item If the eccentricity is fitted, it is frequently biased to a larger value due to a
non-linearity in the Keplerian RV model. This bias moves some exoplanets with small $e$
away from zero, forming a gap about $e=0$ that serves as a background for the central very
narrow peak.
\item There is a well-known physical phenomenon of tidal circularization that settles many
``hot jupiters'' to $e\approx 0$, further increasing the central peak.
\end{enumerate}
We can see that only the third effect is physical and related to the actual eccentricity
distribution, while the other two are only apparent effects imposted by the RV data
fitting.

We did not reveal any other small-scale structures in this distribution. Also, we did not
analyse transit candidates here, since most of them do not have a good eccentricity
estimate.

\subsection{Orbital periastra: no hints of a detection bias}
\label{sec_omega}
We analysed the distribution of the orbital pericenter argument, $\omega$. Our motivation
was to detect possible selection effects. Any detection bias in $e$ should reveal itself in
the $\omega$ too, because the shape of the eccentric Keplerian RV curve depends on $\omega$
dramatically. In the eccentricities we cannot disentangle this bias from the actual
distribution shape, which is unknown. However, the spatial orientation of exoplanetary
orbital periastra should likely be isotropic, meaning a uniform distribution in $\omega$.
Therefore, any deviation from the uniform distribution would immediately indicate a
detection bias.

\begin{figure*}
\begin{tabular}{@{}l@{}l@{}}
\includegraphics[width=0.5\linewidth]{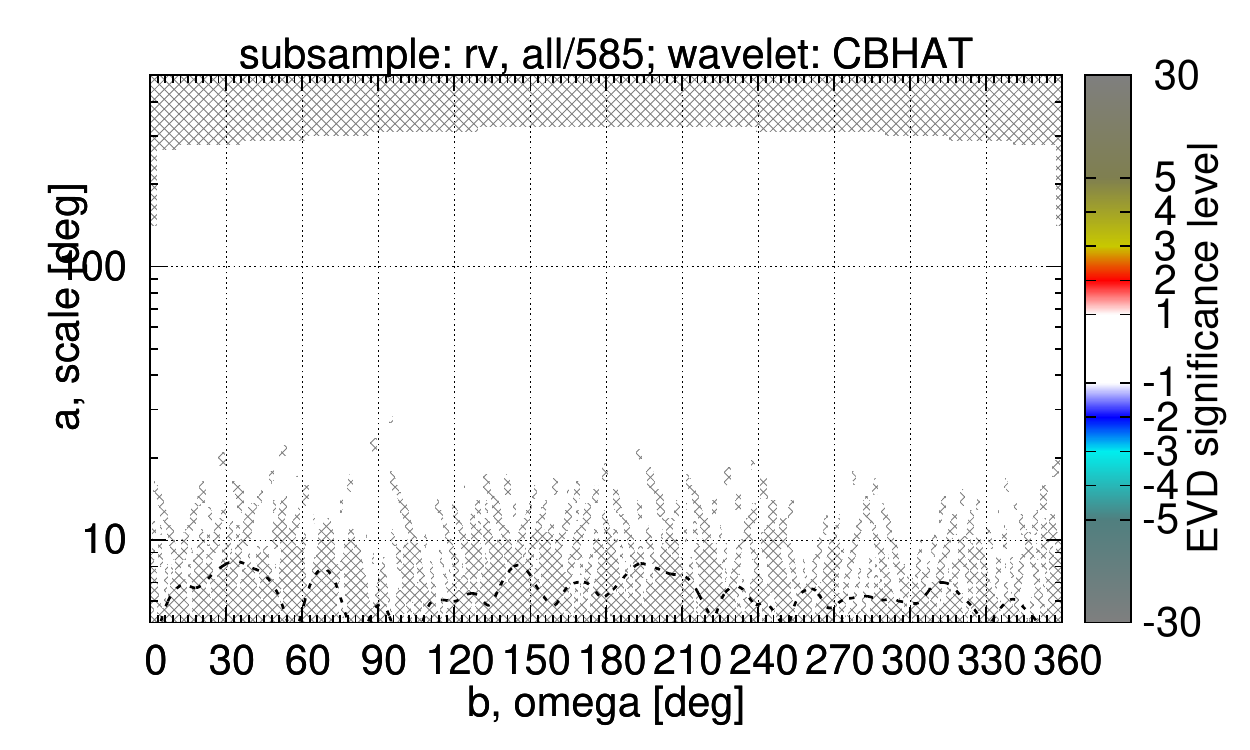} & \includegraphics[width=0.5\linewidth]{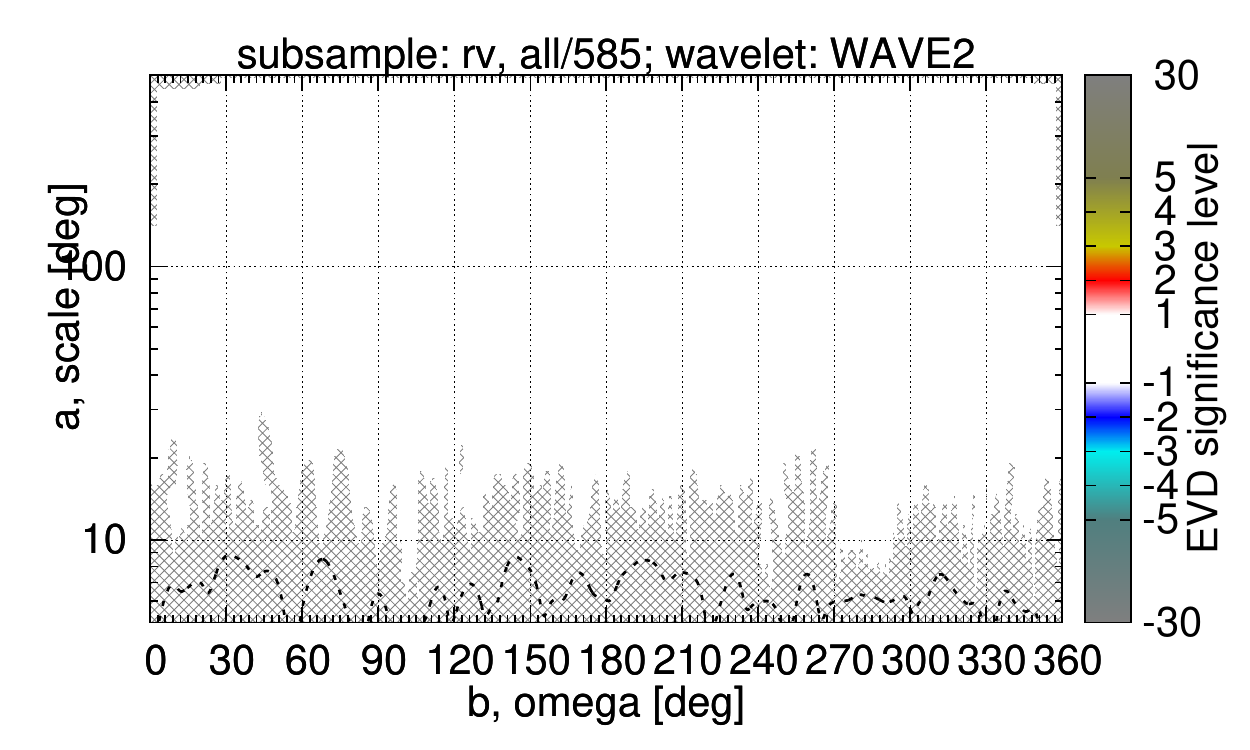}\\
\includegraphics[width=0.5\linewidth]{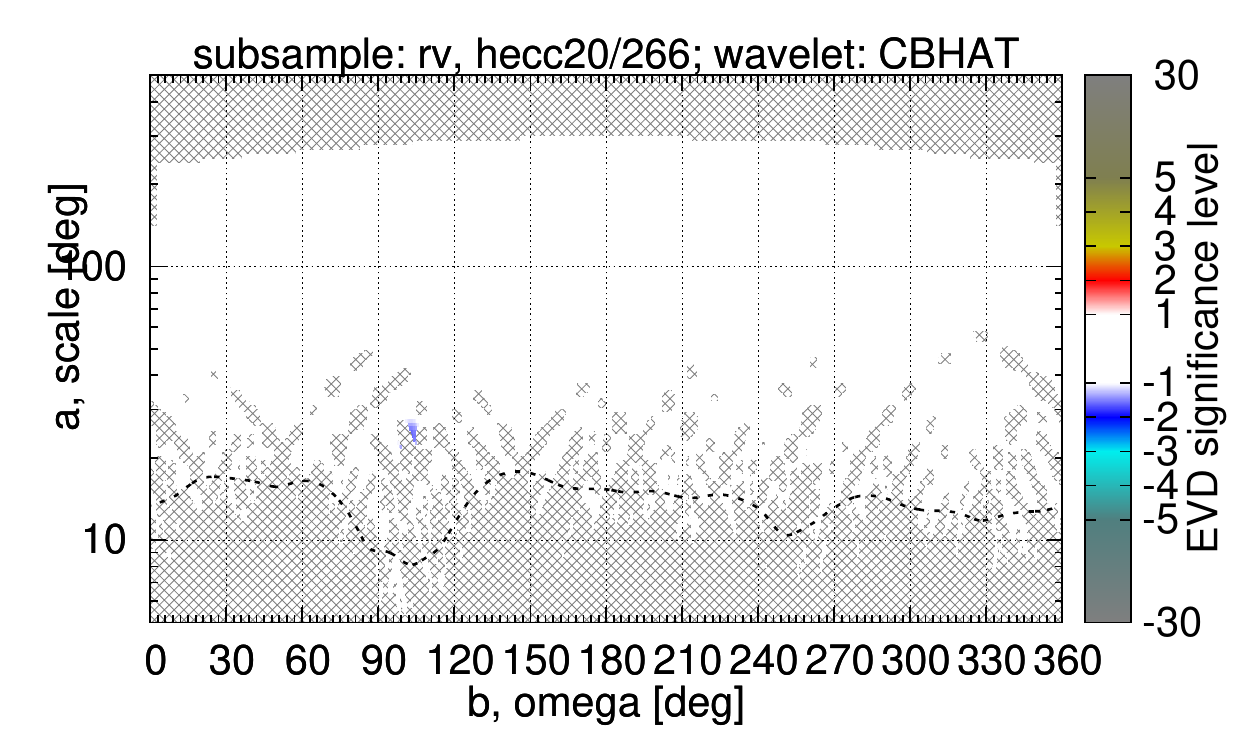} & \includegraphics[width=0.5\linewidth]{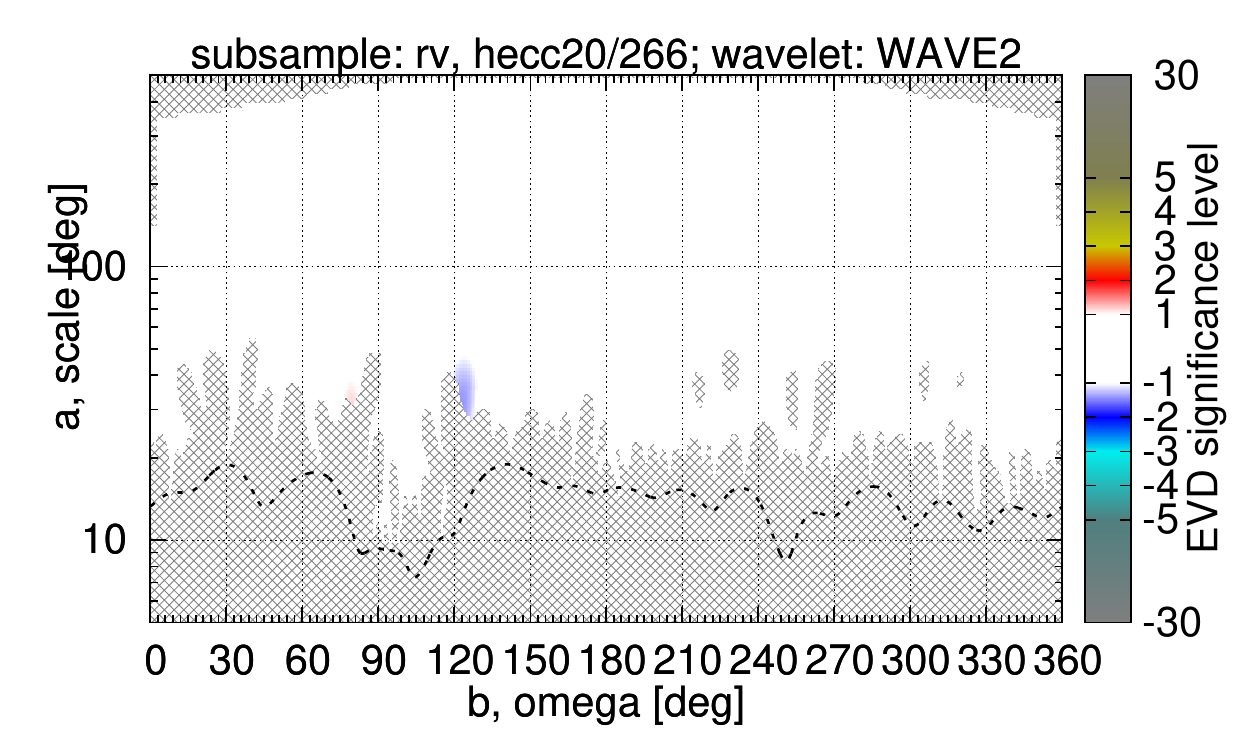}\\
\includegraphics[width=0.5\linewidth]{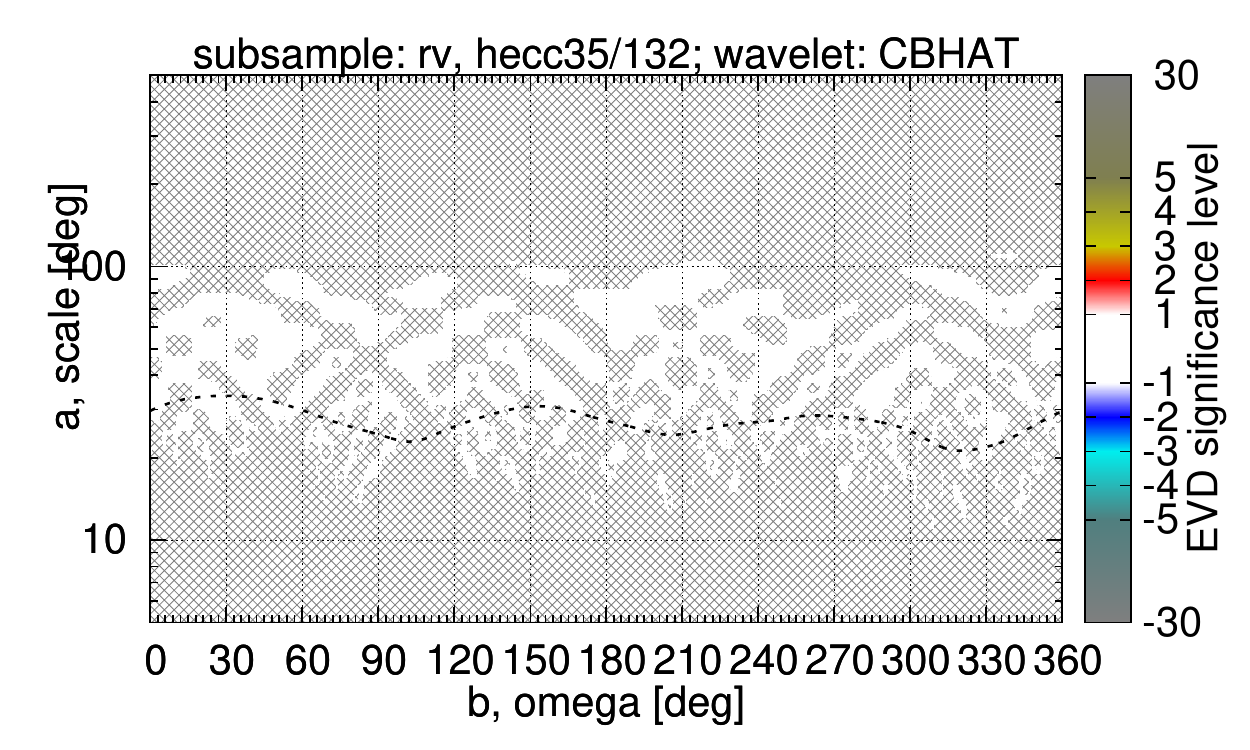} & \includegraphics[width=0.5\linewidth]{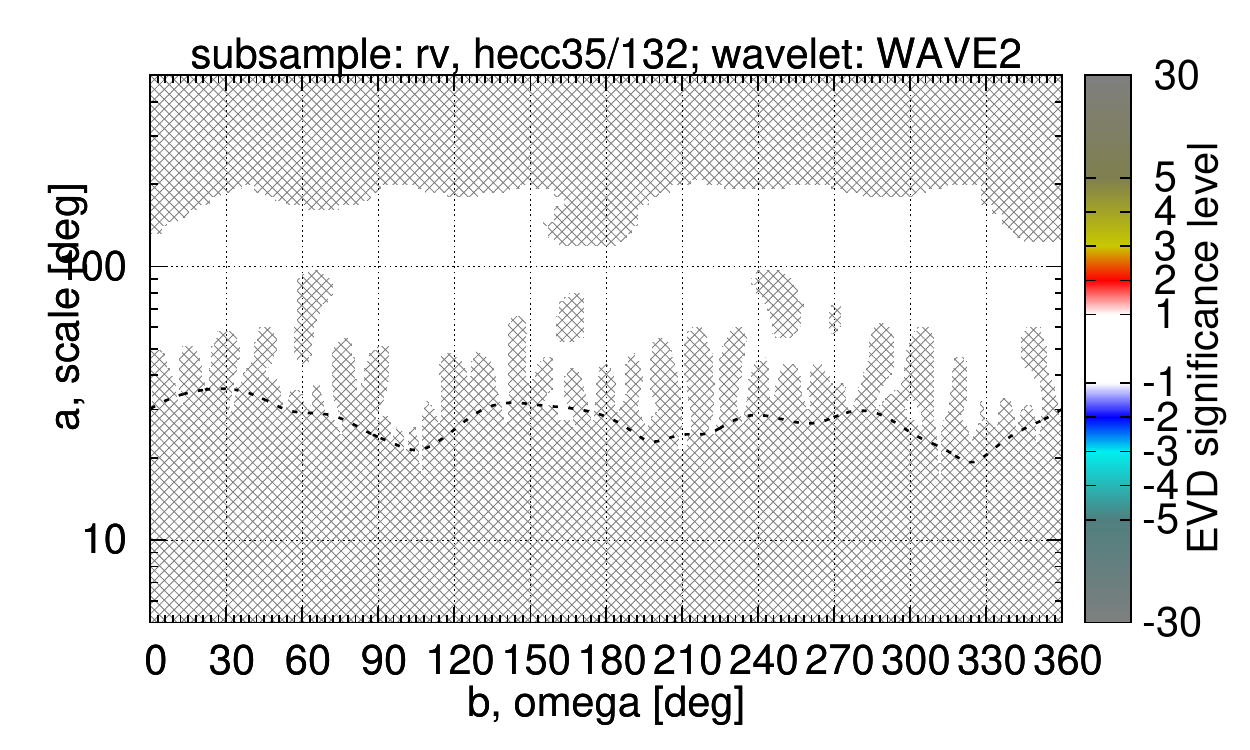}\\
\end{tabular}
\caption{Wavelet analyis of known exoplanetary candidates. Subsample: RV detection method,
cutting higher eccentricities ($e>0.15$, $e>0.20$, $e>0.35$). Variable: argument of the
periastron. The figure layout is the same as in Fig.~\ref{fig_spike10}.}
\label{fig_rv_om}
\end{figure*}

The results for the $\omega$-distribution (the \sampleid{rv} sample) are presented in
Fig.~\ref{fig_rv_om}. Additionally to this full sample, we also computed wavelet maps for
various subsamples by removing low-eccentric exoplanets ($e<0.15, 0.20, 0.35$). In either
case, the distribution is consistent with the uniform one. No significant deviation was
detected.

Note that to avoid non-physical cutoffs at $\omega=0$ and $\omega=2\pi$, we applied an
approach similar to that from Sect.~\ref{sec_ecc}: we replaced every sample value $\omega$
with a triplet $\omega-2\pi,\omega,\omega+2\pi$. This artificially increased the total
sample size to $3N$, but we again used just $N$ in the significance estimates where
appropriate.

\subsection{Planet masses}
\label{sec_mass}
Finally, we present our results for the exoplanetary mass distribution. In this analysis,
we considered the \sampleid{rv.FGK} and \sampleid{pt.FGK} samples. Note that the second (transit)
sample appeared much smaller than e.g. transit sample in Sect.~\ref{sec_rad}, because the
planet mass can be determined only by radial velocities. Most transiting planet candidates
remain unconfirmed by the Doppler method, mainly because of technical limitations on faint
stars. As a result, both the samples appeared to have almost the same moderate size, $N\sim
500$.

\begin{figure*}
\begin{tabular}{@{}l@{}l@{}}
\includegraphics[width=0.5\linewidth]{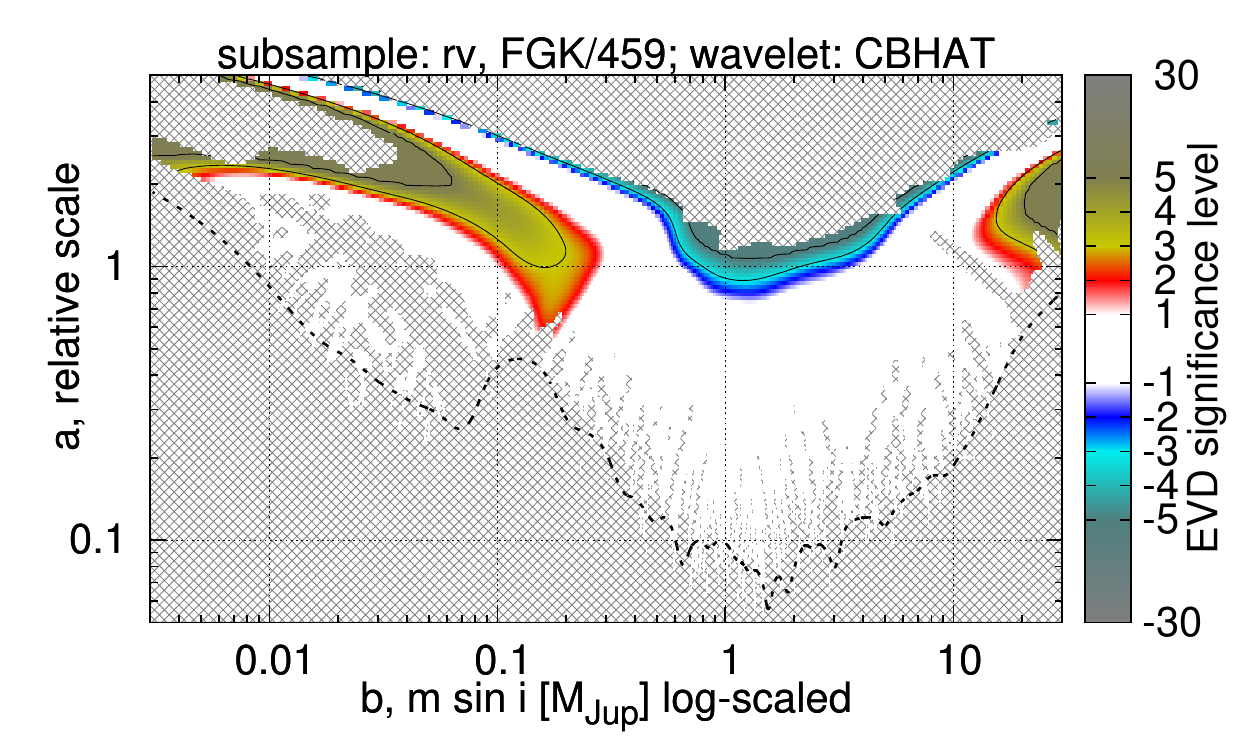} & \includegraphics[width=0.5\linewidth]{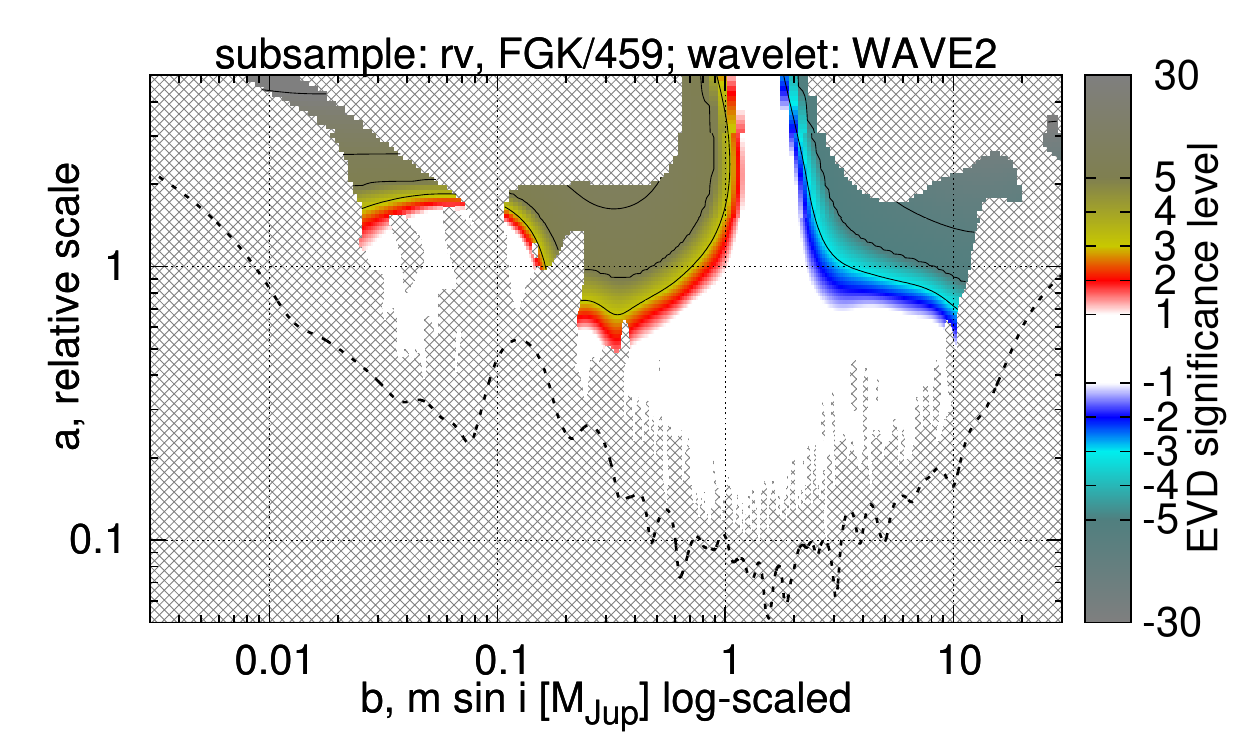}\\
\includegraphics[width=0.5\linewidth]{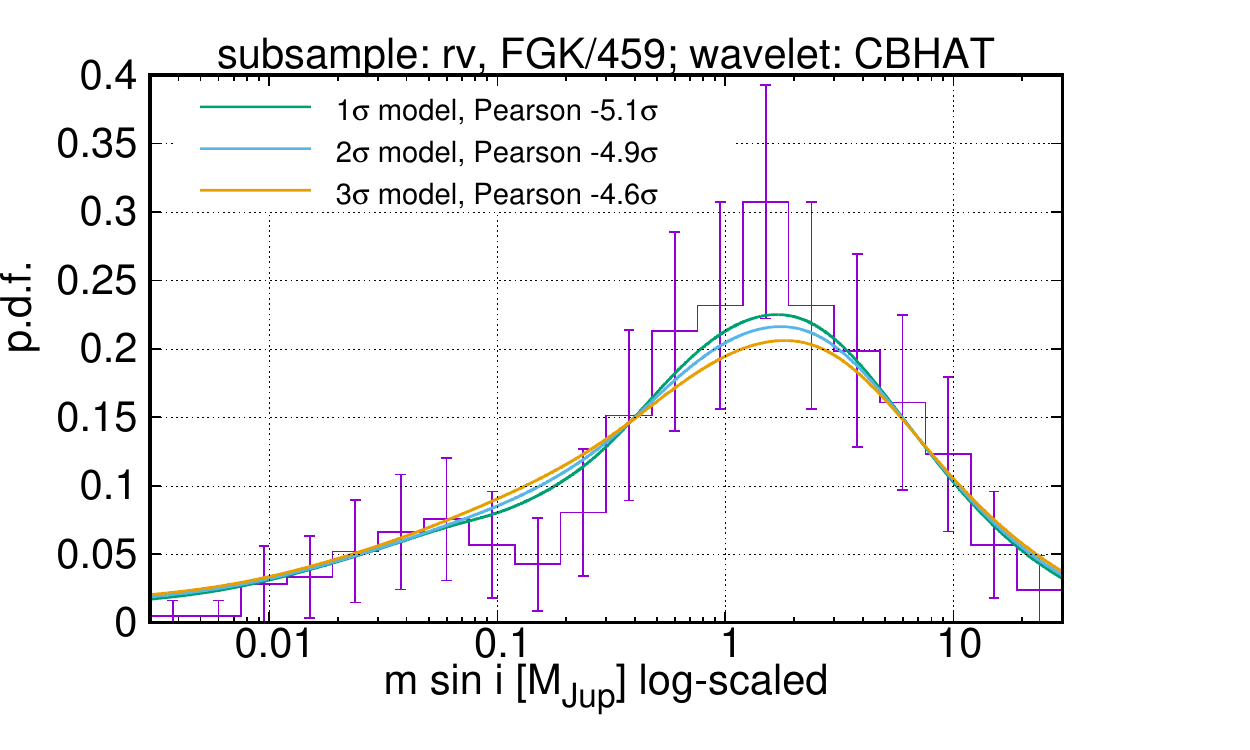} & \includegraphics[width=0.5\linewidth]{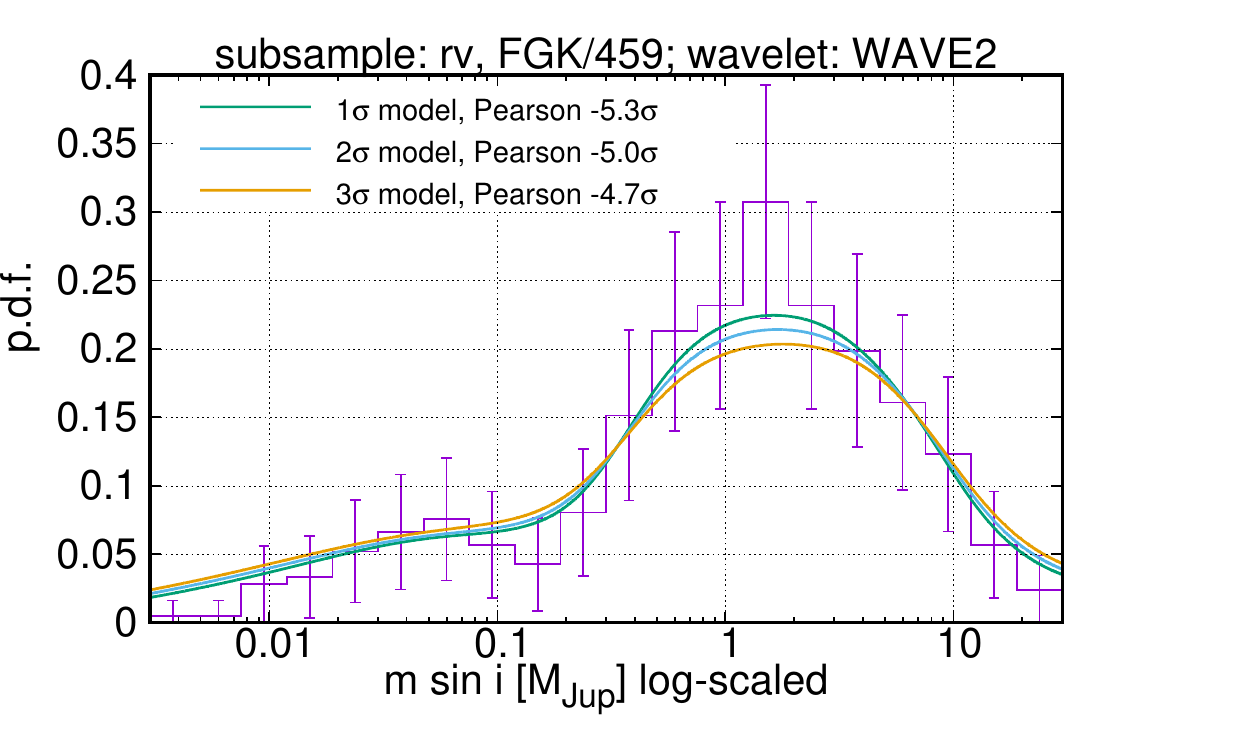}
\end{tabular}
\caption{Wavelet analyis of known exoplanetary candidates. Subsample: RV detection method,
FGK-type host stars, $N=459$. Variable: planet minimum mass $m \sin i$. The figure layout
is the same as in Fig.~\ref{fig_spike10}.}
\label{fig_rv_msi}
\end{figure*}

\begin{figure*}
\begin{tabular}{@{}l@{}l@{}}
\includegraphics[width=0.5\linewidth]{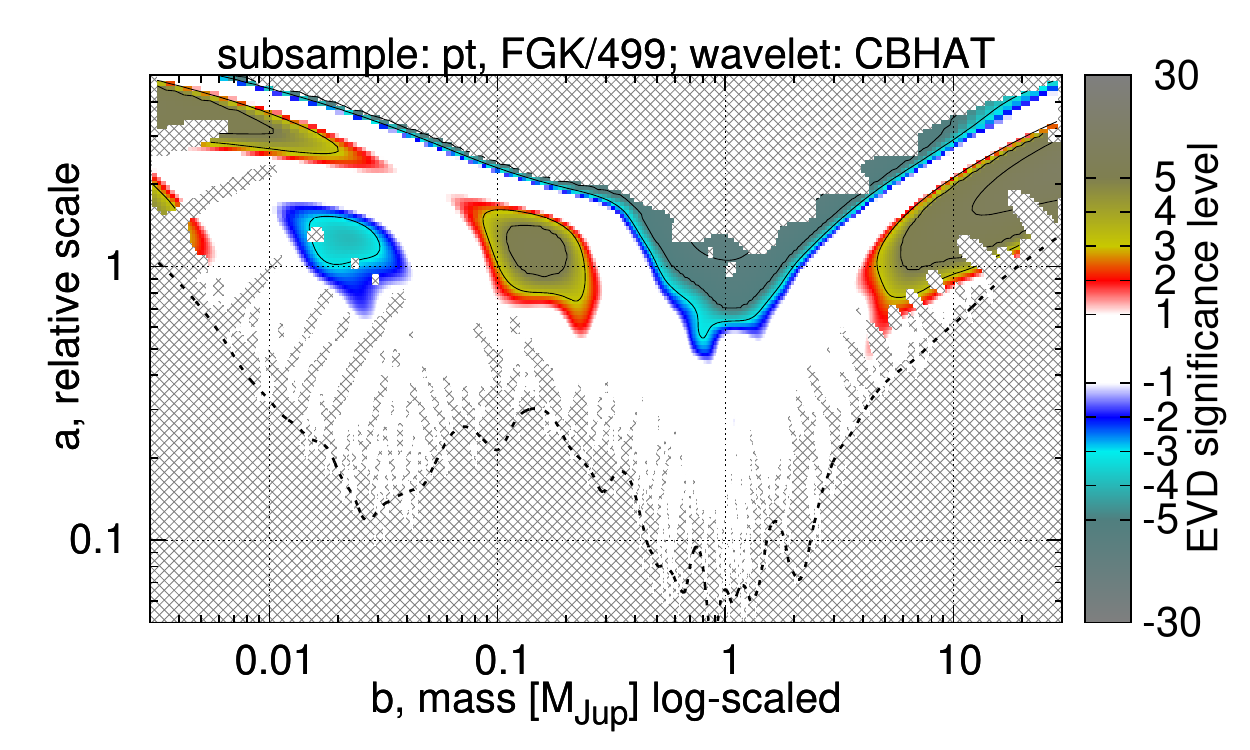} & \includegraphics[width=0.5\linewidth]{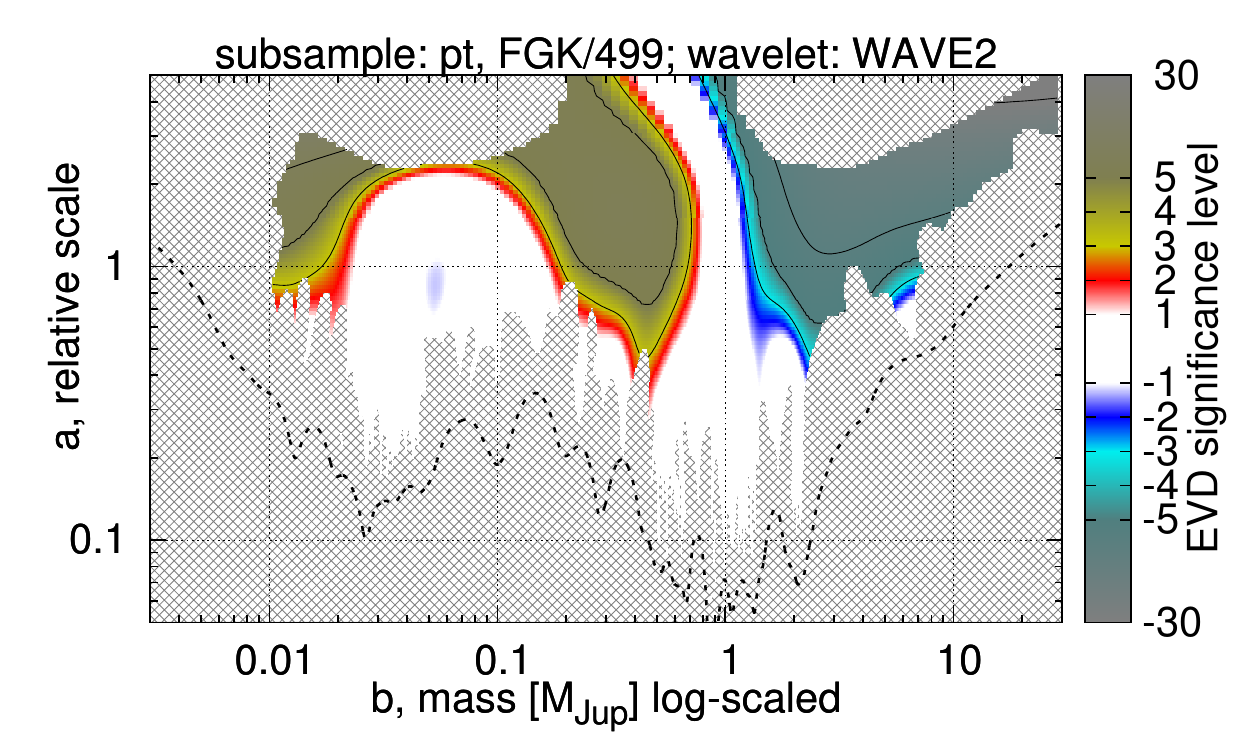}\\
\includegraphics[width=0.5\linewidth]{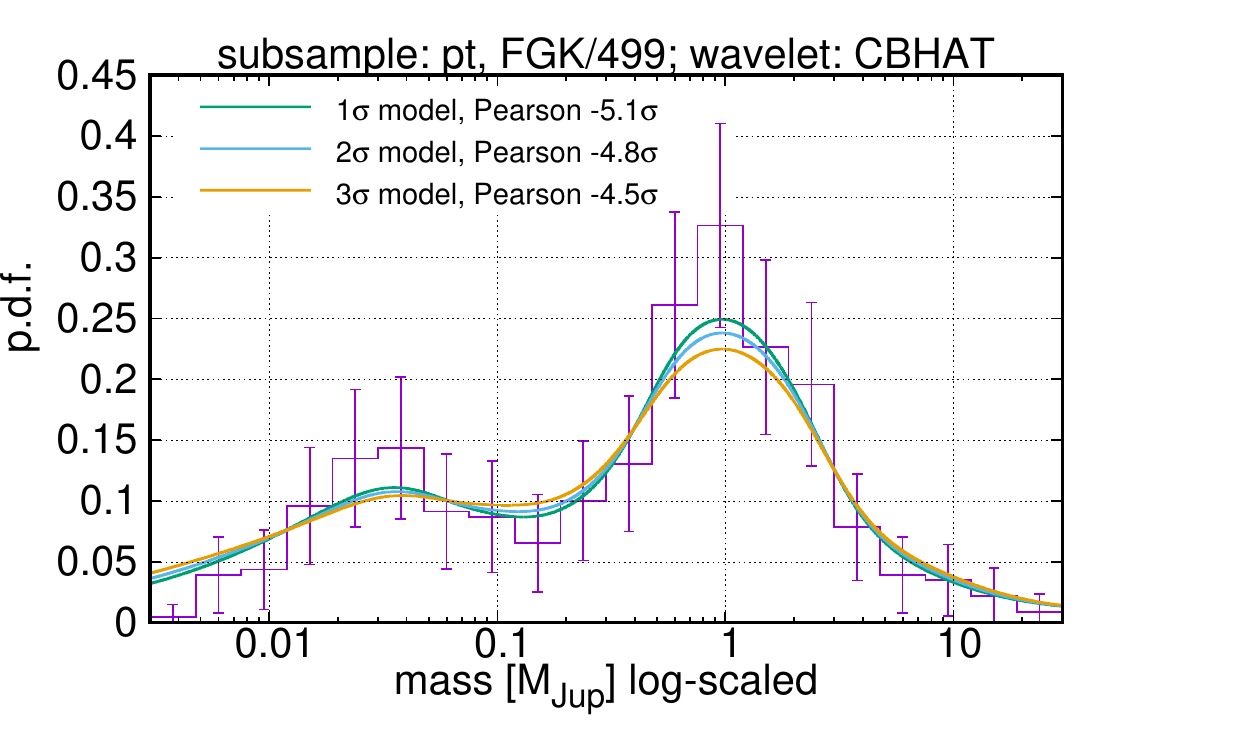} & \includegraphics[width=0.5\linewidth]{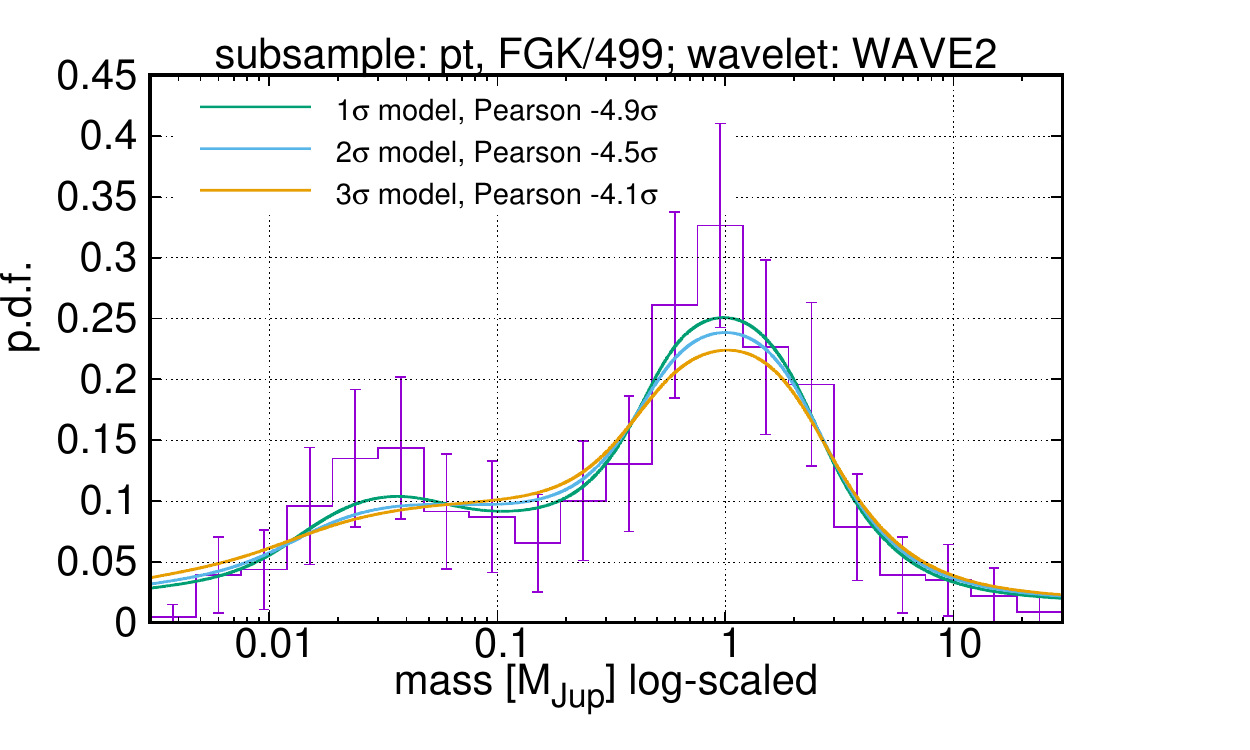}
\end{tabular}
\caption{Wavelet analyis of known exoplanetary candidates. Subsample: Primary transit
detection method, FGK-type host stars, $N=499$. Variable: planet mass $m$. The figure
layout is the same as in Fig.~\ref{fig_spike10}.}
\label{fig_pt_mss}
\end{figure*}

The results are presented in Fig.~\ref{fig_rv_msi} (\sampleid{rv.FGK} sample, distribution of
$m\sin i$) and in Fig.~\ref{fig_pt_mss} (\sampleid{pt.FGK} sample, distribution of $m$). In the
first case, the distribution appears unimodal with a wide maximum covering the range $m\sin
i = 0.5-5 M_{\rm Jup}$. On contrary, the transit sample reveals an undoubtful bimodality,
with a secondary maximum at $m \sim 0.03 M_{\rm Jup}$ (or $\sim 10 M_\oplus$). The first
maximum involves giant exoplanets, while the second one incorporates super-earths and
Neptune-mass planets. The gap between the families covers the range $m=0.1-0.3 M_{\rm
Jup}$.

We may conclude that the transit detection technique is remarkably more sensitive to
low-mass planets. Although all exoplanets in the both samples were observed by the radial
velocity technique, the transit candidates required just an independent follow-up Doppler
\emph{confirmation}, which is much easier than to make a \emph{detection}.

We did not detect any small-scale details in the exoplanetary mass distribution.

\section{Some self-crititicism concerning the significance estimates}
\label{sec_sig}
Unfortunately, the issue of a hidden p-hacking effect often appears very difficult to
overcome completely. The numeric significance levels reported above are still not entirely
free from an overestimation caused by multiple testing.

First of all, we always considered \emph{two} wavelet transforms simultaneously, WAVE2 and
CBHAT. Hence, our analysis hiddenly involved \emph{two} comparison tests, on the
\emph{same} distribution. The number of false alarms should then appear larger than
expected from just a single wavelet transform. In fact, our test involved the maximum
between \emph{two} distinct test statistics, namely $\max(z_{\rm WAVE2},z_{\rm CBHAT})$,
rather than the maximum of just a single $z(a,b)$.

Assuming that the maxima on the WAVE2 and CBHAT wavelet maps are uncorrelated with each
other, we should roughly double the $\FAP$ estimate. This implies the need of a multiple
testing correction, analogous to~(\ref{mtcorr}) with $N_{\rm bin}=2$:
\begin{equation}
g' = \Phi^{-1}\left(\frac{1+\Phi(g)}{2}\right), \quad g=\Phi^{-1}\left[2\Phi(g')-1\right].
\label{gcorr}
\end{equation}
Now, $g'$ is the value actually obtained from the significance map (assuming no multiple
testing), and $g$ is the corrected significance taking into account the double testing
penalty.\footnote{The interpretation of $g$ and $g'$ is swapped here,
because~(\ref{mtcorr}) had the meaning of a \emph{preventive} modification, while
formula~(\ref{gcorr}) is \emph{corrective} instead.}

For the subfamily $P=300-600$~d we obtained in Sect.~\ref{sec_psma} the following detection
significances in our CBHAT analysis: $g=2.0$ (left concavity), $g=3.4$ (central convexity),
and $g=2.1$ (right concavity). Formula~(\ref{gcorr}) transforms these levels to the
following: $g'=1.6$, $g'=3.2$, and $g'=1.8$. The central convexity still remains pretty
convincing, but the side separations appear to have just a marginal significance.

For the CKS radii distribution we obtain with CBHAT: $g>5$ (left maximum), $g=1.5$ (central
gap), $g>5$ (right maximum), and with WAVE2: $g=1.6$ (descent to the gap) and $g=2.0$
(ascent out of the gap). Using~(\ref{gcorr}), this is corrected to: $g'>5$ (left maximum),
$g'=1.1$ (central gap), $g'>5$ (right maximum), $g'=1.2$ (descent to the gap) and $g'=1.6$
(ascent out of the gap). We can see that the gap becomes insignificant.

But from the other side, the WAVE2 and CBHAT wavelet maps should be correlated: if a
pattern is detected by one of the wavelets, the other wavelet should likely reveal
something unusual too. Therefore, the correction~(\ref{gcorr}) is probably too
conservative. Moreover, our $\FAP$ estimate~(\ref{fap}) was made conservative by design,
because should tend to overestimate the actual $\FAP$. It might seem likely that these
effects could in turn compensate each other, so the uncorrected $g$-levels might be more
realistic than $g'$ from~(\ref{gcorr}).

But more difficulties appear because we actually need the significance of \emph{combined}
patterns that involve \emph{several} elementary structures on the wavelet maps. For
example, to ensure that the subfamily $P=300-600$~d does exist, we needed the whole triplet
``concavity-convexity-concavity'' to be confirmed as a whole. Just a single central
convexity without side concavities is notable, but we would not classify it as a
``subfamily'', because a subfamily should be separated in some way from the background
planets. Without a separation, it should be interpreted just as relatively abrupt PV/WJ
transition, but not as a separate family.

Similarly, when detecting the bimodality in the radii distribution, we needed to ensure the
significance of the triplet ``convexity-concavity-convexity'' as a whole (in CBHAT) or of
the doublet ``descent-ascent'' as a whole (in WAVE2). Without that, we cannot claim the
two-family distribution structure.

The $\FAP$ of a doublet, or of a triplet, or of a general $n$-tuple set, cannot be
expressed via $\FAP$s for their individual elements. More detailed discussion of this issue
is given in \citep{Baluev13d} respectively to the period search in time series.
Unfortunately, a more rigorous solution to this issue involved too complicated theory and
heavy computations. This is currently not feasible in our present task, so we did not
incorporate anything beyond the single-component significance.

\section{Conclusions and discussion}
On the basis of the analysis presented above, we draw two main conclusions concerning the
exoplanetary distributions:
\begin{enumerate}
\item There is a remarkable statistical support in favour of a narrow subfamily of giant
exoplanets within the range $P=300-600$~d. This subfamily seems to be separated from the
rest of the sample by p.d.f. concavities on both sides (but this does not necessarily imply
bimodality). Most probably, this subfamily has a connection with the iceline barrier effect
in a protoplanetary disk. But to our knowledge, current theory work on planet formation did
not predict such a family yet. Curiously, this submafily appears to overlap well with the
habitable zone for a solar-type star.

\item There is some statistical support in favour of the ``evaporation valley'' in the
planet radii distribution of the CKS sample. But significance of this feature reported in
other works seems to be overestimated. That high significance appeared irreproducible in
our work.
\end{enumerate}

We highlight additionally, that although any ``absolute'' statistical significance is
always difficult to estimate adequately, the ``relative'' comparison is easier. For
example, it follows from our analysis that the ``evaporation valley'' is notably less
confident than the $300-600$~d subfamily.

The distributions of exoplanetary masses and orbital eccentricities did not reveal new
features. The angular distribution of orbital pericenters is consistent with the uniform
one, suggesting that there is no detectable selection effects with respect to this
parameter.

\acknowledgments
V.Sh.~Shaidulin was supported by the Russian Foundation for Basic Research grant
17-02-00542~A. R.V.~Baluev was supported by the Presidium of Russian Academy of Sciences
programme P-28, subprogramme ``The space: investigating fundamental processes and their
interrelations''.

%% References
%% Please cite all reference entries in the article text using \cite or
%% equivalent command. 

%%  Using BibTeX  (Name-Year style)

 \bibliographystyle{spr-mp-nameyear-cnd}  %% BibTeX style
 \bibliography{wavexo}                %% BibTeX data

\begin{thebibliography}{26}
% BibTex style file: spr-mp.bst (nameyear,cnd), 2011-05-27
\ifx \bisbn   \undefined \def \bisbn  #1{ISBN #1}\fi
\ifx \binits  \undefined \def \binits#1{#1} \fi
\ifx \bauthor  \undefined \def \bauthor#1{#1} \fi
\ifx \batitle  \undefined \def \batitle#1{#1} \fi
\ifx \bjtitle  \undefined \def \bjtitle#1{#1}\fi
\ifx \bvolume  \undefined \def \bvolume#1{\textbf{#1}}\fi
\ifx \byear  \undefined \def \byear#1{#1} \fi
\ifx \bissue  \undefined \def \bissue#1{#1} \fi
\ifx \bfpage  \undefined \def \bfpage#1{#1} \fi
\ifx \blpage  \undefined \def \blpage #1{#1} \fi
\ifx \burl  \undefined \def \burl#1{\textsf{#1}} \fi
\ifx \doiurl  \undefined \def \doiurl#1{\textsf{#1}} \fi
\ifx \betal  \undefined \def \betal{\textit{et al.}} \fi
\ifx \binstitute  \undefined \def \binstitute#1{#1} \fi
\ifx \binstitutionaled  \undefined \def \binstitutionaled#1{#1} \fi
\ifx \bctitle  \undefined \def \bctitle#1{#1} \fi
\ifx \beditor  \undefined \def \beditor#1{#1} \fi
\ifx \bpublisher  \undefined \def \bpublisher#1{#1} \fi
\ifx \bbtitle  \undefined \def \bbtitle#1{#1} \fi
\ifx \bedition  \undefined \def \bedition#1{#1} \fi
\ifx \bseriesno  \undefined \def \bseriesno#1{#1} \fi
\ifx \blocation  \undefined \def \blocation#1{#1} \fi
\ifx \bsertitle  \undefined \def \bsertitle#1{#1} \fi
\ifx \bsnm \undefined \def \bsnm#1{#1} \fi
\ifx \bsuffix \undefined \def \bsuffix#1{#1} \fi
\ifx \bparticle \undefined \def \bparticle#1{#1} \fi
\ifx \barticle \undefined \def \barticle#1{#1} \fi
\ifx \bconfdate \undefined \def \bconfdate #1{#1} \fi
\ifx \botherref \undefined \def \botherref #1{#1} \fi
\ifx \url \undefined \def \url#1{\textsf{#1}} \fi
\ifx \bchapter \undefined \def \bchapter#1{#1} \fi
\ifx \bbook \undefined \def \bbook#1{#1} \fi
\ifx \bcomment \undefined \def \bcomment#1{#1} \fi
\ifx \oauthor \undefined \def \oauthor#1{#1} \fi
\ifx \citeauthoryear \undefined \def \citeauthoryear#1{#1} \fi
\ifx \endbibitem  \undefined \def \endbibitem {}\fi
\ifx \bconflocation  \undefined \def \bconflocation#1{#1} \fi
\ifx \arxivurl  \undefined \def \arxivurl#1{\textsf{#1}} \fi

\bibitem[\protect\citeauthoryear{Baluev}{2008}]{Baluev08a}
\begin{barticle}
\bauthor{\bsnm{Baluev}, \binits{R.V.}}:
\bjtitle{\mnras}
\bvolume{385},
\bfpage{1279}
(\byear{2008})
\end{barticle}
\endbibitem

\bibitem[\protect\citeauthoryear{Baluev}{2013}]{Baluev13d}
\begin{barticle}
\bauthor{\bsnm{Baluev}, \binits{R.V.}}:
\bjtitle{\mnras}
\bvolume{436},
\bfpage{807}
(\byear{2013})
\end{barticle}
\endbibitem

\bibitem[\protect\citeauthoryear{Baluev}{2018}]{Baluev18a}
\begin{barticle}
\bauthor{\bsnm{Baluev}, \binits{R.V.}}:
\bjtitle{Astron. \& Comp.}
\bvolume{23},
\bfpage{151}
(\byear{2018})
\end{barticle}
\endbibitem

\bibitem[\protect\citeauthoryear{Cumming}{2010}]{CummingStat}
\begin{bchapter}
\bauthor{\bsnm{Cumming}, \binits{A.}}:
In: \beditor{\bsnm{Seager}, \binits{S.}} (ed.)
\bbtitle{Exoplanets},
p. \bfpage{191}.
\bpublisher{University of Arizona Press},
\blocation{Tucson}
(\byear{2010}).
\bcomment{Chap. 9}
\end{bchapter}
\endbibitem

\bibitem[\protect\citeauthoryear{Foster}{1995}]{Foster95}
\begin{barticle}
\bauthor{\bsnm{Foster}, \binits{G.}}:
\bjtitle{\aj}
\bvolume{109},
\bfpage{1889}
(\byear{1995})
\end{barticle}
\endbibitem

\bibitem[\protect\citeauthoryear{Foster}{1996}]{Foster96a}
\begin{barticle}
\bauthor{\bsnm{Foster}, \binits{G.}}:
\bjtitle{\aj}
\bvolume{111},
\bfpage{541}
(\byear{1996})
\end{barticle}
\endbibitem

\bibitem[\protect\citeauthoryear{Freedman and
  Diaconis}{1981}]{FreedmanDiaconis81}
\begin{barticle}
\bauthor{\bsnm{Freedman}, \binits{D.}},
\bauthor{\bsnm{Diaconis}, \binits{P.}}:
\bjtitle{Z. Wahrscheinlichkeitstheorie verw. Gebiete}
\bvolume{57},
\bfpage{453}
(\byear{1981})
\end{barticle}
\endbibitem

\bibitem[\protect\citeauthoryear{Fulton et~al.}{2017}]{Fulton17}
\begin{barticle}
\bauthor{\bsnm{Fulton}, \binits{B.J.}},
\bauthor{\bsnm{Petigura}, \binits{E.A.}},
\bauthor{\bsnm{Howard}, \binits{A.W.}},
\bauthor{\bsnm{Isaacson}, \binits{H.}},
\bauthor{\bsnm{Marcy}, \binits{G.W.}},
\bauthor{\bsnm{Cargile}, \binits{P.A.}},
\bauthor{\bsnm{Hebb}, \binits{L.}},
\bauthor{\bsnm{Weiss}, \binits{L.M.}},
\bauthor{\bsnm{Johnson}, \binits{J.A.}},
\bauthor{\bsnm{Morton}, \binits{T.D.}},
\bauthor{\bsnm{Sinukoff}, \binits{E.}},
\bauthor{\bsnm{Crossfield}, \binits{I.J.M.}},
\bauthor{\bsnm{Hirsch}, \binits{L.A.}}:
\bjtitle{\aj}
\bvolume{154},
\bfpage{109}
(\byear{2017})
\end{barticle}
\endbibitem

\bibitem[\protect\citeauthoryear{Gelman and Loken}{2014}]{GelmanLoken14}
\begin{barticle}
\bauthor{\bsnm{Gelman}, \binits{A.}},
\bauthor{\bsnm{Loken}, \binits{E.}}:
\bjtitle{American Scientist}
\bvolume{102},
\bfpage{460}
(\byear{2014})
\end{barticle}
\endbibitem

\bibitem[\protect\citeauthoryear{Hara et~al.}{2017}]{Hara17}
\begin{barticle}
\bauthor{\bsnm{Hara}, \binits{N.C.}},
\bauthor{\bsnm{Bou{\'e}}, \binits{G.}},
\bauthor{\bsnm{Laskar}, \binits{J.}},
\bauthor{\bsnm{Correia}, \binits{A.C.M.}}:
\bjtitle{\mnras}
\bvolume{464},
\bfpage{1220}
(\byear{2017})
\end{barticle}
\endbibitem

\bibitem[\protect\citeauthoryear{Hartigan and Hartigan}{1985}]{Hartigan85}
\begin{barticle}
\bauthor{\bsnm{Hartigan}, \binits{J.A.}},
\bauthor{\bsnm{Hartigan}, \binits{P.M.}}:
\bjtitle{Annals of Statistics}
\bvolume{13},
\bfpage{70}
(\byear{1985})
\end{barticle}
\endbibitem

\bibitem[\protect\citeauthoryear{Hasegawa and Pudritz}{2013}]{Hasegawa13}
\begin{barticle}
\bauthor{\bsnm{Hasegawa}, \binits{Y.}},
\bauthor{\bsnm{Pudritz}, \binits{R.E.}}:
\bjtitle{\apj}
\bvolume{778},
\bfpage{78}
(\byear{2013})
\end{barticle}
\endbibitem

\bibitem[\protect\citeauthoryear{Hayashi}{1981}]{Hayashi81}
\begin{barticle}
\bauthor{\bsnm{Hayashi}, \binits{C.}}:
\bjtitle{Prog. Theor. Phys. Suppl.}
\bvolume{70},
\bfpage{35}
(\byear{1981})
\end{barticle}
\endbibitem

\bibitem[\protect\citeauthoryear{Ida and Lin}{2004}]{IdaLin04}
\begin{barticle}
\bauthor{\bsnm{Ida}, \binits{S.}},
\bauthor{\bsnm{Lin}, \binits{D.N.C.}}:
\bjtitle{\apj}
\bvolume{604},
\bfpage{388}
(\byear{2004})
\end{barticle}
\endbibitem

\bibitem[\protect\citeauthoryear{Ida and Lin}{2008}]{IdaLin08}
\begin{barticle}
\bauthor{\bsnm{Ida}, \binits{S.}},
\bauthor{\bsnm{Lin}, \binits{D.N.C.}}:
\bjtitle{\apj}
\bvolume{685},
\bfpage{584}
(\byear{2008})
\end{barticle}
\endbibitem

\bibitem[\protect\citeauthoryear{Johnson et~al.}{2017}]{Johnson17}
\begin{barticle}
\bauthor{\bsnm{Johnson}, \binits{J.A.}},
\bauthor{\bsnm{Petigura}, \binits{E.A.}},
\bauthor{\bsnm{Fulton}, \binits{B.J.}},
\bauthor{\bsnm{Marcy}, \binits{G.W.}},
\bauthor{\bsnm{Howard}, \binits{A.W.}},
\bauthor{\bsnm{Isaacson}, \binits{H.}},
\bauthor{\bsnm{Hebb}, \binits{L.}},
\bauthor{\bsnm{Cargile}, \binits{P.A.}},
\bauthor{\bsnm{Morton}, \binits{T.D.}},
\bauthor{\bsnm{Weiss}, \binits{L.M.}},
\bauthor{\bsnm{Winn}, \binits{J.N.}},
\bauthor{\bsnm{Rogers}, \binits{L.A.}},
\bauthor{\bsnm{Sinukoff}, \binits{E.}},
\bauthor{\bsnm{Hirsch}, \binits{L.A.}}:
\bjtitle{\aj}
\bvolume{154},
\bfpage{108}
(\byear{2017})
\end{barticle}
\endbibitem

\bibitem[\protect\citeauthoryear{Mayor and Queloz}{1995}]{MayorQueloz95}
\begin{barticle}
\bauthor{\bsnm{Mayor}, \binits{M.}},
\bauthor{\bsnm{Queloz}, \binits{D.}}:
\bjtitle{Nature}
\bvolume{378},
\bfpage{355}
(\byear{1995})
\end{barticle}
\endbibitem

\bibitem[\protect\citeauthoryear{Owen and Wu}{2017}]{OwenYanqin17}
\begin{barticle}
\bauthor{\bsnm{Owen}, \binits{J.E.}},
\bauthor{\bsnm{Wu}, \binits{Y.}}:
\bjtitle{\apj}
\bvolume{847},
\bfpage{29}
(\byear{2017})
\end{barticle}
\endbibitem

\bibitem[\protect\citeauthoryear{Petigura et~al.}{2017}]{Petigura17}
\begin{barticle}
\bauthor{\bsnm{Petigura}, \binits{E.A.}},
\bauthor{\bsnm{Howard}, \binits{A.W.}},
\bauthor{\bsnm{Marcy}, \binits{G.W.}},
\bauthor{\bsnm{Johnson}, \binits{J.A.}},
\bauthor{\bsnm{Isaacson}, \binits{H.}},
\bauthor{\bsnm{Cargile}, \binits{P.A.}},
\bauthor{\bsnm{Hebb}, \binits{L.}},
\bauthor{\bsnm{Fulton}, \binits{B.J.}},
\bauthor{\bsnm{Weiss}, \binits{L.M.}},
\bauthor{\bsnm{Morton}, \binits{T.D.}},
\bauthor{\bsnm{Winn}, \binits{J.N.}},
\bauthor{\bsnm{Rogers}, \binits{L.A.}},
\bauthor{\bsnm{Sinukoff}, \binits{E.}},
\bauthor{\bsnm{Hirsch}, \binits{L.A.}},
\bauthor{\bsnm{Crossfield}, \binits{I.J.M.}}:
\bjtitle{\aj}
\bvolume{154},
\bfpage{107}
(\byear{2017})
\end{barticle}
\endbibitem

\bibitem[\protect\citeauthoryear{Roberts et~al.}{1987}]{Roberts87}
\begin{barticle}
\bauthor{\bsnm{Roberts}, \binits{D.H.}},
\bauthor{\bsnm{Lehar}, \binits{J.}},
\bauthor{\bsnm{Dreher}, \binits{J.W.}}:
\bjtitle{\aj}
\bvolume{93},
\bfpage{968}
(\byear{1987})
\end{barticle}
\endbibitem

\bibitem[\protect\citeauthoryear{Roques and Schneider}{1995}]{Schneider}
\begin{botherref}
\oauthor{\bsnm{Roques}, \binits{F.}},
\oauthor{\bsnm{Schneider}, \binits{J.}}:
1995
The Extrasolar Planets Encyclopaedia.
www.exoplanet.eu
\end{botherref}
\endbibitem

\bibitem[\protect\citeauthoryear{Schlaufman et~al.}{2009}]{Schlaufman09}
\begin{barticle}
\bauthor{\bsnm{Schlaufman}, \binits{K.C.}},
\bauthor{\bsnm{Lin}, \binits{D.N.C.}},
\bauthor{\bsnm{Ida}, \binits{S.}}:
\bjtitle{\apj}
\bvolume{691},
\bfpage{1322}
(\byear{2009})
\end{barticle}
\endbibitem

\bibitem[\protect\citeauthoryear{Schneider et~al.}{2011}]{Schneider11}
\begin{barticle}
\bauthor{\bsnm{Schneider}, \binits{J.}},
\bauthor{\bsnm{Dedieu}, \binits{C.}},
\bauthor{\bsnm{Sidaner}, \binits{P.L.}},
\bauthor{\bsnm{Savalle}, \binits{R.}},
\bauthor{\bsnm{Zolotukhin}, \binits{I.}}:
\bjtitle{\aap}
\bvolume{532},
\bfpage{79}
(\byear{2011})
\end{barticle}
\endbibitem

\bibitem[\protect\citeauthoryear{Schwarzenberg-Czerny}{1998}]{SchwCzerny98b}
\begin{barticle}
\bauthor{\bsnm{Schwarzenberg-Czerny}, \binits{A.}}:
\bjtitle{Baltic Astron.}
\bvolume{7},
\bfpage{43}
(\byear{1998})
\end{barticle}
\endbibitem

\bibitem[\protect\citeauthoryear{Skuljan et~al.}{1999}]{Skuljan99}
\begin{barticle}
\bauthor{\bsnm{Skuljan}, \binits{J.}},
\bauthor{\bsnm{Hearnshaw}, \binits{J.B.}},
\bauthor{\bsnm{Cottrell}, \binits{P.L.}}:
\bjtitle{\mnras}
\bvolume{308},
\bfpage{731}
(\byear{1999})
\end{barticle}
\endbibitem

\bibitem[\protect\citeauthoryear{S{\"u}veges}{2014}]{Suveges14}
\begin{barticle}
\bauthor{\bsnm{S{\"u}veges}, \binits{M.}}:
\bjtitle{\mnras}
\bvolume{440},
\bfpage{2099}
(\byear{2014})
\end{barticle}
\endbibitem

\end{thebibliography}

%% Non-BibTeX  (Name-Year style)
%
% \begin{thebibliography}{}
% \bibitem[\protect\citeauthoryear{<author>}{<year>]{ref:?}
%    <ref. entry>
% \bibitem[\protect\citeauthoryear{<author>}{<year>]{ref:?}
%    <ref. entry>
% \end{thebibliography}

\end{document}